\documentclass[11pt]{article} 

\usepackage[utf8]{inputenc} 

\usepackage{geometry} 
\usepackage{graphicx} 

\usepackage{amssymb,amsmath}
\usepackage{algorithm}
\usepackage{algpseudocode}
\def\NoNumber#1{{\def\alglinenumber##1{}\State #1}\addtocounter{ALG@line}{-1}}

\usepackage{booktabs} 
\usepackage{array} 
\usepackage{paralist} 
\usepackage{verbatim} 
\usepackage{subfig} 

\usepackage{caption}
\usepackage[numbers]{natbib}

\usepackage{fancyhdr} 
\usepackage{sectsty}
\allsectionsfont{\sffamily\mdseries\upshape} 

\usepackage{xcolor}
\usepackage[multiple]{footmisc}
\usepackage[hyperfootnotes=false,colorlinks,filecolor=blue,citecolor=cyan]{hyperref}

\providecommand\abstractname{Abstract}
\def\abstract{}

\renewenvironment{abstract}{%
  \centering\small
  \textbf\abstractname
  \list{}{\leftmargin1.0cm \rightmargin\leftmargin}
  \item\relax
}{%
  \endlist \par\bigskip
}

\usepackage{soul}
\usepackage{cancel}
\usepackage{enumitem}

\graphicspath{{./}}

\newcommand{\fnsep}{$^{,}$}
\newcommand{\fnsept}{$\!$}

\newcommand{\referenceSet}{X}
\newcommand{\targetSet}{Y}
\newcommand{\referenceDensity}{\rho}
\newcommand{\referenceMeasure}{\nu_{\referenceDensity}}
\newcommand{\globalTargetDensity}{\Psi} 
\newcommand{\targetDensity}{\psi} 
\newcommand{\targetMeasure}{\nu_{\targetDensity}}
\newcommand{\di}[1]{\mathrm{d}{{#1}}} 
\newcommand{\priorClassDensity}{\widehat{\targetDensity}}
\newcommand{\rv}[1]{\mathcal{{#1}}}

\newcommand{\gsedataset}{GSE43151}
\newcommand{\pathwayzlSD}{hsa04650}
\newcommand{\Yt}{\rv{Y}} 
\newcommand{\rvsubset}{(\rv{Y}_{j})_{j=1}^{6}}
\newcommand{\sigmap}[1]{\sigma_{{#1}}}
\newcommand{\sigmapinv}[1]{\sigma_{{#1}}^{-1}}
\newcommand{\ps}[1]{\sigma_{{#1}}(\Yt)}

\newcommand{\psInv}[2]{\sigma_{{#1}}^{-1}({#2})}
\newcommand{\rvSubset}[2]{(\rv{Y}_{{#1}})_{{#2}=1}^{6}}

\newcommand{\nperm}{n_{\mathrm{perm}}}
\newcommand{\mymatrix}[1]{\mathrm{{#1}}}
\newcommand{\FCperm}[2]{\mymatrix{B}_{{#1}}^{{#2}}}
\newcommand{\FCpermId}[2]{\mymatrix{P}_{{#1}}^{{#2}}}
\newcommand{\FCall}[2]{\mymatrix{Q}_{{#1}}^{{#2}}}
\newcommand{\TkNr}{T_{k}^{(N,r)}}
\newcommand{\invTkNr}{\left(T_{k}^{(N,r)}\right)^{-1}}
\newcommand{\targetkNr}{\targetDensity_{k}^{(N,r)}}
\newcommand{\jj}{\tilde{\mbox{\textit{\j}}}}
\newcommand{\ii}{\tilde{\mbox{\textit{\i}}}}
\newcommand{\acomment}[1]{\small{#1}}
\newcommand{\mto}{\textbf{to}}
\newcommand{\cf}{\textit{cf}}

\title{Density Estimation via Measure Transport: Outlook for Applications in the Biological Sciences\footnote{ Work supported by the U.S. Department of Energy, Office of Science, RadBio program under Award KP1601011/FWP CC121. The views and opinions of authors expressed herein do not necessarily state or reflect those of the United States Government or any agency thereof.}}
\author{Vanessa L{\'o}pez-Marrero\thanks{Corresponding author. E-mail: vlopezmar@bnl.gov} \fnsep
               \footnote{Computational Science Initiative, Brookhaven National Laboratory} \fnsep
               \footnote{Institute for Advanced Computational Science (IACS), Stony Brook University}  
    \and Patrick R. Johnstone\footnotemark[3] \fnsep \thanks{Current affiliation: Meta}
    \and Gilchan Park\footnotemark[3]\fnsept
    \and Xihaier Luo\footnotemark[3]\fnsept
} 

\begin{document}
\maketitle

\begin{abstract}
One among several advantages of measure transport methods is that they allow for a unified framework for processing and analysis of data distributed according to a wide class of probability measures.  Within this context, we present results from computational studies aimed at assessing the potential of measure transport techniques, specifically, the use of triangular transport maps, as part of a workflow intended to support research in the biological sciences.   Scenarios characterized by the availability of limited amount of sample data, which are common in domains such as radiation biology, are of particular interest.  We find that when estimating a distribution density function given limited amount of sample data, adaptive transport maps are advantageous.  In particular, statistics gathered from computing series of adaptive transport maps,  trained on a series  of randomly chosen subsets of the set of available data samples, leads to uncovering information hidden in the data.  As a result, in the radiation biology application considered here, this approach provides a tool for  generating hypotheses about gene relationships and their dynamics under radiation exposure.
\vspace*{2ex} \\
\textit{Keywords:} density estimation, computational measure transport, triangular transport maps, gene expression data, biostatistics, statistical learning
\end{abstract}

\section{Introduction}
\label{sec:introduction}

The problem of estimating a probability distribution density from samples (e.g., observations, measurements, or simulation data) is ubiquitous in data science, uncertainty quantification, clustering and classification, and probabilistic modeling and inference tasks. Moreover, it is common among various scientific and engineering fields, including biology \cite{bk:silverman1998density, bk:hastie2009elements, agnelli2010, tabak_turner_2013, boluki2017, deTorrente_2020, ghanegolmohammadi2022, spantini2022, theiler2022}.  Often, well-known parametric density functions (dependent on few parameters), such as the Gaussian or Weibull density distribution functions, are adopted.  While this may simplify certain tasks (e.g., computational ones), many of these known density distribution functions are not necessarily suitable for characterizing data that exhibit complex features, such as (spatial and/or temporal) correlations and non-Gaussian characteristics.  

Furthermore, in practice it is often not known beforehand which is the underlying probability distribution density that typifies a given set of samples.  Having a common density estimation framework, that can seamlessly process Gaussian and non-Gaussian data alike, is thus clearly advantageous.  For instance, as reported in \cite{deTorrente_2020}, accounting for differences in the distribution densities of gene expressions can lead to improved interpretation of cancer transcriptomic data.  This is just one situation (and one can envision many) where it would be beneficial to avoid the task of having to ``hand-pick'', for each of a number of different data sets, the best underlying density among a list of candidate (parametric or non-parametric) distribution densities.  Hence, a unified density estimation framework capable of characterizing a diverse range of properties is highly desirable.  A measure transport approach \cite{bk:villani_old_new, bk:santambrogio2015optimal, bk:comput_opt_transport} offers this possibility.

Optimal measure transport, broadly defined, deals with the problem of minimizing the cost of transporting one (probability) measure to another. The subject has a long history with significant impact in many fields of mathematics, such as probability theory, optimization, and partial differential equations, among others \cite{bk:villani_old_new, bk:santambrogio2015optimal}. Current interest on the topic stems from its potential in numerous areas, including density estimation, data science, machine learning, Bayesian inference, sampling, and data assimilation \cite{agnelli2010, tabak_vanden-eijnden_2010, tabak_turner_2013, marzouk2016, RealNVP, spantini2018inference, bk:comput_opt_transport, papamakarios2021, gabrie2022, spantini2022, yang_tabak_2022, katzfuss2023}. Among various measure transport techniques under active research, we mention the use of triangular transport maps, normalizing flows, gradient flows, diffusion maps, and invertible neural networks. Although differing in the particulars, affording a unified framework for processing and analysis of data from a range of distributions (Gaussian, non-Gaussian, uni-modal, multi-modal, etc.) is a powerful property common to all.

In this paper we consider measure transport for density estimation from data, in particular the use of triangular transport maps \cite{kolesnikov2004,bogachev2005,marzouk2016,baptista2022_ATM}, envisioning its potential within a computational workflow for tasks associated with biological applications.  Contributions of this paper are summarized in Section \ref{sec:contributions} below.  As described briefly next and in more detail in Sections \ref{sec:basics_measure_transport} and \ref{sec:TM_computation}, adoption of a measure transport framework for density estimation (given data samples) provides several advantages.  We have already mentioned that it avoids the need to ``hand-pick'' among a list of candidate densities and that it can be used, seamlessly, for Gaussian and non-Gaussian data alike.  In addition:
\begin{itemize}[topsep=0pt, partopsep=0pt, itemsep=-3pt]
    \item It yields an explicit formula for the unknown probability distribution density we wish to estimate.  As will be described in Sections \ref{sec:basics_measure_transport} and \ref{sec:TM_computation}, once a suitable \textit{transport map} is computed, or \textit{learned}, evaluation of the unknown density at a given sample is simply done via said formula. (This will be defined precisely in Section \ref{sec:basics_measure_transport}, equation \eqref{eq:pullback_reference}; see also equations \eqref{eq:target_density_intro1}--\eqref{eq:target_density_intro2}  below.)
    \item Generative modeling capabilities are enabled in the unknown probability space because, via the learned transport map, one can transport samples from a known probability measure to the unknown probability measure of interest.
    \item If new samples from the unknown measure become available, data assimilation is facilitated without the need to learn (or re-train) from scratch the density estimation model (which, within the present context, is the transport map).
\end{itemize}
Moreover, some techniques, such as the adoption of triangular transport maps, allow one to infer information regarding the dependence structure among random variables in the unknown probability space.  We make use of this property in one methodology we propose and develop, which is briefly described further below in Section \ref{sec:contributions} and discussed in detail in Section \ref{sec:ATM_tool_scientific_discovery}.

The mathematical foundations of measure transport, as they relate to the problem of density estimation we are concerned with, will be discussed in Sections \ref{sec:basics_measure_transport} and \ref{sec:TM_computation}.  Here we introduce the setting, in a nutshell.  Let $\rv{Y}=\left(\rv{Y}_{1},\rv{Y}_{2},\dots,\rv{Y}_{m}\right) \in \mathbb{R}^{m}$ be a vector of $m$ random variables jointly distributed according to an \textit{unknown} probability measure $\targetMeasure$ defined on the $\sigma$-algebra $\mathcal{B}(\mathbb{R}^{m})$ of Borel sets of $\mathbb{R}^{m}$ and absolutely continuous with respect to the Lebesgue measure.  Let $\targetDensity$ denote the (unknown) distribution density function of $\targetMeasure$.  It is the unknown measure $\targetMeasure$ and its density $\targetDensity$ that we wish to estimate.   We thus refer to $\targetMeasure$ and $\targetDensity$, respectively, as the \textit{target measure} and \textit{target density}.  Under the assumption that $\targetMeasure$ is absolutely continuous with respect to the Lebesgue measure,\footnote{As will be discussed in Section \ref{sec:basics_measure_transport},  we consider only measures that are absolutely continuous with respect to the Lebesgue measure.  However, the subject of measure transport is not restricted to such.  It is outside the scope of this paper to treat other types of measures (such as discrete ones), but the reader may refer to \cite{bk:villani_old_new, bk:santambrogio2015optimal,bk:comput_opt_transport} and references therein for more information.} one may identify $\targetMeasure$ with its probability density $\targetDensity$.  In other words, we have 
\begin{eqnarray}  \label{eq:acwrtLebesgue}
    \int_{B}  \targetMeasure(\di{y})  &  = &   \int_{B}  \targetDensity(y) \, \di{y} \, ,
\end{eqnarray}
for every $B\in\mathcal{B}(\mathbb{R}^{m})$.  Say we are given a set $\targetSet = \{y \in \mathbb{R}^{m} \,:\, y \sim \targetMeasure\}$ of samples $y \in \mathbb{R}^{m}$, which are values assumed by the random vector $\rv{Y}=\left(\rv{Y}_j\right)_{j=1}^{m}$.  The general problem is to estimate the unknown target measure $\targetMeasure$, given the set of samples $\targetSet$.  By \eqref{eq:acwrtLebesgue}, to estimate the unknown target measure $\targetMeasure$, it thus suffices to estimate its unknown probability density function $\targetDensity$.

Now, to estimate the unknown target probability density function $\targetDensity$ using a measure transport approach, let $\rv{X}=\left(\rv{X}_{1},\rv{X}_{2},\dots,\rv{X}_{m}\right) \in \mathbb{R}^{m}$ be a vector of $m$ random variables jointly distributed according to a \textit{known} probability measure $\referenceMeasure$ on $\mathcal{B}(\mathbb{R}^{m})$ (assumed to be absolutely continuous with respect to the Lebesgue measure).  Call $\referenceMeasure$ the \textit{reference measure} and let $\referenceDensity$ represent the (known) probability density function of $\referenceMeasure$.   As will be discussed  in Sections \ref{sec:basics_measure_transport} and \ref{sec:TM_computation}, for practical purposes the reference measure $\referenceMeasure$ should be chosen so that both evaluating its density $\referenceDensity$ and drawing samples from $\referenceMeasure$ can be done efficiently.  A common candidate for the reference measure is the $m$-variate standard normal (i.e., Gaussian) measure $\mathcal{N}(0,I)$, and that is our choice in the present work.

\begin{figure}[htbp]
    \centering
    \includegraphics[width=0.9\textwidth]{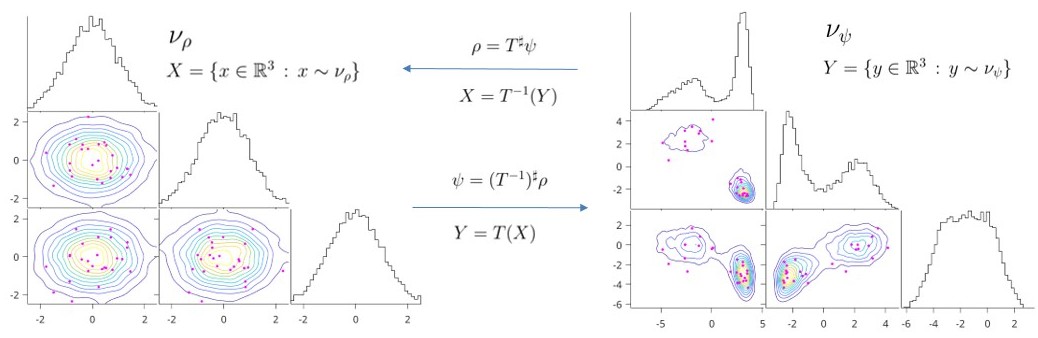}
    \caption{The map $T\!:\! \mathbb{R}^{3} \!\rightarrow\! \mathbb{R}^{3}$ transports samples from $\referenceMeasure$, a three-variate standard normal (i.e., Gaussian) measure $\mathcal{N}(0,I)$ with density $\referenceDensity(x) = (2\pi)^{-3/2}\,e^{{-||x||^{2}}/2}$, to samples from $\targetMeasure$, a three-variate measure, with multi-modal density $\targetDensity$. The density $\referenceDensity$ is the pullback $T^{\sharp}\targetDensity$ of the density $\targetDensity$, and so we have $\referenceDensity(x) =  T^{\sharp}\targetDensity(x)  =  \targetDensity(T(x)) \, \det( J_{T}(x))$, where $J_{T}$ is the Jacobian matrix of $T$. Conversely, the map $T^{-1}$ transports samples from $\targetMeasure$ to samples from $\referenceMeasure$. The density $\targetDensity$ is the pullback $(T^{-1})^{\sharp}\referenceDensity$ of the density $\referenceDensity$, and so $\targetDensity(y)  =  (T^{-1})^{\sharp}\referenceDensity(y)  =  \referenceDensity(T^{-1}(y)) \, \det( J_{T^{-1}}(y))$. Contour plots of the corresponding one- and two-dimensional marginal distribution densities also are shown.}
    \label{fig:measure_transport_nutshell}
\end{figure}

The central problem then becomes to compute, or learn, an invertible map $T\!:\! \mathbb{R}^{m} \!\rightarrow\! \mathbb{R}^{m}$ that ``couples'' the measures $\referenceMeasure$ and $\targetMeasure$, where the meaning of the term ``couples'' will be made precise, mathematically, in Section \ref{sec:basics_measure_transport} (equation \eqref{eq:measure_transport}).  Here is suffices to note that the map $T$ and its inverse $T^{-1}$ \textit{transport} samples between $\referenceMeasure$ and $\targetMeasure$.  That is, $T$ maps any given sample $x$ from $\referenceMeasure$ to a sample $y = T(x)$ from $\targetMeasure$ and, conversely, $T^{-1}$ maps any given sample $y$ from $\targetMeasure$ to a sample $x = T^{-1}(y)$ from $\referenceMeasure$.  Furthermore, the measure transport framework yields explicit formulas for the densities (more on this in Section \ref{sec:basics_measure_transport}, equations \eqref{eq:pullback_target}--\eqref{eq:pullback_reference}; also \eqref{eq:target_density_intro1}--\eqref{eq:target_density_intro2} below).  These fundamental concepts are illustrated pictorially in Figure \ref{fig:measure_transport_nutshell} and expanded upon in Section \ref{sec:basics_measure_transport}.  Details regarding the mathematical representation and numerical computation of transport maps will be provided in Section \ref{sec:TM_computation} as well as in Appendix \ref{app:ATM_framework}.

As will be discussed in Section \ref{sec:basics_measure_transport}, within the measure transport framework the unknown target density $\targetDensity$ of interest to us is given explicitly by the formula
\begin{eqnarray}  \label{eq:target_density_intro1}
     \targetDensity(y)  &  = &  \referenceDensity(T^{-1}(y)) \, \det(J_{T^{-1}}(y)) \, ,
\end{eqnarray}
where $J_{T^{-1}}$ is the Jacobian matrix of $T^{-1}$ and $\referenceDensity$ is the chosen, known reference density function.  In particular, when the reference measure $\referenceMeasure$ is chosen to be the $m$-variate standard normal (i.e., Gaussian) measure $\mathcal{N}(0,I)$ with density $\referenceDensity(x) = (2\pi)^{-m/2}\,e^{-||x||^{2}/2}$, where $||\boldsymbol{\cdot}||$ denotes the Euclidean norm on $\mathbb{R}^{m}$, formula \eqref{eq:target_density_intro1} for the target density $\targetDensity$ becomes
\begin{eqnarray}  \label{eq:target_density_intro2}
     \targetDensity(y)  &  = &  (2\pi)^{-m/2} \, e^{-||T^{-1}(y)||^{2}\,/\, 2} \, \det(J_{T^{-1}}(y)) \, .
\end{eqnarray}
Hence, once the transport map $T$ is learned, given a (new) sample $y$, the value $\targetDensity(y)$ of the target density is thus obtained via equation \eqref{eq:target_density_intro1}, since all terms on the right-hand-side of \eqref{eq:target_density_intro1} are known and computable.  In addition, as previously noted, a sample $x$ drawn from the (known) reference measure $\referenceMeasure$ yields a sample $y = T(x)$ from the (unknown) target measure $\targetMeasure$.  If the reference measure $\referenceMeasure$ is purposefully chosen so that sampling from it can be done efficiently, and since one can now transport samples from the reference to the target via the learned transport map $T$, samples from the target measure $\targetMeasure$ can be generated efficiently as well (as long as the transport map $T$ can be evaluated efficiently).  Hence, by construction, tasks requiring probabilistic modeling, inference, and statistical analysis in the target space are enabled via a measure transport framework.

\subsection{Contributions of this Paper}
\label{sec:contributions}

As previously noted, we consider the problem of measure transport for density estimation from data, in particular the use of triangular transport maps, envisioning its potential within a computational workflow for tasks associated with biological applications.  The context is density estimation (or prior/posterior construction) when the number of  data samples for model training is limited in amount.  More precisely, we are concerned with sizes of data sets of order $O(1)$--$O(10)$ samples, which are common in the radiation biology domain, yet pose challenges when faced with the problem of estimating distribution densities from data (whether or not measure transport methods are employed for the purpose).  The motivation for our work will be described in Section \ref{sec:motivating_application}, along with the proposed computational workflow for biological research. We note, however, that the methods we investigate and develop are not restricted to the particular application motivating our research.

We show that when estimating a probability density function given limited amount of sample data for model training, \emph{and without any a priori assumptions about the structure of the target probability space}, adaptive transport maps  \cite{baptista2022_ATM} are advantageous.  As will be explained in Section \ref{sec:ATM_experiments}, due to the small number ($O(1)$--$O(10)$) of samples at our disposal for model training, it was not sufficient to rely on learning a single (adaptive) transport map using the totality of samples available for model training.  However, we introduce a method whereby statistics gathered from learning series of adaptive transport maps, trained on a series of randomly chosen subsets of the set of available data samples, offers a viable alternative in robustly revealing the structure of the target measure by distinguishing a statistically dominant triangular transport map with a specific dominant component structure.   Precisely, we find that with our proposed randomized technique it is indeed possible to infer information about the dependence structure among the random variables in the unknown (multi-variate) target space.  As a result, in the radiation biology application considered in Section \ref{sec:ATM_experiments}, this approach provides a tool for  generating hypotheses about gene relationships and their dynamics under radiation exposure.  Details are expanded upon in Section \ref{sec:ATM_tool_scientific_discovery} and future research directions stemming from our studies are discussed in Section \ref{sec:discussion}.

The measure transport techniques we consider (and to be described in Sections \ref{sec:basics_measure_transport}--\ref{sec:TM_computation}) are at present actively researched but, to the best of our knowledge, there are few studies in the literature that have examined scenarios with the limited number of samples for model training which characterizes our motivating application.  The adaptive transport maps approach proposed in \cite{baptista2022_ATM} was shown by its authors to perform well for density estimation problems when training on data sets of various sizes (mostly $O(100)$ samples or larger).  In particular, good performance was reported for data sets with $O(100)$ samples, when compared to the conditional kernel density estimation method and the $\epsilon$-neighborhood kernel density estimation method, using four data sets from the UC Irvine Machine Learning Repository \cite{UCI_ML_repository}.   It is thus one of the measure transport techniques we evaluate and build upon for our purposes.

\subsection{Organization of the Paper}
\label{sec:paper_organization}

The rest of the paper is organized as follows.  In Section \ref{sec:motivating_application} we introduce the application motivating our research.  A discussion of the foundational mathematical and computational concepts from the subject of measure transport is provided in Sections \ref{sec:basics_measure_transport}--\ref{sec:TM_computation}, as well as in Appendix \ref{app:ATM_framework}.  The mathematical theory is covered in Section \ref{sec:basics_measure_transport}, whereas Section \ref{sec:TM_computation} and Appendix \ref{app:ATM_framework} focus on numerical methods and computational aspects.  Section \ref{sec:density_estimation_from_data} includes generic examples (i.e., not specific to the application motivating our work) which illustrate the measure transport framework for density estimation, seamlessly used to process both Gaussian and non-Gaussian data sets.

In Section \ref{sec:application_GSE_dataset} and the accompanying Appendix \ref{app:ATM_experiments} we present results from our computational experiments, aimed at assessing the performance of select measure transport techniques for density estimation for a gene expression data set from the radiation biology domain.  Sections \ref{sec:DAG_experiments} and \ref{sec:ATM_classification} focus on phenotypic classification using measure transport densities estimated from the data set, whereas Section \ref{sec:ATM_tool_scientific_discovery} is concerned with the problem of inferring the dependence structure between random variables (which, within the context, are associated with genes) in the unknown target probability space.  (In other words, the problem of inferring the structure of the unknown target measure and its marginals.)  Finally, Section \ref{sec:discussion} contains a discussion of our results and future research directions.

\section{Motivating Application}
\label{sec:motivating_application}

\begin{figure}[t] 
    \centering
    {{\includegraphics[trim=0.0cm 0.0cm 0.0cm 0.0cm, clip=true, width=0.8\textwidth]{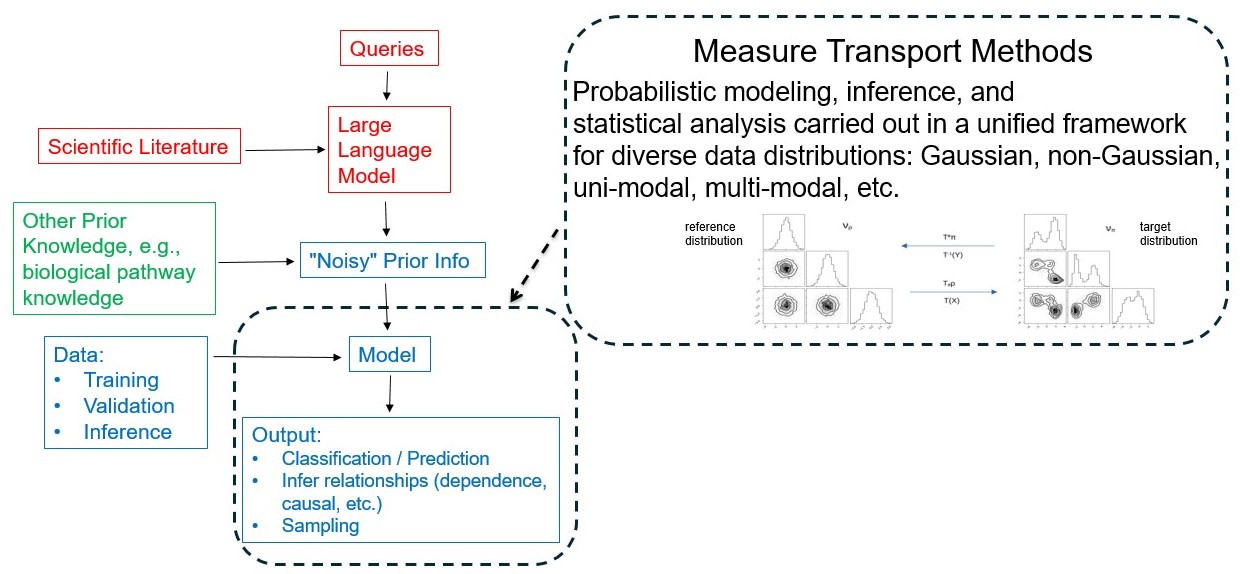}}}
    \caption{Computational workflow for probabilistic modeling, inference, and statistical analysis.  To account for availability of limited number of data samples, model training can be augmented with prior knowledge.  Such prior knowledge may come from large language models or human experts, for instance.  Measure transport techniques allow for a unified framework for density estimation and processing of data exhibiting diverse characteristics.}
    \label{fig:proposed_workflow}
\end{figure}

The present work stems from the conception of a comprehensive computational workflow intended to support research in the biological sciences, in particular the radiation biology domain.  Within this effort, our interest is in enabling a \textit{unified} framework where probabilistic modeling, inference, and statistical analysis can be carried out for a wide range of data distributions.  As noted in the Introduction (Section \ref{sec:introduction}), measure transport offers the possibility of such unified approach.  We envision measure transport techniques comprising one component within the larger scope, as depicted in Figure \ref{fig:proposed_workflow}.

One motivating problem within the workflow from Figure \ref{fig:proposed_workflow} is that of probability distribution density estimation for phenotypic classification problems, for example, for inferring class labels, such as radiation exposure levels or cancer types, given sample data.  Normalizing flows (one kind among the various measure transport methods) have been used successfully for clustering and classification in the cancer domain \cite{agnelli2010}.  Now, when the quantity of data samples available for model training is limited in amount, regardless of the density estimation technique adopted, prior knowledge (associated with the particular application) may be used in conjunction with data to estimate the class-conditional densities.  In the work \cite{boluki2017}, the authors proposed incorporation of prior knowledge from biological pathways for class-conditional density estimation as one approach for optimal Bayesian classifier design when data availability is limited.  Here we also consider the use of prior knowledge from biological pathways, along with measure transport techniques, in the computational experiments reported in Sections \ref{sec:DAG_experiments} and \ref{sec:ATM_classification} for phenotypic classification problems, for which we employ the classification technique described shortly below (via equation \eqref{eq:classify} and the associated explanation).  In the workflow from Figure \ref{fig:proposed_workflow}, we envision prior knowledge coming from other sources as well, such as large language models \cite{park2023automated,park2023comparative} and human experts.  This is the subject of future work.

It is outside the scope of the present paper to cover the subject of classification itself and the various techniques available for classifying samples into different classes.  The reader is referred to \cite{bk:hastie2009elements} for a thorough treatment of the topic of classification via probability density distribution functions (kernel density classification, per the terminology from \cite{bk:hastie2009elements}), among many other classification methods.  Following \cite{bk:hastie2009elements}, we summarize next the technique we employed for the classification results reported in Sections \ref{sec:density_estimation_from_data}, \ref{sec:DAG_experiments}, and \ref{sec:ATM_classification}.  The particular technique was selected because it relies on evaluation of class-conditional densities given samples and thus fits into our proposed framework, for which explicit formulas for the class-conditional densities would be available for evaluation (refer to equation \eqref{eq:target_density_intro1} and the associated discussion).

Assume we have $K$ classes to infer sample membership from.  Upon computing, separately (and with any suitable density estimation method), an estimated probability density distribution function $\targetDensity_{k}$ for each class $k$, where $k = 0,\ldots,K-1$, the probability $\mathrm{Pr}(C = k \, | \, \rv{Y} = y)$ that a given sample $y$ belongs to class $k$ can be obtained from the class-conditional densities $\{\targetDensity_{l}\}_{l = 0}^{K-1}$ via
\begin{eqnarray}  \label{eq:classify}
    \mathrm{Pr}(C = k \, | \, \rv{Y} = y)  & = &  \frac{\priorClassDensity_{k} \,\targetDensity_{k}(y)}{\sum_{l=0}^{K-1} \priorClassDensity_{l} \,\targetDensity_{l}(y)} \, ,
\end{eqnarray}
where $\{\priorClassDensity_{l}\}_{l = 0}^{K-1}$, $\sum_{l = 0}^{K-1} \priorClassDensity_{l} = 1$, are estimates for the class prior probabilities.  The estimates for the class prior probabilities may be determined from the proportion of training samples, for instance, or from other prior knowledge.  In Sections \ref{sec:density_estimation_from_data}, \ref{sec:DAG_experiments}, and \ref{sec:ATM_classification}, we compare our classification results with those obtained using a naive Bayes classifier, for which the class-conditional densities in \eqref{eq:classify} were estimated by fitting the training data to a Gaussian distribution, and with a support vector machine classifier, an effective classification method (not for density estimation) that constructs hyperplanes that approximate the decision boundaries separating the different underlying classes in order to perform classification of samples \cite{bk:hastie2009elements}.

Finally, in Section \ref{sec:ATM_tool_scientific_discovery} we discuss how, as part of the computational framework from Figure \ref{fig:proposed_workflow},  measure transport methods have the potential to address other problems, in addition to the above-mentioned classification one.  More precisely, by using the randomization technique we propose in Section \ref{sec:ATM_tool_scientific_discovery} to infer the dependence structure of the underlying unknown probability measure (i.e., dependence relations between the random variables), adaptive transport maps \cite{baptista2022_ATM} could provide a means for generating hypotheses regarding gene relationships for the radiation biology application under consideration.  Before getting into such details, we now continue with an overview of required topics from measure transport theory.

\section{Basics of Measure Transport}
\label{sec:basics_measure_transport}

As mentioned in the Introduction (Section \ref{sec:introduction}), optimal measure transport addresses the problem of minimizing the cost of transporting one probability measure to another.\footnote{Recall from Section \ref{sec:introduction} that we consider only measures that are absolutely continuous with respect to the Lebesgue measure.}  We now give a summary of the basic concepts, as they relate to the problem of density estimation which is of interest to us.  For a thorough treatment of the subject of measure transport, the reader may refer to \cite{bk:villani_old_new,bk:santambrogio2015optimal,kolesnikov2004,bogachev2005} (theory), \cite{marzouk2016, bk:comput_opt_transport, baptista2022_ATM} (computational), and references therein.

Informally, given two probability measures $\referenceMeasure$ and $\targetMeasure$ defined on the $\sigma$-algebra of Borel sets of $\mathbb{R}^{m}$, $m \ge 1$, and absolutely continuous with respect to the Lebesgue measure $\di{x}$ on $\mathbb{R}^{m}$, with densities $\referenceDensity$ and $\targetDensity$, respectively, an invertible map $T\! : {\mathbb{R}^{m}} \rightarrow {\mathbb{R}^{m}}$ \textit{transports}  $\referenceMeasure$ to $\targetMeasure$ if and only if
\begin{eqnarray}  \label{eq:measure_transport}
    \targetMeasure({B}) & = & \referenceMeasure(T^{-1}({B}))
\end{eqnarray}
for every Borel set $B\subset\mathbb{R}^{m}$.
One says the map $T$ \textit{pushes forward} $\referenceMeasure$ to $\targetMeasure$, denoted $T_{\sharp}\referenceMeasure \ = \ \targetMeasure$.  Given any sample $x \in\mathbb{R}^{m}$ from $\referenceMeasure$, $T$ maps $x$ to a sample $y = T(x)$ from $\targetMeasure$.  Conversely, the inverse $T^{-1}$ of $T$ pushes forward $\targetMeasure$ to $\referenceMeasure$, denoted $(T^{-1})_{\sharp}\targetMeasure = \referenceMeasure$, and so $T^{-1}$ maps any given sample $y \in\mathbb{R}^{m}$ from $\targetMeasure$ to a sample $x = T^{-1}(y)$ from $\referenceMeasure$.

Now, since we assume that both probability measures $\referenceMeasure$ and $\targetMeasure$ are absolutely continuous with respect to the Lebesgue (volume) measure $\di{x}$ on $\mathbb{R}^{m}$, one may identify them with their probability densities $\referenceDensity$ and $\targetDensity$.  That is,
$\referenceMeasure(\di{x}) = \referenceDensity\, \di{x}$ and $\targetMeasure(\di{x}) = \targetDensity\, \di{x}$ (\cf. equation \eqref{eq:acwrtLebesgue}).
Moreover, we restrict the class of maps $T$ that we consider to the class of differentiable or even infinitely differentiable invertible maps.  With these assumptions one can write the measure transport equation \eqref{eq:measure_transport} in terms of densities. Namely, the density $\referenceDensity$ is the \textit{pullback} $T^{\sharp}\targetDensity$ of the density $\targetDensity$,
\begin{equation}  \label{eq:pullback_target}
    \referenceDensity(\cdot) \ = \  T^{\sharp}\targetDensity(\cdot) \ = \ \targetDensity(T(\cdot)) \, \det( J_{T}(\cdot)) \, ,
\end{equation}
where $J_{T}$ is the Jacobian matrix of $T$.  Conversely, the density $\targetDensity$ is the pullback $(T^{-1})^{\sharp}\referenceDensity$ of the density $\referenceDensity$:
\begin{equation}  \label{eq:pullback_reference}
    \targetDensity(\cdot) \ = \ (T^{-1})^{\sharp}\referenceDensity(\cdot) \ = \ \referenceDensity(T^{-1}(\cdot)) \, \det( J_{T^{-1}}(\cdot)) \, ,
\end{equation}
where $J_{T^{-1}}$ is the Jacobian matrix of $T^{-1}$.  An illustration of these fundamental concepts appeared previously in Section \ref{sec:introduction}, Figure \ref{fig:measure_transport_nutshell}.

Recall that we are concerned with the problem of estimating an unknown probability density function characterizing some given data set of samples or observations, for instance, gene expression data.  Denote the corresponding unknown probability measure by $\targetMeasure$ and call it the \textit{target} measure. On the other hand, let $\referenceMeasure$ be a probability measure amenable to computations, for example, the standard normal (i.e., Gaussian) measure $\mathcal{N}(0,I)$, so one can easily draw samples from it and evaluate its density $\referenceDensity$. Call this the \textit{reference} measure.  Now, let $\rv{X}=\left(\rv{X}_{1},\rv{X}_{2},\dots,\rv{X}_{m}\right)$ and $\rv{Y}=\left(\rv{Y}_{1},\rv{Y}_{2},\dots,\rv{Y}_{m}\right)$ be random vectors with probability distribution functions $F_{\rv{X}}$ and $F_{\rv{Y}}$, respectively, such that $\di{F_{\rv{X}}}=\referenceMeasure$ and $\di{F_{\rv{Y}}}=\targetMeasure$.  The random vectors $\rv{X}$, $\rv{Y}$ assume, respectively, values $x,y \in \mathbb{R}^{m}$ (also denoted by $x \sim \referenceMeasure$ and $y \sim \targetMeasure$). Then if  $\referenceSet = \{x \in \mathbb{R}^{m} \,:\, x \sim \referenceMeasure\}$ and $\targetSet = \{y \in \mathbb{R}^{m} \,:\, y \sim \targetMeasure\}$ are sets of samples formed from the values of the random vectors $\rv{X}$ and $\rv{Y}$  (i.e., the elements in $\referenceSet$ are distributed according to $\referenceMeasure$, and those in $\targetSet$ are distributed according to $\targetMeasure$) the transport map should satisfy $\targetMeasure(\targetSet) \ = \ \referenceMeasure(T^{-1}(\targetSet))$ (by virtue of the measure transport equation \eqref{eq:measure_transport}). 

Note that the set $\targetSet$ consists of samples (usually taken from experiment) of the random vector $\rv{Y}$ distributed according to an \textit{a priori unknown} target probability measure $\targetMeasure$, while the reference probability measure $\referenceMeasure$ is \textit{a priori known} (e.g., Gaussian) and the set $\referenceSet$ is generated from samples of a random vector $\rv{X}$ distributed according to this known reference measure $\referenceMeasure$.  Thus, given the sets of samples $\referenceSet$ and $\targetSet$, our goal is to \textit{learn} (i.e., compute) the transport map $T$ and use it later as a predictor for the values of $\rv{Y}$ and for evaluation of the target density $\targetDensity$ (and therefore to serve as an estimator of the target measure $\targetMeasure$).  Under the assumption that one can compute such a map $T$ transporting between the target and reference measures, the framework outlined herein and depicted in Figure \ref{fig:measure_transport_nutshell} yields a method for approximating the desired unknown density function that characterizes the given sample data set $\targetSet$. Then, the central problem is to compute, or learn, such a map $T$. This problem is further addressed in Section \ref{sec:TM_computation} below.

Upon learning such a map $T$ from data, evaluation of the (unknown) density $\targetDensity$ at a given (new) sample $y \sim \targetMeasure$ can be done via equation \eqref{eq:pullback_reference} as, by construction, all the functions on the right-hand side of equation \eqref{eq:pullback_reference} are known and can be evaluated. Moreover, because the reference measure is chosen to make sampling from it an easy and efficient task, sampling from the target measure can be done efficiently by drawing samples from the reference and transporting them to the target via the learned map $T$. Thus, as previously noted in Section \ref{sec:introduction}, probabilistic modeling, inference, and statistical analysis in the (unknown) target space are enabled by such measure transport framework.

\section{Transport Map Computation}
\label{sec:TM_computation}

To compute or, equivalently, learn the transport maps we seek, we use software \cite{TransportMaps_library,ATM_library} made available by the {MIT} Uncertainty Quantification (MUQ) Group \cite{MIT_UQ}.  The reader is referred to the original publications \cite{elmoselhy2012,marzouk2016,spantini2018inference,baptista2022_ATM} for complete descriptions of the numerical methods and software libraries.  A summary is provided next, to make the present manuscript self-contained.

Again, let $\referenceMeasure$ and $\targetMeasure$ be two probability measures defined on the $\sigma$-algebra of Borel sets of $\mathbb{R}^{m}$, $m \ge 1$. Suppose that $T:\mathbb{R}^{m} \rightarrow \mathbb{R}^{m}$ is a transport map pushing forward $\referenceMeasure$ to $\targetMeasure$. That is, it satisfies the equation (see the discussion surrounding equation \eqref{eq:measure_transport} in Section \ref{sec:basics_measure_transport})
\begin{eqnarray}  \label{eq:pushforward_reference}
    \targetMeasure & = & T_{\sharp}\referenceMeasure 
\end{eqnarray}
for $T$.  By definition,
\begin{eqnarray}  \label{eq:pushforward_target}
    \referenceMeasure & = & (T^{-1})_{\sharp}\targetMeasure 
\end{eqnarray}
also holds.  
The transport map $T$ has $m$ components
\begin{eqnarray}  \label{eq:TM_general_x}
    T_{i}(x_{1},x_{2},\ldots,x_{m}), \quad i = 1, \ldots, m, 
\end{eqnarray}
and one has that 
\begin{eqnarray}  \label{eq:y_eq_Tx_general}
    y_{i}  & = &  T_{i}(x_{1},x_{2},\ldots,x_{m}), \quad i = 1, \ldots, m,
\end{eqnarray}
where $y = (y_{1},y_{2},\ldots,y_{m}) \sim \targetMeasure$ and $x = (x_{1},x_{2},\ldots,x_{m}) \sim \referenceMeasure$.  In other words, $T$ maps any given sample $x$ from the reference measure $\referenceMeasure$ to a sample $y = T(x)$ from the target measure $\targetMeasure$.  Conversely, the inverse transport map $T^{-1}$ of $T$ has $m$ components
\begin{eqnarray}  \label{eq:TM_general_y}
    (T^{-1})_{i}(y_{1},y_{2},\ldots,y_{m}), \quad i = 1, \ldots, m, 
\end{eqnarray}
and one has that 
\begin{eqnarray}  \label{eq:x_eq_Tinvy_general}
    x_{i} = (T^{-1})_{i}(y_{1},y_{2},\ldots,y_{m}) \quad i = 1, \ldots, m.
\end{eqnarray}
(That is, $T^{-1}$ maps any given sample $y$ from the target measure $\targetMeasure$ to a sample $x = T^{-1}(y)$ from the reference measure $\referenceMeasure$.)

A transport map $T$ is computed via the solution of a minimization problem with cost or objective function $C_{\referenceDensity}(T)$ generally defined as 
\begin{eqnarray}  \label{eq:TM_cost_function}
    C_{\referenceDensity}(T)  & = &  \int_{\mathbb{R}^{m}} c(x,T(x)) \, \referenceMeasure(\di{x}) \, ,
\end{eqnarray}
subject to the measure transport equation constraint \eqref{eq:pushforward_reference} (or \eqref{eq:pushforward_target}).  (More on this minimization problem in Section \ref{sec:triangular_TMs} below.)  The structure of the map $T$ depends on $\referenceDensity$, $\targetDensity$, and $c(\cdot,\cdot)$.  Many types of maps (not necessarily optimal), or couplings, exist between measures \cite{bk:villani_old_new, bk:comput_opt_transport}.  In this study we consider the use of triangular transport maps, as they enable practical numerical computation of a transport map and its inverse \cite{marzouk2016, baptista2022_ATM}.

\subsection{Triangular Transport Maps}
\label{sec:triangular_TMs}

A transport map $T:\mathbb{R}^{m} \rightarrow \mathbb{R}^{m}$, satisfying the measure transport equation $\targetMeasure =  T_{\sharp}\referenceMeasure$, is \textit{triangular} if 
\begin{eqnarray}  \label{eq:triangular_TM}
    y_{i}  & = &  T_{i}(x_{1},x_{2},\ldots,x_{i}), \quad i = 1, \ldots, m
\end{eqnarray}
(\cf. equation \eqref{eq:y_eq_Tx_general}).  In other words, the $i$-th component $T_{i}$ of a triangular transport map $T$ depends only on the values $\{x_{j}\}_{j=1}^{i}$ of the first $i$ random variables $\{\rv{X}_{j}\}_{j=1}^{i}$. Thus, the map components are
\begin{eqnarray} \label{eq:triangularTM2}
                   \hspace*{-2.0cm}
    T(x)  & = & \left [
            \begin{array}{l}
            T_{1}(x_{1}) \\
            T_{2}(x_{1},x_{2}) \\
            T_{3}(x_{1},x_{2},x_{3})  \\
              \qquad \vdots   \\
            T_{m}(x_{1},x_{2},\ldots,x_{m})
            \end{array}
            \right ] \, , 
\end{eqnarray}
from where one can observe the triangular structure of the map.  Conversely, the inverse map $T^{-1}$ of $T$ is also triangular and, so,
\begin{eqnarray}  \label{eq:triangular_TM_inverse}
    x_{i}  & = &  (T^{-1})_{i}(y_{1},y_{2},\ldots,y_{i}), \quad i = 1, \ldots, m
\end{eqnarray}
(\cf. equation \eqref{eq:x_eq_Tinvy_general}).

To ensure (the required) positivity of the densities \eqref{eq:pullback_target} and \eqref{eq:pullback_reference}, each map component $T_{i}$ in equation \eqref{eq:triangular_TM} must be differentiable with respect to $x_{i}$, and it must hold that $\partial_{x_{i}} T_{i} > 0$, for every $i=1,\ldots,m$. We refer to the class of maps with the latter property as the class of increasing triangular maps. Similarly, each component $(T^{-1})_{i}$ from equation \eqref{eq:triangular_TM_inverse} of the inverse map $T^{-1}$ of $T$ must satisfy said properties with respect to the variables $y_{i}$ \cite{bogachev2005, marzouk2016}.
Furthermore, when both $\referenceMeasure$ and $\targetMeasure$ (with densities, respectively, $\referenceDensity$ and $\targetDensity$) are absolutely continuous with respect to the Lebesgue measure $\di{x}$ on $\mathbb{R}^{m}$, if in the cost function \eqref{eq:TM_cost_function} we have $c(x,T(x)) = ||x - T(x)||^{2}$,  where $||\cdot||$ is the Euclidean distance on $\mathbb{R}^{m}$, and $\referenceMeasure$ is Gaussian, instead of the cost function $C_{\referenceDensity}(T)$ in \eqref{eq:TM_cost_function} one may seek to minimize the relative entropy, or Kullback-Leibler divergence,
\begin{eqnarray}  \label{eq:KL_divergence}
     \mathrm{D_{KL}}(\targetMeasure \, || \, \referenceMeasure)  & = &  \int_{\mathbb{R}^{m}} \targetDensity \log(\targetDensity/\referenceDensity) \, \di{x}  \ = \  \mathbb{E}_{\targetDensity}(\log(\targetDensity/\referenceDensity)) \, ,
\end{eqnarray}
as $C_{\referenceDensity}(T) \le 2A \, \mathrm{D_{KL}}(T_{\sharp}\referenceMeasure \, || \, \referenceMeasure)$, where $A > 0$ is a constant \cite{ottovillani2000,kolesnikov2004,bogachev2005}.
Under the assumption that the map $T$ is differentiable or even smooth, one may thus look at the problem of learning a transport map $T$ as a solution of the optimization problem
\begin{equation}\label{eq:KL_divergence_optimization}
\arg\min_{T}\mathrm{D_{KL}}(T_{\sharp}\referenceMeasure \, || \, \referenceMeasure),
\end{equation}
(conversely, $\arg\min_{T}\mathrm{D_{KL}}(\targetMeasure \, || \, (T^{-1})_{\sharp}\targetMeasure)$) subject to the partial differential equation constraint \eqref{eq:pullback_target} for the map $T$ (conversely, constraint \eqref{eq:pullback_reference} for the map $T^{-1}$).

\begin{figure}[t]
    \centering
    \subfloat[][Dense map]{{{\includegraphics[trim=0.0cm 0.0cm 0.0cm 0.0cm, clip=true, height=0.29\textwidth]{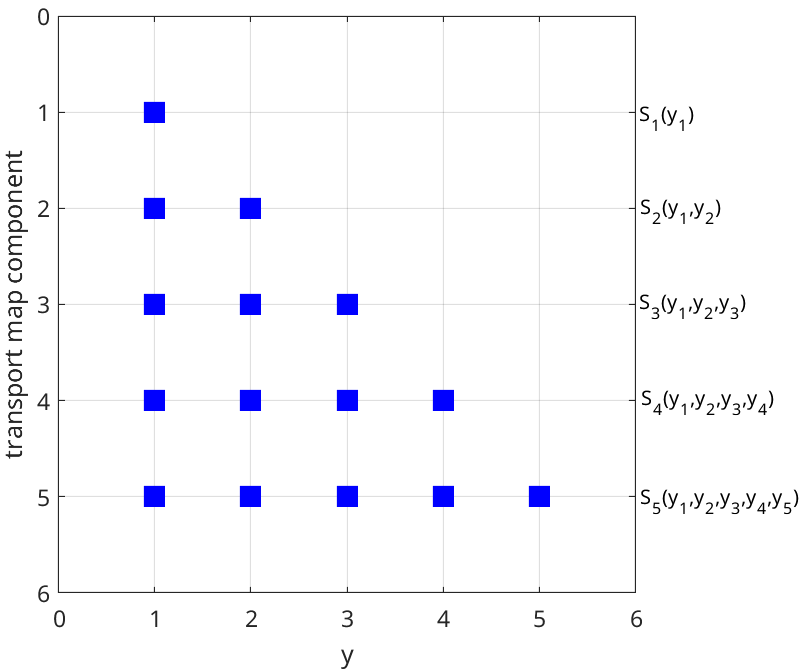}}}
    \label{fig:illustration_dense_map}}
    \subfloat[][Sparse map]{{{\includegraphics[trim=0.0cm 0.0cm 0.0cm 0.0cm, clip=true, height=0.29\textwidth]{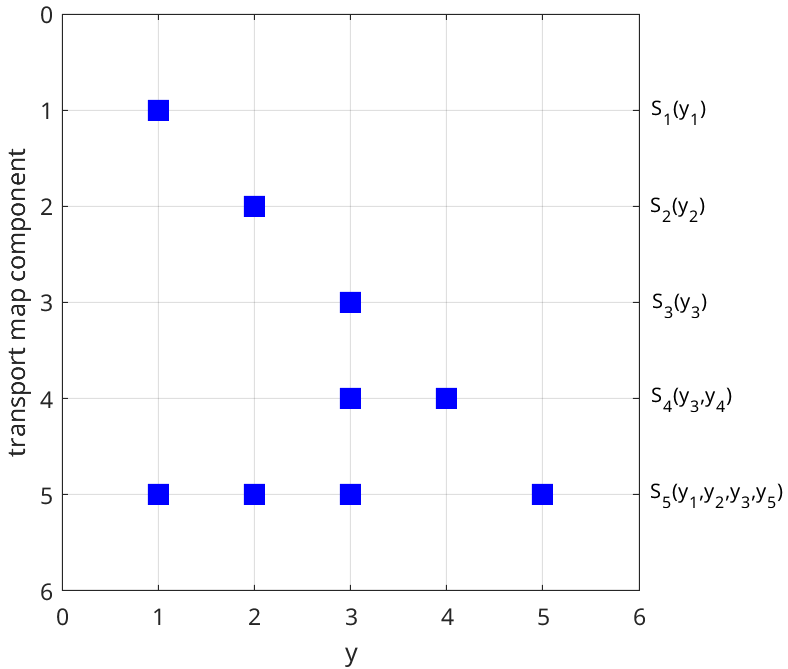}}}
    \label{fig:illustration_sparse_map}}
    \subfloat[][Diagonal map]{{{\includegraphics[trim=0.0cm 0.0cm 0.0cm 0.0cm, clip=true, height=0.29\textwidth]{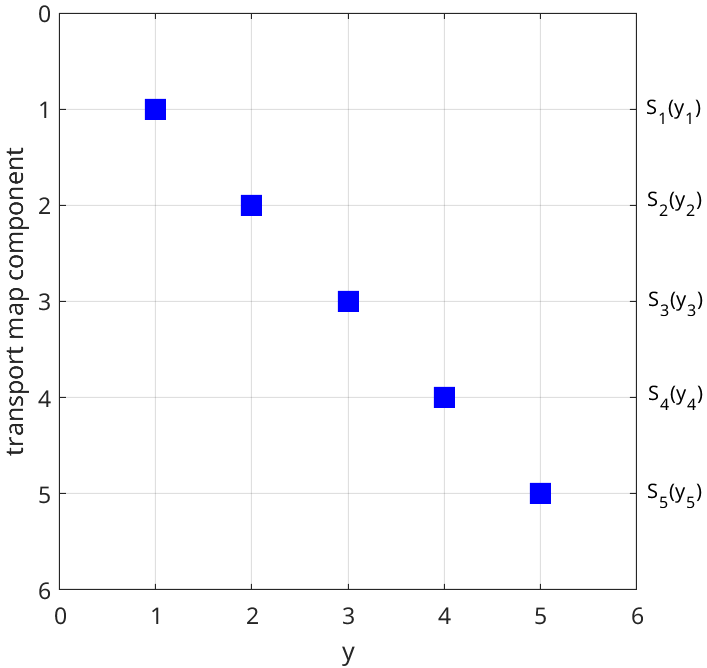}}}
    \label{fig:illustration_diagonal_map}}
    \caption{Pictorial representation of triangular transport maps sparsity patterns.  To simplify notation in Subfigures \ref{fig:illustration_dense_map}--\ref{fig:illustration_diagonal_map}, we denote $S \equiv T^{-1}$ for the transport map $T^{-1}$ from \eqref{eq:triangular_TM_inverse}.  In this example, $S\!:\!\mathbb{R}^{5} \!\rightarrow\! \mathbb{R}^{5}$. In each plot, the horizontal axis indexes the random variables $\{\rv{Y}_{j}\}_{j=1}^{5}$ and the vertical axis indexes the map components $\{S_{i}\}_{i=1}^{5}$. A square appears at the intersection of grid point $(j,i)$ if the $i$-th map component $S_{i}$ depends on the value $y_{j}$ of the $j$-th random variable $\rv{Y}_{j}$.  For any given map component, the set of active variables is the set of random variables that the map component depends on.  Collectively, the active variables define a sparsity pattern for the transport map.}
    \label{fig:TM_sparsity_example}
\end{figure}

Triangular maps enable efficient numerical computation of a transport map and its inverse, which is also triangular.  Moreover, triangular transport maps are well-suited for conditional density estimation and sampling \cite{elmoselhy2012, marzouk2016, spantini2018inference, baptista2022_ATM, katzfuss2023}. Techniques for computational efficiency and scalability to high-dimensional problems include, among others, composition of low order maps \cite{elmoselhy2012} and the use of \textit{sparse triangular transport maps} \cite{morrison2017, spantini2018inference,baptista2022_ATM,katzfuss2023}.  
For such sparse maps, any given map component $(T^{-1})_{i}$ in equation \eqref{eq:triangular_TM_inverse} may depend only on a proper subset (possibly empty) of the variables $\{y_j\}_{j=1}^{i-1}$ and on $y_{i}$.  Dependence on $y_{i}$ is required.  A \textit{diagonal map}, in particular, has $x_{i} = (T^{-1})_{i}(y_{i})$ for each $i = 1,\ldots,m$.  The random variables that a given map component depends on is referred to as the set of \textit{active variables} for the map component.  Collectively, the active variables define a \textit{sparsity pattern} for the transport map.  For a summary of these notions, see Figure \ref{fig:TM_sparsity_example}.  We note that the notion of sparsity pattern characterizes the triangular transport map and reflects the dependence structure of the target probability measure (i.e., by virtue of being triangular, it represents dependence relations between the random variables jointly distributed according to the target probability measure). Thus, ``sparsity'' here does not refer to the number of samples available for learning a triangular transport map.

As explained in detail in the original publications \cite{marzouk2016,baptista2022_ATM} and summarized in Appendix \ref{app:ATM_framework} of the present paper, the algorithms we use and build upon rely on finite-dimensional approximations to the functional representations for the transport map components, which may be described generally to depend on functions $f_{i} $ of the form
\begin{eqnarray}  \label{eq:f_expansion_main_text}
   f_{i}(y_{1},\ldots,y_{i})  &  \ = \ &  \sum_{\alpha^{(i)} \in \Lambda^{(i)}} c_{\alpha^{(i)}} \widetilde{\mathcal{H}}_{\alpha^{(i)}}(y_{1},\ldots,y_{i}),
\end{eqnarray}
where, for each $i = 1, \ldots, m$,
\begin{itemize}[topsep=0pt, itemsep=-3pt] 
    \item $\Lambda^{(i)}\subset\mathbb{Z}_{+}^{i}$ is a finite set  of multi-indices, which identifies the active variables of the $i$-th transport map component and therefore encodes information about the sparsity pattern of the transport map,
    \item $c_{\alpha^{(i)}} \in \mathbb{R}$ are unknown real-valued coefficients, which are to be learned from the data, and
    \item $\widetilde{\mathcal{H}}_{\alpha^{(i)}}:\mathbb{R}^{i} \rightarrow \mathbb{R}$ are suitably defined basis functions (refer to Appendix \ref{app:ATM_framework} and \cite{marzouk2016,baptista2022_ATM}).
\end{itemize}
Sparse triangular transport maps can be learned by explicitly specifying a sparsity pattern for the transport map and enforcing the pattern (via the sets $\Lambda^{(i)}$ of multi-indices from equations \eqref{eq:f_expansion_main_text}) when learning the transport map.  Alternatively, (sparse) triangular transport maps may be learned without a sparsity pattern being specified a priori, by learning the sets $\Lambda^{(i)}$ of multi-indices in equations \eqref{eq:f_expansion_main_text} along with the unknown coefficients $c_{\alpha^{(i)}}$, using specifically designed algorithms which have been shown to perform well for density estimation problems that considered a wide range of values (mainly $O(100)$ or larger) for the number of training samples used to learn the transport maps \cite{baptista2022_ATM}. 

Recall from Sections \ref{sec:contributions} and \ref{sec:motivating_application} that one motivation for our research is density estimation from data when the number of samples available for learning the model (within this context, learning a transport map) is limited in amount (more precisely, of order $O(1)$--$O(10)$ number of samples).  For the purpose we consider the use of triangular transport maps, first by explicitly enforcing transport map sparsity patterns (Section \ref{sec:DAG_experiments}) and second by  not enforcing transport map sparsity patterns a priori at all (Section \ref{sec:ATM_experiments}). In order to avoid possible confusion, we emphasize again that the sparsity pattern of a triangular transport map is a property of the target probability measure and not a property related to the number of samples used for transport map learning.  Next, we conclude the current Section \ref{sec:TM_computation} with some brief comments on density estimation via triangular transport maps and several illustrative examples.

\subsection{Density Estimation from Data}
\label{sec:density_estimation_from_data}

Referring to equations \eqref{eq:pushforward_reference}--\eqref{eq:pushforward_target}, recall from Section \ref{sec:basics_measure_transport} that we designate the probability measure $\referenceMeasure$ as the reference measure and the probability measure $\targetMeasure$ as the target measure.  The target density $\targetDensity$ of interest to us is unknown, but samples from the target measure $\targetMeasure$ are available.  On the other hand, the reference measure $\referenceMeasure$ is known and purposefully chosen so that its density $\referenceDensity$ can be evaluated in an efficient manner and so that sampling from $\referenceMeasure$ can be done efficiently as well.  The reference measure $\referenceMeasure$ is often selected to be the standard normal (i.e., Gaussian) measure $\mathcal{N}(0,I)$, and that is our choice in all of our computational experiments.

Given a set of samples from the unknown target measure $\targetMeasure$, we use software libraries \cite{TransportMaps_library,ATM_library} made available by the MIT Uncertainty Quantification (MUQ) Group \cite{MIT_UQ} to learn a transport map (and its inverse) between the target and reference measures.  The transport map is learned as the solution of an optimization problem with objective function minimizing \eqref{eq:KL_divergence}, using the Sample Average Approximation (SAA) algorithm.   It is outside the scope of this paper to cover the details pertaining to the solution of the optimization problem (although we provide a summary of some aspects in Appendix \ref{app:ATM_framework}), but the interested reader is referred to \cite{marzouk2016,baptista2022_ATM} for a full account.  

Recall from Section \ref{sec:basics_measure_transport} that upon learning a transport map $T$ that ``couples'' the reference and target measures, the target density $\targetDensity$ can be evaluated via equation \eqref{eq:pullback_reference} (which, within the context of this paper, corresponds to equation \eqref{eq:target_density_intro2}) and, so, given a (new) sample $y$, the value $\targetDensity(y)$ can thus be obtained and used as needed.  Hence our density estimation problem is solved by the measure transport framework.  The MUQ software libraries \cite{TransportMaps_library,ATM_library} used in our computations provide functionality for evaluating the target density $\targetDensity$ and sampling from the target measure $\targetMeasure$, among many other capabilities.

\begin{figure}[t]
    \centering
    \subfloat[][]{{{\includegraphics[trim=1.0cm 8.3cm 15.0cm 3.6cm, clip=true, width=0.40\textwidth]{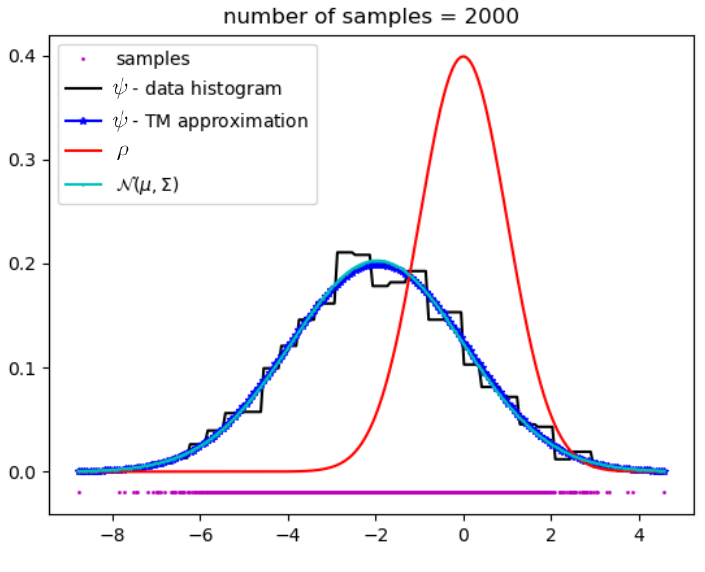}}}
    \label{fig:example_1d_gaussian}}
    \hspace{2ex}
    \subfloat[][]{{{\includegraphics[trim=1.0cm 8.3cm 15.0cm 3.6cm, clip=true, width=0.40\textwidth]{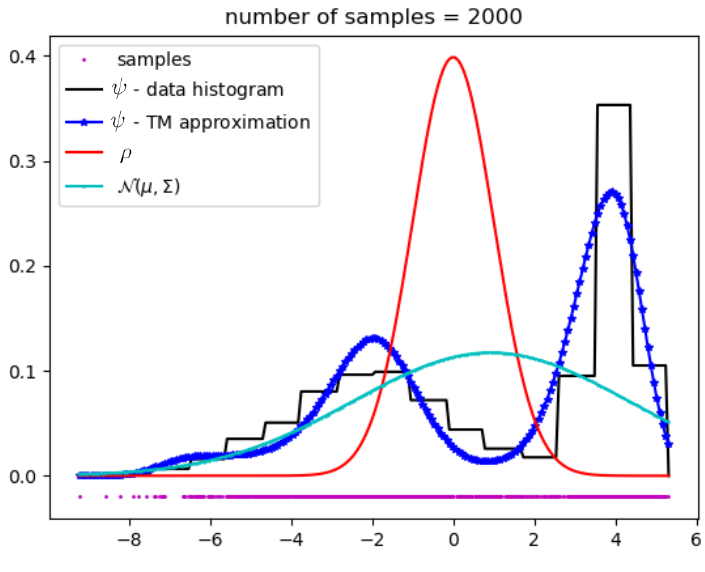}}}
    \label{fig:example_1d_non_gaussian}}
    \caption{Approximations to univariate density functions via triangular transport maps.  In each of the Subfigures \ref{fig:example_1d_gaussian} and \ref{fig:example_1d_non_gaussian}, (i) the red curve is the reference density $\referenceDensity$ -- which is the density of a standard normal (i.e., Gaussian) measure $\mathcal{N}(0,1)$, (ii) a histogram of samples from the target measure $\targetMeasure$ is depicted with black lines, (iii) the blue curve is the transport map (TM) approximation to the target density $\targetDensity$, and (iv) the cyan curve is the density for the normal (i.e., Gaussian) measure $\mathcal{N}(\mu,\Sigma)$ with mean and standard deviation being that of the sample data.  For Subfigure \ref{fig:example_1d_gaussian}, the data set was created by drawing samples from a normal (i.e., Gaussian) measure $\mathcal{N}(\mu,\Sigma)$ with $\mu \ne 0$ and $\Sigma \ne 1$.  As can be seen, the TM approximation to the target density (blue curve) agrees with that of the true density (cyan curve).  The samples for the data set from Subfigure \ref{fig:example_1d_non_gaussian} were drawn from a multi-modal target measure $\targetMeasure$.  The resulting TM approximation to the target density (blue curve) reflects this and follows the data histogram (black curve) more closely, as opposed to the density for $\mathcal{N}(\mu,\Sigma)$ (cyan curve).}
    \label{fig:example_1d_data}
\end{figure}

To describe an application of the measure transport framework, in Figure \ref{fig:example_1d_data} are displayed transport map computed approximations to density functions for univariate measures.  The triangular transport maps for these approximations were learned using the TransportMaps library \cite{TransportMaps_library}, version 2.0b3.  The figure shows how the measure transport framework can be used for uni-modal and multi-modal data distributions alike.  In each case, we learned the transport maps using synthetic data sets (i.e., data sets generated by us).  In Subfigure \ref{fig:example_1d_gaussian}, the target is a Gaussian measure $\mathcal{N}(\mu,\Sigma)$ with mean $\mu \ne 0$ and standard deviation $\Sigma \ne 1$, and so we know the formula $\targetDensity_{true}(y) = (\Sigma)^{-1} (2\pi)^{-\frac{1}{2}} \exp(-\frac{1}{2}(\frac{y-\mu}{\Sigma})^{2})$ for the true target density.  We can thus verify that the transport map approximation \eqref{eq:pullback_reference} to the target $\targetDensity$ (blue curve in Subfigure \ref{fig:example_1d_gaussian}) agrees with that of the true density $\targetDensity_{true}(y)$ (cyan curve in Subfigure \ref{fig:example_1d_gaussian}).  On the other hand, the target density $\targetDensity$ in Subfigure \ref{fig:example_1d_non_gaussian} is multi-modal and, hence, the density for a Gaussian measure (cyan curve in Subfigure \ref{fig:example_1d_non_gaussian}) with mean and standard deviation being that of the sample data is a poor approximation to the data histogram (black curve in Subfigure \ref{fig:example_1d_non_gaussian}).  The transport map approximation to the target density $\targetDensity$ (blue curve in Subfigure \ref{fig:example_1d_non_gaussian}) captures the multi-modality and better approximates the data histogram (black curve in Subfigure \ref{fig:example_1d_non_gaussian}).  Further details are provided in the caption for Figure \ref{fig:example_1d_data}.  

\begin{figure}[htbp]
    \centering
    \subfloat[][Class 0 (762 samples in data set \cite{UCI_banknote_authentication_267})]{{\includegraphics[trim=0.0cm 13.5cm 0.0cm 2.5cm, clip=true, width=0.90\textwidth]{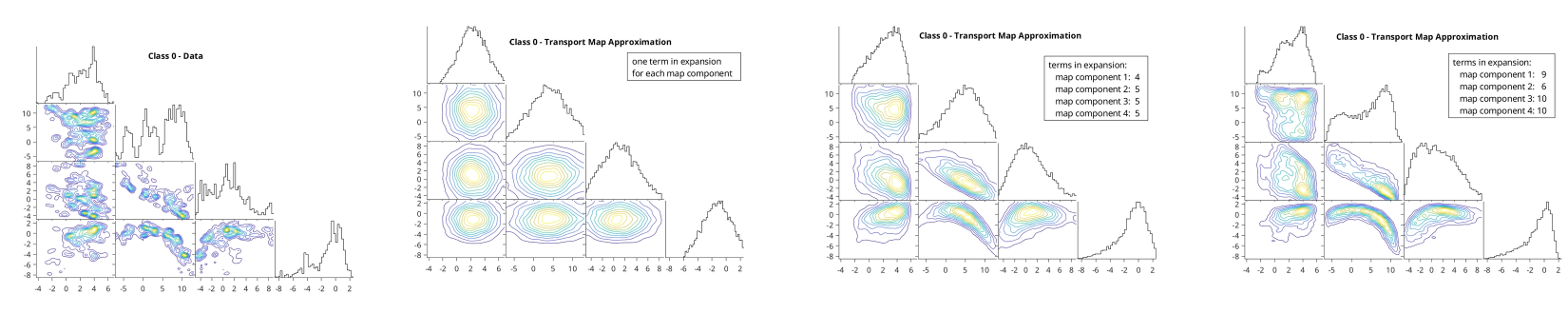}}
    \label{fig:bn_class0_ATM}}
    \\
    \subfloat[][Class 1 (610 samples in data set \cite{UCI_banknote_authentication_267})]{{\includegraphics[trim=0.0cm 8.5cm 0.0cm 8.0cm, clip=true, width=0.90\textwidth]{banknote_dataset_TM_approximations.pdf}}
    \label{fig:bn_class1_ATM}}
    \\
    \subfloat[][Transport maps reference probability density (left) and confusion matrices corresponding to the classification results obtained using the classification formula \eqref{eq:classify}, along with the TM class-conditional density approximations from \ref{fig:bn_class0_ATM} and \ref{fig:bn_class1_ATM}]{{\includegraphics[trim=0.0cm 2.8cm 0.0cm 13.0cm, clip=true, width=0.90\textwidth]{banknote_dataset_TM_approximations.pdf}}
    \label{fig:bn_classification_ATM}}
    \\
    \subfloat[][Confusion matrices from Matlab's naive Bayes (left), support vector machine (middle), and neural network (right) classifiers]{{\hspace*{1.20cm}\includegraphics[trim=7.50cm 0.0cm 0.0cm 7.25cm, clip=true, width=0.5\textwidth]{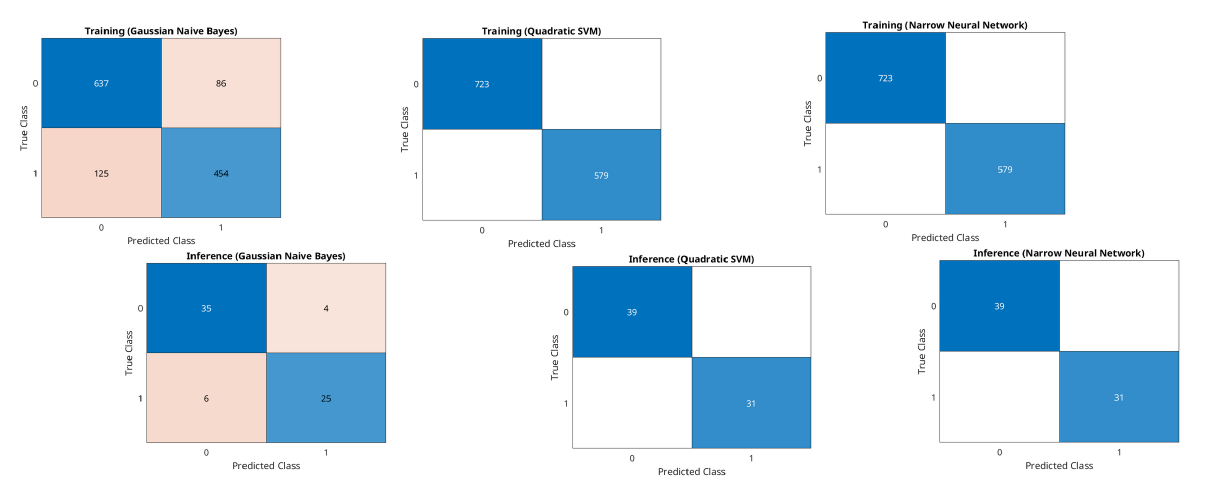}}
    \label{fig:bn_classification_matlab}}
    \caption{Transport map (TM) approximations to class-conditional densities for the data set \cite{UCI_banknote_authentication_267}.  
    \textit{Subfigure \ref{fig:bn_class0_ATM}}: Class 0 marginal distribution densities from data (1st subplot, far left) and three different TM approximations (2nd through 4th subplots), with an increasing number of terms in the basis function expansions \eqref{eq:f_expansion} for the TM components.   Clearly, the data density distribution (1st subplot in Figure \ref{fig:bn_class0_ATM}) is not Gaussian.  The first TM approximation (2nd subplot in Figure \ref{fig:bn_class0_ATM}), with only one term in each basis function expansion \eqref{eq:f_expansion}, approximates the target density (1st subplot in Figure \ref{fig:bn_class0_ATM}) poorly.  The accuracy of the approximations improves (3rd and 4th subplots in Figure \ref{fig:bn_class0_ATM}, compared to 1st subplot in Figure \ref{fig:bn_class0_ATM}) with an increased number of terms in the basis function expansions \eqref{eq:f_expansion}.
    \textit{Subfigure \ref{fig:bn_class1_ATM}:} As in Subfigure \ref{fig:bn_class0_ATM}, but for class 1.
    \textit{Subfigure \ref{fig:bn_classification_ATM}:} Left: reference probability density, which is a standard multivariate normal for all TM approximations. Afterwards: confusion matrices showing improved classification result as the accuracy of the TM density approximations increases.  The top row confusion matrices are from training data and the bottom row ones are from inference data. 
    \textit{Subfigure \ref{fig:bn_classification_matlab}:} As the data is not Gaussian, a naive Bayes classifier (left) performs poorly.  A support vector machine (middle) and neural network (right) classifier perform as well as the classification resulting from the most accurate TMs approximation (far right, Subfigures \ref{fig:bn_class0_ATM}-\ref{fig:bn_classification_ATM}).
    }  
    \label{fig:banknote_data}
\end{figure}
Finally, transport map approximations to densities for a multivariate data set chosen from the UC Irvine Machine Learning Repository \cite{UCI_ML_repository} appear in Figure \ref{fig:banknote_data}.  Although not from the biological domain, we selected the data set \cite{UCI_banknote_authentication_267} because it yielded a good, non-contrived example for illustrating the benefits of a measure transport approach to density estimation.  The data set \cite{UCI_banknote_authentication_267} was created by its author from images taken from genuine and forged banknote-like specimens, creating two classes for classification problems.  For the purpose of discussion, we identify  the classes as class 0 and class 1 and compute approximations to the densities for each of the two classes.  The triangular transport maps for the approximated densities were obtained using the AdaptiveTransportMaps library \cite{ATM_library}, which allows for computing adaptive parameterizations of transport maps \cite{baptista2022_ATM}. 

We provide a summary of the framework for such \textit{adaptive transport maps} in Appendix \ref{app:ATM_framework}.  Here we note that among the several parameters defining the adaptive transport maps is the number of terms in the basis function expansions \eqref{eq:f_expansion} used in the functional representation of each transport map component.  Figure \ref{fig:banknote_data} illustrates the effect of varying this allowed maximum number of terms and shows that, for each of the class 0 and class 1 data, increasingly good approximations to the target densities can be obtained.  Classification results obtained via the transport map approximated density distribution functions, using the classification formula \eqref{eq:classify}, are also shown.  For comparison, classification results obtained using  a neural network classifier from Matlab  and the naive Bayes and support vector machine Matlab classifiers \cite{matlab_ml_toolbox} are included in Figure \ref{fig:banknote_data} as well.

\section{Application to Gene Expression Data}
\label{sec:application_GSE_dataset}

As we discussed in Section \ref{sec:motivating_application}, one of the problems motivating the present research is that of probability distribution density estimation for phenotypic classification problems, for example, for inferring class labels, such as radiation exposure levels, given sample data.  Recall from Section \ref{sec:contributions} that we are concerned with data sets containing a small number of samples, specifically of order $O(1)$--$O(10)$ samples, which are common in the radiation biology domain.

To conduct an initial assessment of the potential of measure transport techniques for phenotypic classification within the above mentioned context, we use the \gsedataset\ data set \cite{GSE43151_dataset} available from the Gene Expression Omnibus (GEO) database \cite{GEO_database}, as in previous work \cite{luo2022comprehensive,luo2023_frontiers}.  This data set contains gene expression data for 121 human blood samples, some with zero exposure to radiation, and others exposed to low-dose radiation or high-dose radiation.  Out of the total number of samples, 18 are in the zero-dose category, 87 are in the low-dose category, and 16 are in the high-dose category.  For more details on the \gsedataset\ data set, the reader may refer to \cite{GSE43151_dataset}. The data set, as used in our experiments and the previous works \cite{luo2022comprehensive,luo2023_frontiers}, is available from \cite{xihaier_github}.

Genes correspond to the random variables jointly distributed according to the unknown target probability densities we seek to estimate.  For feature (i.e., random variables) selection, we consider genes in biological pathways relevant to the \gsedataset\ data set, as determined by the pathway ranking procedure from \cite{luo2022comprehensive,luo2023_frontiers}.  As previously mentioned at the end of Section \ref{sec:triangular_TMs}, we investigate two different measure transport techniques, namely, explicitly enforcing sparsity patterns in the transport maps (i.e., explicitly enforcing a pre-determined dependence structure among the random variables in the target measure) and adaptive transport maps (for which a dependence structure among the set of random variables in the target measure will not be enforced a priori).  Details appear in Section \ref{sec:DAG_experiments} for the former approach and in Section \ref{sec:ATM_experiments} for the latter one.
In each case, the resulting transport maps are sparse, as defined towards the end of Section \ref{sec:triangular_TMs} and depicted in Figure \ref{fig:TM_sparsity_example}.  Recall from the discussion in Section \ref{sec:triangular_TMs}, in particular, that sparsity of a triangular transport map is a property of the target probability measure and not a property of the number of samples used for learning the transport map.

As discussed next in Section \ref{sec:DAG_experiments} below, prior knowledge extracted from known biological pathways, such as relationships among genes, could be used to fix sets of active variables (i.e., sets of variables that the map components depend on), and thereby explicitly enforce sparsity patterns in the transport maps when learning the maps.  This is advantageous computationally, particularly if the extracted relationships among genes are known with (reasonable) certainty.  However, this approach could end up being biased and too restrictive (for example, missing possible dependencies between genes that are a priori unknown as well as obscuring their change when exposed to radiation). Hence,  in Section \ref{sec:ATM_experiments}, we also consider adaptive transport maps without any a priori imposed restrictions on the sparsity patterns of the transport map components.  With adaptive transport maps, via specifically designed algorithms \cite{baptista2022_ATM}, sparse transport maps can be constructed and learned without the need to fix the set of active variables a priori. (Refer to the discussion surrounding equation \eqref{eq:f_expansion_main_text} and also to Appendix \ref{app:ATM_framework}.)  For our application, adaptive transport maps may thus offer more flexibility and would be more advantageous when prior knowledge about gene relationships is lacking or uncertain. 

\subsection{Prescribed Sparsity Patterns from Gene Relationships}
\label{sec:DAG_experiments}  

One simple way to incorporate prior information into transport map construction is by building a sparse transport map using knowledge of the dependence structure of the set of random variables from the target measure of interest. If the random variables adhere to a Bayesian network, then the active variables of the transport map (as defined in Section \ref{sec:triangular_TMs} -- see Figure \ref{fig:TM_sparsity_example}) may be obtained from this network in a straightforward way \cite{spantini2018inference}. In doing so, the number of parameters that need to be fit in order to learn the transport map can be significantly reduced compared with triangular dense maps. 

For the gene expression data under radiation exposure, we used the Kyoto Encyclopedia of Genes and Genome (KEGG) database \cite{KEGG_database} to extract Bayesian networks describing the relationships between the genes for a given pathway. While KEGG provides a variety of relationships between genes, we focused on protein-protein (PPrel) activation and inhibition relationships. In particular, if there was a directed edge from gene A to gene B in the KEGG graph of type PPrel-activation/inhibition, then an edge from A to B was added to the Bayesian network. We also added edges between genes that are part of a gene complex. This information was extracted from KEGG Markup Language (KGML) files and read into Python using the BioPython package. We only considered pathways that resulted in directed acyclic graphs (DAGs) under these PPrel relationships.  Software to extract this information is available from \cite{patrick_kegg_github}.  The transport maps for the experiments reported in this section were computed using the TransportMaps library \cite{TransportMaps_library}, version 2.0b3.

\begin{figure}[t]
    \centering
    \includegraphics[height=.55\linewidth]{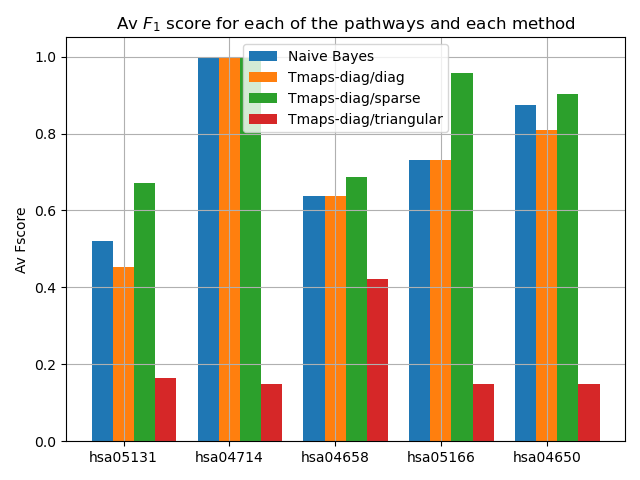}
    \caption{Classification results with KEGG-informed sparse transport map}
    \label{classResults}
\end{figure}
To test out the performance of the sparse transport maps framework within this context, we selected 5 gene pathways from the KEGG database that performed well on the pathway activation scoring performed in \cite{luo2023_frontiers} and resulted in DAGs as well, through the above analysis. We considered the following pathways: hsa05131, hsa04714, hsa04658, hsa05166, and  hsa04650. We performed the classification task of discriminating zero-dose versus low-dose samples using the subset of genes from the \gsedataset\ data set corresponding to each pathway. We were interested in the ``small-data'' (i.e., small number of samples) regime where prior information can provide the biggest benefit, so we used only 25\% of the total number of data samples for the training set and reserved the rest of the samples for the inference set. Since there is so little data in the zero-dose case (only 4 training samples) it made sense to use a diagonal transport map for this case, because other approaches performed poorly. In the low-dose case, there are 21 training samples and we consider using the following maps:
\begin{itemize}[topsep=0pt, partopsep=0pt, itemsep=-3pt]
     \item \textit{Diagonal map}: The $i$-th component of the transport map $T^{-1}$ from equation \eqref{eq:triangular_TM_inverse} only depends on the value $y_{i}$ of the $i$-th random variable $\rv{Y}_{i}$.
     \item \textit{Dense triangular map}:  The $i$-th component of the transport map $T^{-1}$ depends on the values $\{y_{j}\}_{j=1}^{i}$ of the random variables $\{\rv{Y}_{j}\}_{j=1}^{i}$.
     \item \textit{KEGG-informed sparse map}: The $i$-th component of the transport map $T^{-1}$ depends on the value $y_{i}$ of the random variable $\rv{Y}_{i}$ and the values of a proper subset $Z_{i}$ (possibly empty) of the random variables $\{\rv{Y}_{j}\}_{j=1}^{i-1}$.  For the experiments reported here, $Z_{i}=\emptyset$ typically.
     The set of active variables for each map component is derived from biological prior knowledge in the form of a topological sort of the KEGG graph \cite{patrick_kegg_github}.
\end{itemize}
Refer to Figure \ref{fig:TM_sparsity_example} from Section \ref{sec:triangular_TMs} for a pictorial representation of the above listed types of maps.  In each of the above cases, we have $T^{-1}\!:\mathbb{R}^{m} \rightarrow \mathbb{R}^{m}$, where $m$ is the number of genes extracted from the pathway. 

Figure \ref{classResults} displays on the vertical axis the $F_1$ classification score \cite{f1_score_scikit-learn} 
\begin{eqnarray}   \label{eq:f1_score}
     F_{1}  & = &   2 \,\times\, \frac{\mathrm{precision} \times \mathrm{recall}}{\mathrm{precision} +\mathrm{recall}} \, ,
\end{eqnarray}
averaged over both classes for each of the pathways (note that pathways are indexed on the horizontal axis in Figure \ref{classResults}) and each transport map method considered, and including naive Bayes classification as a baseline.  In equation \eqref{eq:f1_score}, for each class $k \in \{0,1\}$ (0: zero-dose class; 1: low-dose class), $\mathrm{precision} = \mathrm{TP}/(\mathrm{TP+FP})$ and $\mathrm{recall} = \mathrm{TP}/(\mathrm{TP+FN})$, where $\mathrm{TP}$ (True Positive) is the total number of samples correctly classified as belonging to class $k$, $\mathrm{FP}$ (False Positive) is the total number of samples incorrectly classified as belonging to class $k$, and $\mathrm{FN}$ (False Negative) is the total number of samples incorrectly classified as not belonging to class $k$.  From Figure \ref{classResults}, we see that transport maps with the sparsity pattern derived from prior knowledge provides an improvement in performance across all pathways except one. Note the poor performance of the dense triangular map, for which we were unable to fit adequately such a small number of data samples.

\subsection{Adaptive Transport Maps Approach}
\label{sec:ATM_experiments}

\newcommand{\pathwayzl}{hsa04120}
\newcommand{\pathwayzh}{hsa05202}
\newcommand{\geneszl}{108}
\newcommand{\geneszh}{98}
\newcommand{\totalsampleszero}{18}
\newcommand{\totalsampleslow}{87}
\newcommand{\totalsampleshigh}{16}
\newcommand{\atm}{T^{-1}}
\newcommand{\atmcomponent}{(\atm)}
\newcommand{\fscores}{${F}_{1}$ scores}

We now move on to adaptive transport maps \cite{baptista2022_ATM}, for which transport map sparsity patterns (i.e., the dependence structure of the set of random variables jointly distributed according to the target measure) are not being prescribed a priori.
First, we continue with classification experiments in Section \ref{sec:ATM_classification} below.  Afterwards, in Section \ref{sec:ATM_tool_scientific_discovery} we discuss a different application which shows the potential of adaptive transport maps to serve as a tool for scientific discovery.

\subsubsection{Further Classification Experiments}
\label{sec:ATM_classification}
To design classification experiments using probability distribution densities approximated via adaptive transport maps (as opposed to Section \ref{sec:DAG_experiments} above, where transport maps sparsity patterns were prescribed \textit{a priori}), we selected two pathways from the KEGG database \cite{KEGG_database}, namely pathways \pathwayzl\ and \pathwayzh.  This pathway selection was based on the pathway ranking procedure from \cite{luo2023_frontiers}, using the \gsedataset\ data set, where pathway \pathwayzl\ ranked as the top pathway showing significant differential activation in response to low-dose radiation, as compared to zero-dose, and pathway \pathwayzh\ ranked as the top pathway showing significant differential activation in response to high-dose radiation, as compared to zero-dose.  After accounting for genes present in both the \gsedataset\ data set and in the selected pathways, we were left with \geneszl\ genes for pathway \pathwayzl\ and \geneszh\ genes for pathway \pathwayzh.

\begin{figure}[h!]
    \centering
    \subfloat[][Averaged \fscores, all experiments]{{{\includegraphics[trim=0.0cm 0.0cm 0.0cm 0.0cm, clip=true, width=0.40\textwidth]{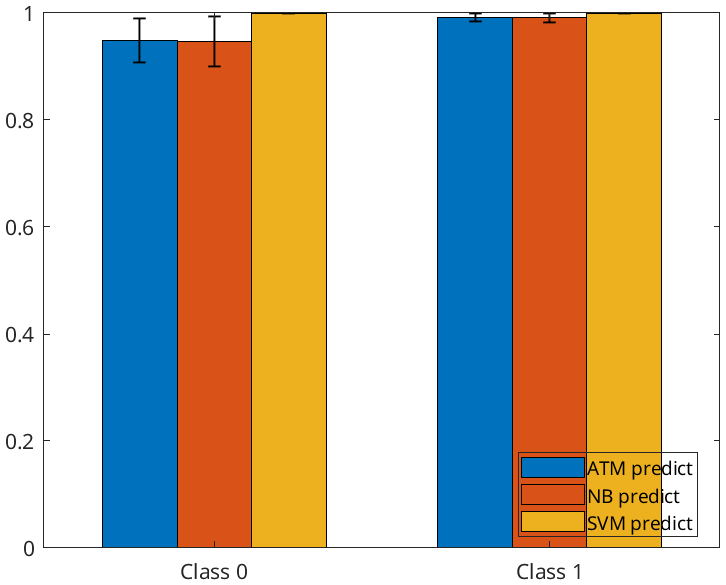}}}
    \label{fig:fscores_zl_all}}
    \hspace{2ex}
    \subfloat[][Averaged \fscores, subset of 20 experiments.  Maximum number of terms allowed in ATM basis function expansions: 1 for class 0, 1 for class 1.]{{{\includegraphics[trim=0.0cm 0.0cm 0.0cm 0.0cm, clip=true, width=0.40\textwidth]{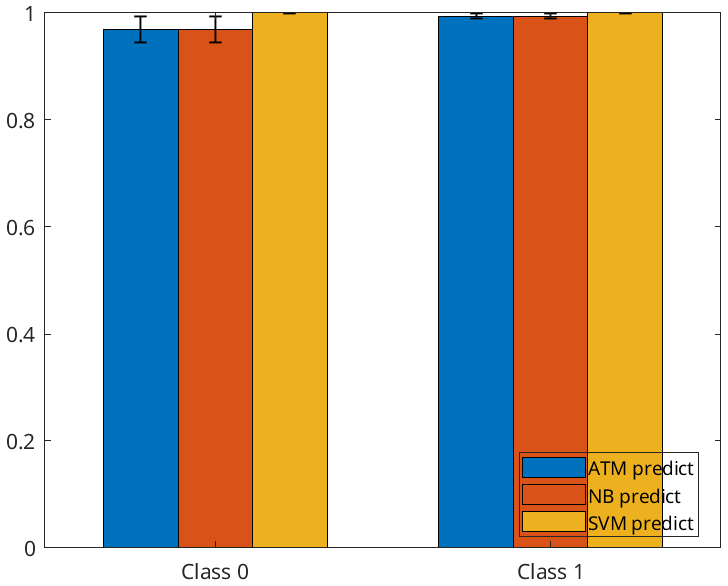}}}
    \label{fig:fscores_zl_test4}}
    \\
    \subfloat[][Averaged \fscores, subset of 20 experiments.  Maximum number of terms allowed in ATM basis function expansions: 2 for class 0, 5 for class 1.]{{{\includegraphics[trim=0.0cm 0.0cm 0.0cm 0.0cm, clip=true, width=0.40\textwidth]{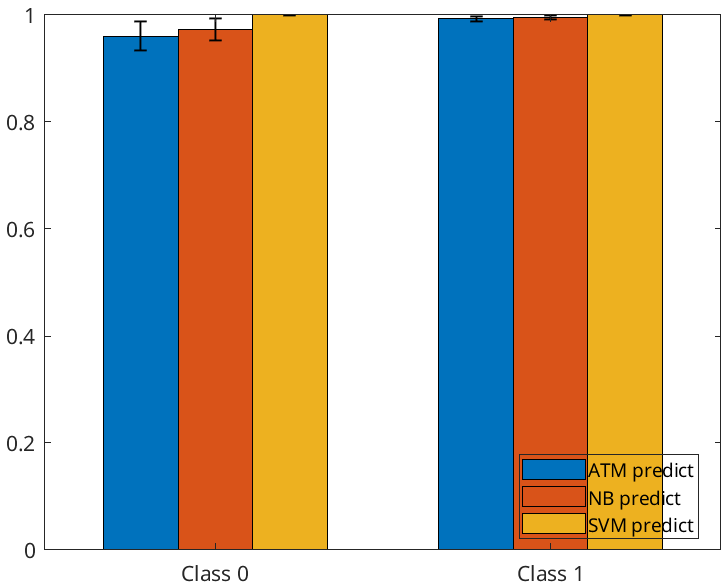}}}
    \label{fig:fscores_zl_test9}}
    \hspace{2ex}
    \subfloat[][Averaged \fscores, subset of 20 experiments.  Maximum number of terms allowed in ATM basis function expansions: 2 for class 0, 6 for class 1.]{{{\includegraphics[trim=0.0cm 0.0cm 0.0cm 0.0cm, clip=true, width=0.40\textwidth]{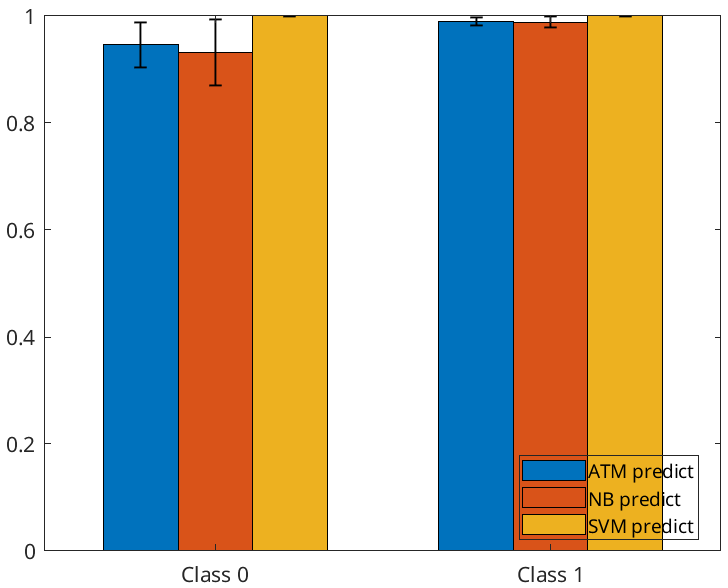}}}
    \label{fig:fscores_zl_test10}}
    \caption{Summary of classification results from adaptive transport maps (ATM) approximations to the class-conditional densities, naive Bayes (NB) classifier, and support vector machine (SVM) classifier.  Zero-dose radiation is class 0; Low-dose radiation is class 1.  The available number of samples were \totalsampleszero\ for class 0 and \totalsampleslow\ for class 1.  For each experiment (i.e., run), the available samples were divided randomly into 50\% for training and 50\% for inference.  The total number of experiments was 100.  For this data set, all methods performed well, with SVM achieving perfect results in most experiments.  On average, the classification results  for the ATM density approximations and for naive Bayes were equal, as seen in Subfigure \ref{fig:fscores_zl_all}.  For disjoint subsets of experiments (Subfigures \ref{fig:fscores_zl_test4}-\ref{fig:fscores_zl_test10}), the ATM and naive Bayes classifiers performed equally, or one slightly better than the other.  For all experiments, classification results were more accurate for class 1.  This is not surprising since there were more data samples available for transport map training for class 1.
    } 
    \label{fig:fscores_zl}
\end{figure}

\begin{figure}[h!]
    \centering
    \subfloat[][Averaged \fscores, all experiments]{{{\includegraphics[trim=0.0cm 0.0cm 0.0cm 0.0cm, clip=true, width=0.40\textwidth]{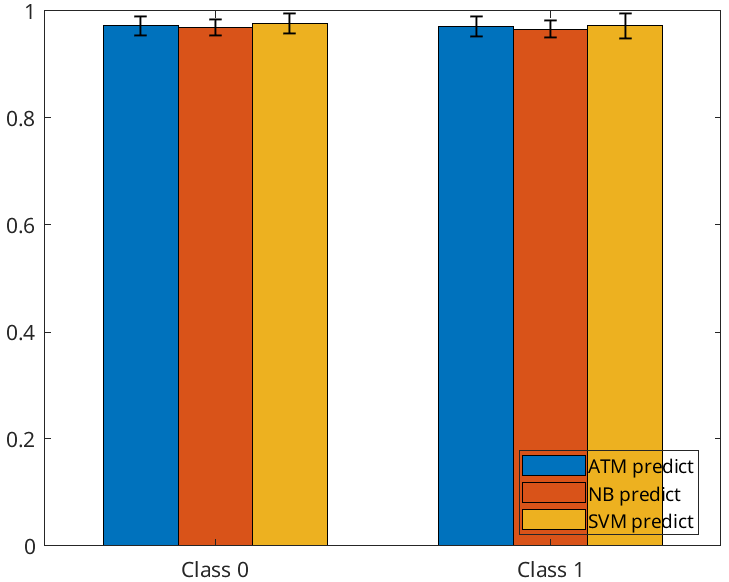}}}
    \label{fig:fscores_zh_all}}
    \hspace{2ex}
    \subfloat[][Averaged \fscores, subset of 20 experiments.  Maximum number of terms allowed in ATM basis function expansions: 1 for class 0, 1 for class 1.]{{{\includegraphics[trim=0.0cm 0.0cm 0.0cm 0.0cm, clip=true, width=0.40\textwidth]{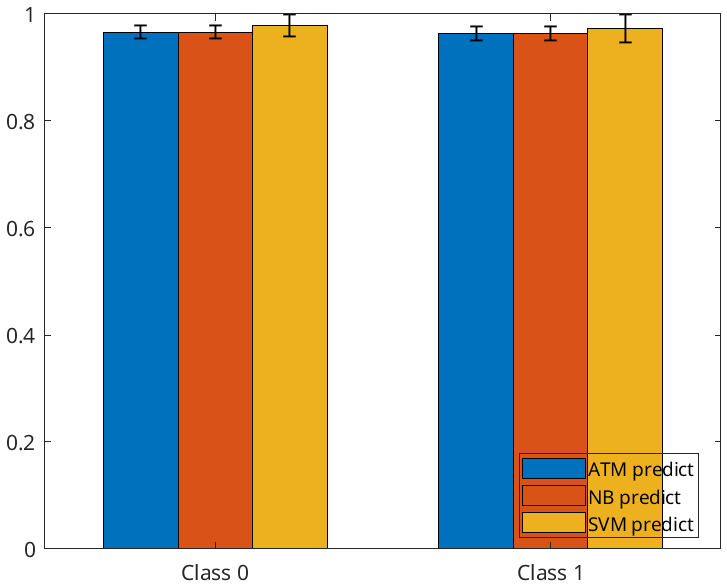}}}
    \label{fig:fscores_zh_test5}}
    \\
    \subfloat[][Averaged \fscores, subset of 20 experiments.  Maximum number of terms allowed in ATM basis function expansions: 2 for class 0, 1 for class 1.]{{{\includegraphics[trim=0.0cm 0.0cm 0.0cm 0.0cm, clip=true, width=0.40\textwidth]{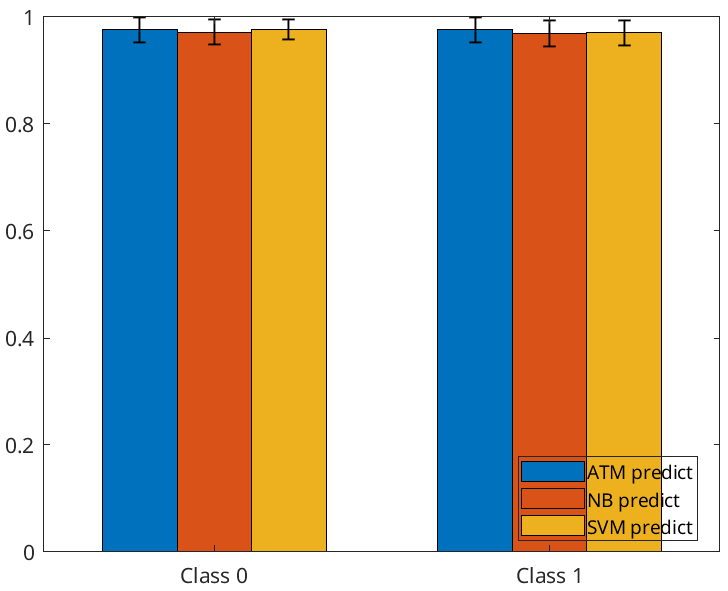}}}
    \label{fig:fscores_zh_test6}}
    \hspace{2ex}
    \subfloat[][Averaged \fscores, subset of 20 experiments.  Maximum number of terms allowed in ATM basis function expansions: 2 for class 0, 2 for class 1.]{{{\includegraphics[trim=0.0cm 0.0cm 0.0cm 0.0cm, clip=true, width=0.40\textwidth]{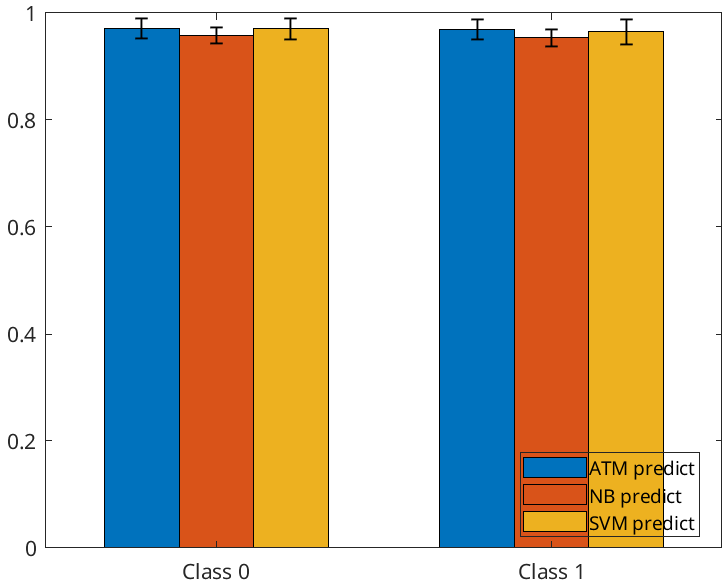}}}
    \label{fig:fscores_zh_test8}}
    \caption{Summary of classification results from adaptive transport maps (ATM) approximations to the class-conditional densities, naive Bayes (NB) classifier, and support vector machine (SVM) classifier.  Zero-dose radiation is class 0; High-dose radiation is class 1.  The available number of samples were \totalsampleszero\ for class 0 and \totalsampleshigh\ for class 1.  For each experiment (i.e., run), the available samples were divided randomly into 50\% for training and 50\% for inference.  The total number of experiments was 100.  For this data set, all methods performed well, and nearly equal, on average.  Classification scores were around the same for class 0 and class 1.
    } 
    \label{fig:fscores_zh}
\end{figure}

Within the context of the classification problem via formula \eqref{eq:classify} from Section \ref{sec:motivating_application}, we performed two groups of experiments.  The first was for classifying samples between zero-dose (class 0) and low-dose (class 1) radiation, using the set of \geneszl\ genes common to both the \gsedataset\ data set and the \pathwayzl\ pathway to define the set of random variables.  In the second group of experiments, classification was  between zero-dose (class 0) and high-dose (class 1) radiation, using the set of \geneszh\ genes common to both the \gsedataset\ data set and the \pathwayzh\ pathway as the set of random variables.  For each of these two ``class 0 - class 1 scenarios'' we then decided on a number of maximum terms allowed in the basis function expansions \eqref{eq:f_expansion} for each map component (refer to the discussion at the end of Section \ref{sec:density_estimation_from_data} and also to Figure \ref{fig:banknote_data}) and, given a maximum number of terms allowed (more on our choices later), a sequence of runs, say $N$ of them, was carried out.  For each of these runs:
\begin{enumerate}[topsep=0pt, partopsep=0pt, itemsep=-3pt]
\item A random ordering of the data samples from the \gsedataset\ data set was performed.
\item For each of class 0 and class 1, the first half of the samples from the resulting ordering was selected as a training data set and the second half was withheld for inference.
\item An adaptive transport map approximation to each target class-conditional density was computed.
\item Classification was done, per the classification formula \eqref{eq:classify}, using the resulting transport map approximations to the class-conditional target densities and the samples held for inference.
\item For comparison, classification was done with Matlab's naive Bayes and support vector machine classifiers using the same training and inference data sets from item 2 above.
\end{enumerate}
This is formally described in Algorithm \ref{alg:ATM_classify}, using the mathematical notation introduced previously in Section \ref{sec:basics_measure_transport}.
The outcomes from the above described runs are summarized in the form of \fscores\ \eqref{eq:f1_score} for the classification results.  These are shown in Figure \ref{fig:fscores_zl} for the case when class 0 and class 1 were set, respectively, as the zero-dose and low-dose radiation classes.  For the zero-dose and high-dose classification scenario, results are summarized in Figure \ref{fig:fscores_zh}.

\begin{algorithm}
\begin{algorithmic}
\State \textbf{Notation:}
\State \hspace*{2ex} $N$: number of runs 
\State \hspace*{2ex} $m \ge 1$: number of genes in subset of genes extracted from  \gsedataset\ data set
\State \hspace*{2ex} $y\in\mathbb{R}^{m}$: values (gene expression data) of a random vector $\mathcal{Y}=\left(\mathcal{Y}_{1},\mathcal{Y}_{2},\dots,\mathcal{Y}_{m}\right)$
\State \hspace*{2ex} $K$: number of classes
\State \hspace*{2ex} $\targetDensity_{k}$: (unknown) target probability density for class $k \in \{0, \ldots, K-1\}$
\State \hspace*{2ex} $\targetSet_{k} = \{y \in\mathbb{R}^{m}\,:\, y \sim {\nu_{\targetDensity_{k}}}\}$: set of all samples for class $k \in \{0,1,\ldots,K-1\}$
\State \hspace*{2ex} $n_{k}=\mathrm{Card}\left(\targetSet_{k}\right)$: number of samples in ${\targetSet_{k}}$, $k \in \{0,1,\ldots,K-1\}$
\State \hspace*{2ex} $\mathcal{S}_{n_{k}}$: group of permutations on $n_{k}$ elements, acting on $\targetSet_{k}$, with $\mathrm{Card}\left(\mathcal{S}_{n_{k}}\right)=n_{k}!$
\State \textbf{Algorithm:}
\end{algorithmic}
\begin{algorithmic}[1]
\For{$n=1$ \mto\ $N$}
    \For{\textbf{each} $k\in\{0,1,\ldots,K-1\}$} 
        \State Choose an order on the set $\targetSet_{k}$ 
        \State Choose randomly an element $\sigma\in\mathcal{S}_{n_{k}}$.
        \State Compute $\sigma\left(\targetSet_{k}\right)$, i.e., act on $\targetSet_{k}$ by $\sigma$ and reorder it.
        \State Split the set $\sigma\left(\targetSet_{k}\right)=\sigma\left(\targetSet_{k}\right)^{\mathrm{tr}}\sqcup\sigma\left(\targetSet_{k}\right)^{\mathrm{inf}}$ along this order into a disjoint  \label{alg:split_set}
        \NoNumber union of the set $\sigma\left(\targetSet_{k}\right)^{\mathrm{tr}}$ of its first $\lfloor\frac{n_{k}}{2}\rfloor$ elements (designated as training  
        \NoNumber set) and the set $\sigma\left(\targetSet_{k}\right)^{\mathrm{inf}}$ of the remaining $n_{k}-\lfloor\frac{n_{k}}{2}\rfloor$ elements (designated
        \NoNumber as inference set).
        \State Learn an adaptive transport map approximation $T_{k}$ from the
        \NoNumber reference $\referenceDensity$ to the target density ${\targetDensity_{k}}$ using only  $\sigma\left(\targetSet_{k}\right)^{\mathrm{tr}}$ as a training set.  
        \NoNumber That is, $T_{k}$ is computed to satisfy the condition 
        \NoNumber \ $\nu_{\targetDensity_{k}}\left(\sigma\left(\targetSet_{k}\right)^{\mathrm{tr}}\right)=\referenceMeasure({T_{k}^{-1}}\left(\sigma\left(\targetSet_{k}\right)^{\mathrm{tr}}\right))$. (\cf. equation \eqref{eq:measure_transport})
    \EndFor  \label{alg:break_line} 
    \For{\textbf{each} $k\in\{0,1,\ldots,K-1\}$} 
        \State Compute $\priorClassDensity_{k}=\lfloor\frac{n_{k}}{2}\rfloor\diagup\sum_{s=0}^{K-1}\lfloor\frac{n_{s}}{2}\rfloor$, the estimates for the class prior 
        \NoNumber probabilities. (\cf. equation \eqref{eq:classify})
    \EndFor  
    \For{\textbf{each} $k\in\{0,1,\ldots,K-1\}$} 
        \For{\textbf{each} $y\in\sigma\left(\targetSet_{k}\right)^{\mathrm{inf}}$ (the inference set)} 
            \State Compute the probability $\mathrm{Pr}(C = k \, | \, \mathcal{Y} = y) =   {\priorClassDensity_{k} \,\targetDensity_{k}(y)}\diagup{\sum_{s=0}^{K-1} \priorClassDensity_{s} \,\targetDensity_{s}(y)}$ 
            \NoNumber that sample $y$ belongs to class $k$ (\cf. equation \eqref{eq:classify}), where $\targetDensity_{s}(y)$ is 
            \NoNumber estimated via equation \eqref{eq:pullback_reference} using the already computed (in step 7 above)
            \NoNumber transport map $T_{s}$, for each $s\in\{0,1,\ldots,K-1\}$.
        \EndFor  
    \EndFor  
    \State (Optional) For comparison, perform classification with Matlab's naive Bayes 
    \NoNumber and support vector machine classifiers using the same training and inference
    \NoNumber data sets from line \ref{alg:split_set} above.
\EndFor  
\end{algorithmic}
\caption{Classification experiments with adaptive transport maps}
\label{alg:ATM_classify}
\end{algorithm}


We note that for the experiments conducted, all classifiers ended up performing well.  This was the case also in other preliminary experiments with the selected \gsedataset\ data set.  A small number of terms (1-2 for the zero- and high-dose radiation classes; 1-6 for the low-radiation class) in the adaptive transport maps basis function expansions \eqref{eq:f_expansion} (refer to the discussion at the end of Section \ref{sec:density_estimation_from_data} and also to Figure \ref{fig:banknote_data}) were sufficient for the approximations.  Varying the maximum number of allowed terms in the transport maps basis function expansions \eqref{eq:f_expansion} did not yield appreciable differences in the results, contrary to the (non-biology) example from Figure \ref{fig:banknote_data}.  However, our experiments (as previous works from others) illustrate that the same measure transport framework can be used successfully (in practice) for diverse data sets.

\subsubsection{Adaptive Transport Maps as a Tool for Scientific Discovery}
\label{sec:ATM_tool_scientific_discovery}

We now depart from the phenotypic classification problems of Sections \ref{sec:DAG_experiments} and \ref{sec:ATM_classification} and consider the problem of inferring the dependence structure among random variables in the unknown target measure, which was previously mentioned in Section \ref{sec:motivating_application} as being also of interest to us.

As shown in the original publication \cite{baptista2022_ATM}, adaptive transport maps approximations can reveal (unknown) dependence structure among the set of random variables jointly distributed according to the target probability density.  Thus, adaptive transport maps have the potential to serve as tools for scientific discovery.  Recall from Section \ref{sec:triangular_TMs} (refer also to Figure \ref{fig:TM_sparsity_example}) that information about the dependence structure of the random variables in the target measure is encoded in the triangular transport map sparsity patterns learned from data samples. (That is, triangular transport map sparsity patterns reflect structure in the target space, and should not be thought of as related to a statement about the number of training samples used to learn the transport maps.) For our radiation biology application, a natural question would then be whether the sparsity patterns resulting from the transport maps approximations to the target densities contain or reveal any biological information, for example, relationships between the genes associated with the transport map sparsity patterns.  This is another task within the workflow introduced in Section \ref{sec:motivating_application}, independent of the classification problems already explored in Sections \ref{sec:DAG_experiments} and \ref{sec:ATM_classification}, where the measure transport approach can have an impact. 

For instance, the transport map sparsity pattern shown in Figure \ref{fig:illustration_sparse_map} resulted from learning a transport map for estimating the class-conditional density for low-dose radiation, trained using a subset of five genes from the \gsedataset\ data set as the random variables jointly distributed according to the target measure.  From the sparsity pattern in Figure \ref{fig:illustration_sparse_map}, one question that arises naturally is whether within the radiation biology domain there are known relationships that correspond to the exhibited transport map sparsity pattern, or whether there may exist unknown relationships, among the genes associated with random variables $\rv{Y}_{3}$ and $\rv{Y}_{4}$ (see row for 4th map component in Figure \ref{fig:illustration_sparse_map}) or between the genes associated with the random variables $\rv{Y}_{1}$, $\rv{Y}_{2}$, $\rv{Y}_{3}$ and $\rv{Y}_{5}$ (see row for 5th map component in Figure \ref{fig:illustration_sparse_map}).  Thus there is potential for hypotheses generation and subsequent expert verification.  In addition, within the context of the computational workflow from Figure \ref{fig:proposed_workflow}, the resulting transport map sparsity patterns may conceivably be useful for designing queries (regarding gene relationships) for large language models as well \cite{park2023automated,park2023comparative}.

\begin{figure}[t] 
    \centering
    \subfloat[][]{{{\includegraphics[trim=0.0cm 0.0cm 0.0cm 0.0cm, clip=true, width=0.25\textwidth]{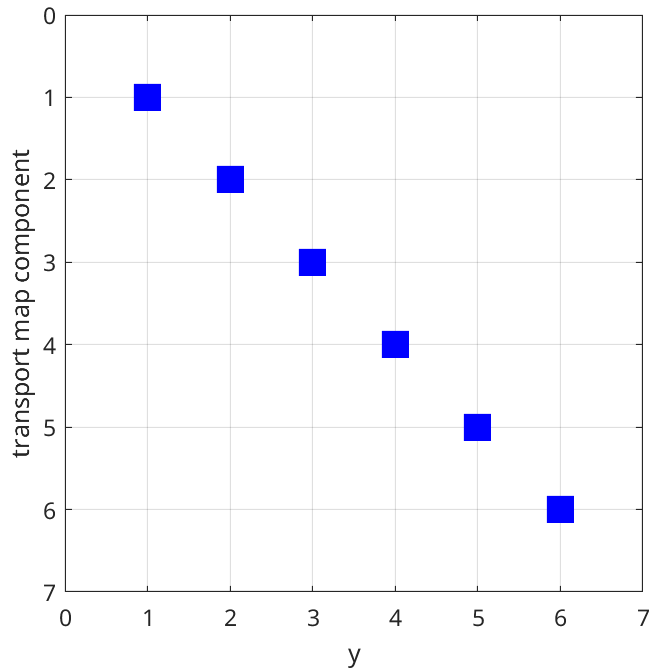}}}
    \label{fig:sp_ex1_C0}}
    \hspace{2ex}
    \subfloat[][]{{{\includegraphics[trim=0.0cm 0.0cm 0.0cm 0.0cm, clip=true, width=0.25\textwidth]{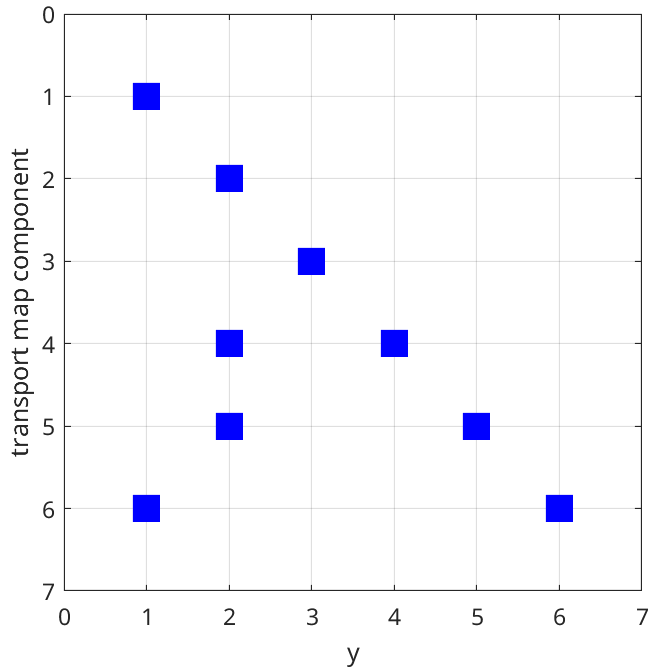}}}
    \label{fig:sp_ex2_C0}}
    \hspace{2ex}
    \subfloat[][]{{{\includegraphics[trim=0.0cm 0.0cm 0.0cm 0.0cm, clip=true, width=0.25\textwidth]{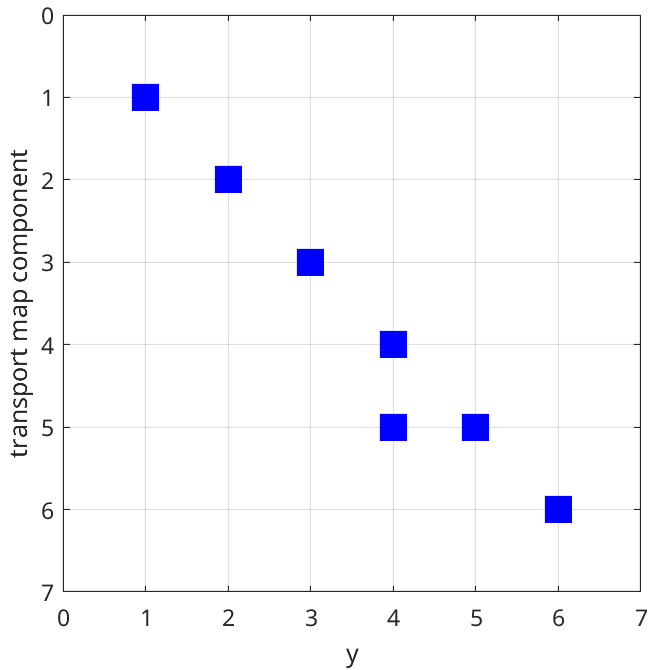}}}
    \label{fig:sp_ex3_C0}}
    \\
    \subfloat[][]{{{\includegraphics[trim=0.0cm 0.0cm 0.0cm 0.0cm, clip=true, width=0.25\textwidth]{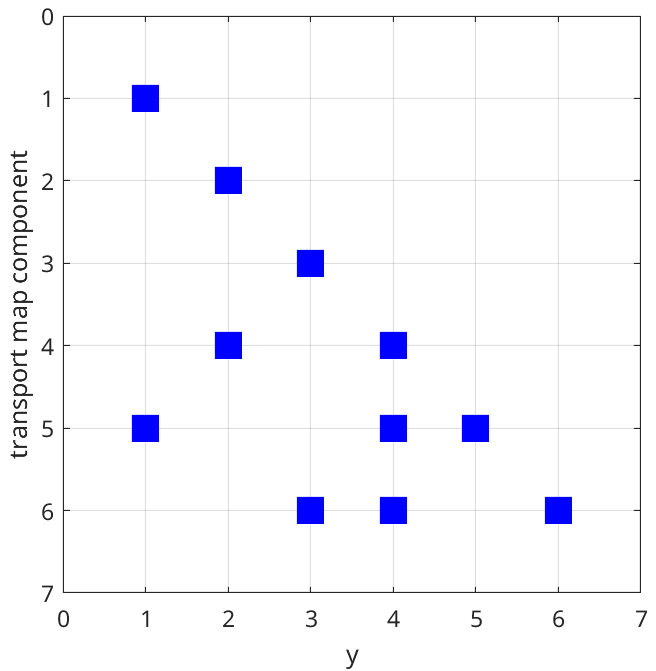}}}
    \label{fig:sp_ex1_C1}}
    \hspace{2ex}
    \subfloat[][]{{{\includegraphics[trim=0.0cm 0.0cm 0.0cm 0.0cm, clip=true, width=0.25\textwidth]{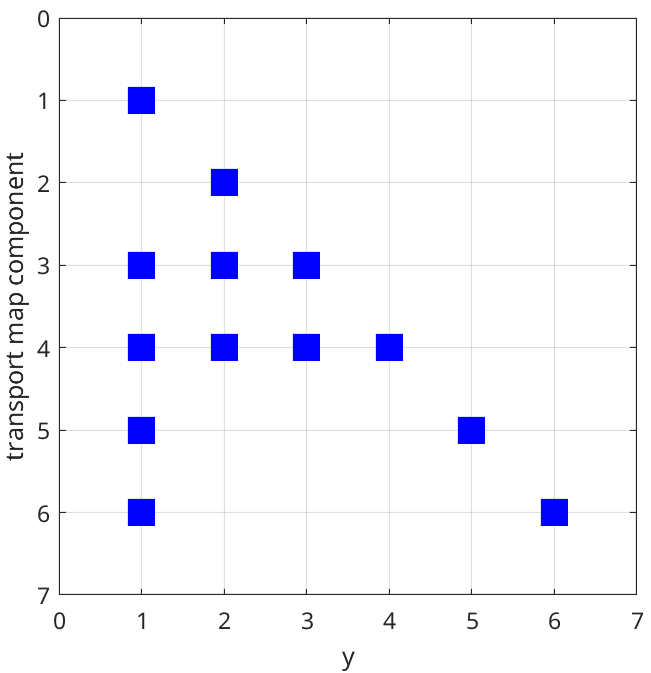}}}
    \label{fig:sp_ex2_C1}}
    \hspace{2ex}
    \subfloat[][]{{{\includegraphics[trim=0.0cm 0.0cm 0.0cm 0.0cm, clip=true, width=0.25\textwidth]{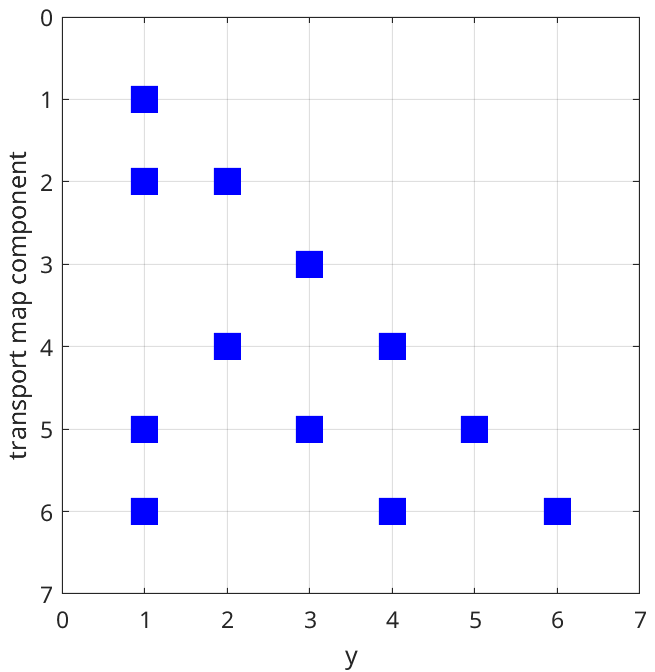}}}
    \label{fig:sp_ex3_C1}}
    \caption{Due to the small number of total samples available for training, resulting adaptive transport map sparsity patterns may differ when different training subsets of the total samples set are used to learn adaptive transport maps.  Subfigures \ref{fig:sp_ex1_C0}--\ref{fig:sp_ex3_C0} exemplify this for three select subsets (of 9 training samples each) from the zero-dose radiation class.  Similarly for the low-dose radiation class in Subfigures \ref{fig:sp_ex1_C1}--\ref{fig:sp_ex3_C1}, where each subset had 44 training samples for learning the corresponding adaptive transport map.  Statistics collected from sparsity pattens from a series of learned adaptive transport maps may nevertheless allow for inferring from data information about the dependency structure among random variables in the target measure, as shown in Figures \ref{fig:sparsity_aggregated_class0}--\ref{fig:sparsity_aggregated_genes}.
    } 
    \label{fig:sparsity_individual}
\end{figure}

Now, given a sufficient number of data samples, one can expect that a triangular transport map sparsity pattern resulting from an adaptive transport map density approximation to the data distribution will accurately uncover information hidden in the data about dependencies or associations among the random variables. However, when the number of training samples is limited in amount ($O(1)$--$O(10)$ samples in our case), it may be difficult to infer dependence structure in the target measure by learning a single adaptive transport map, as we discuss in Figure \ref{fig:sparsity_individual}.  Nevertheless, even when the number of training samples is limited in amount, we hypothesize that one may be able to use adaptive transport map sparsity patterns for inferring dependency structure among random variables in the target measure.

\begin{figure}
    \centering
    \subfloat[][]{{{\includegraphics[trim=0.0cm 0.0cm 0.0cm 0.0cm, clip=true, width=0.30\textwidth]{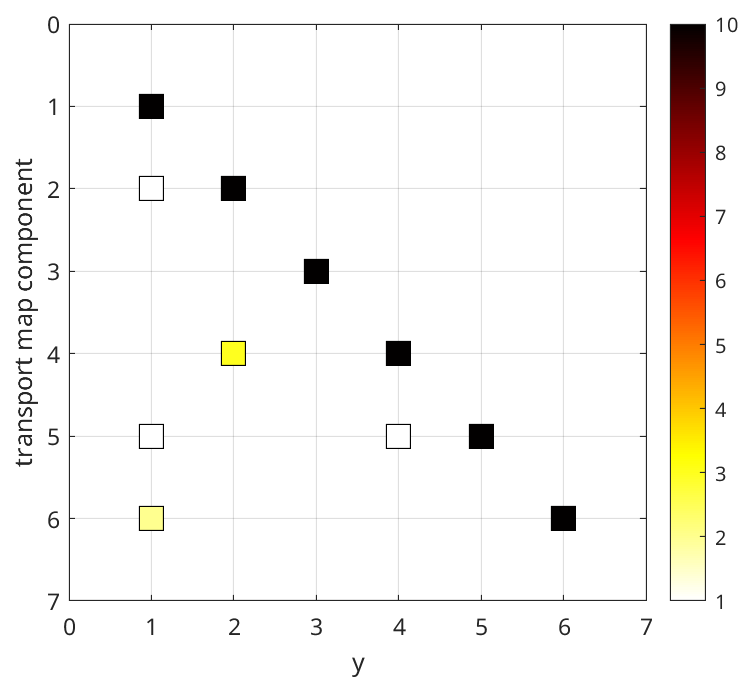}}}
    \label{fig:SP0_10}}
    \hspace{2ex}
    \subfloat[][]{{{\includegraphics[trim=0.0cm 0.0cm 0.0cm 0.0cm, clip=true, width=0.30\textwidth]{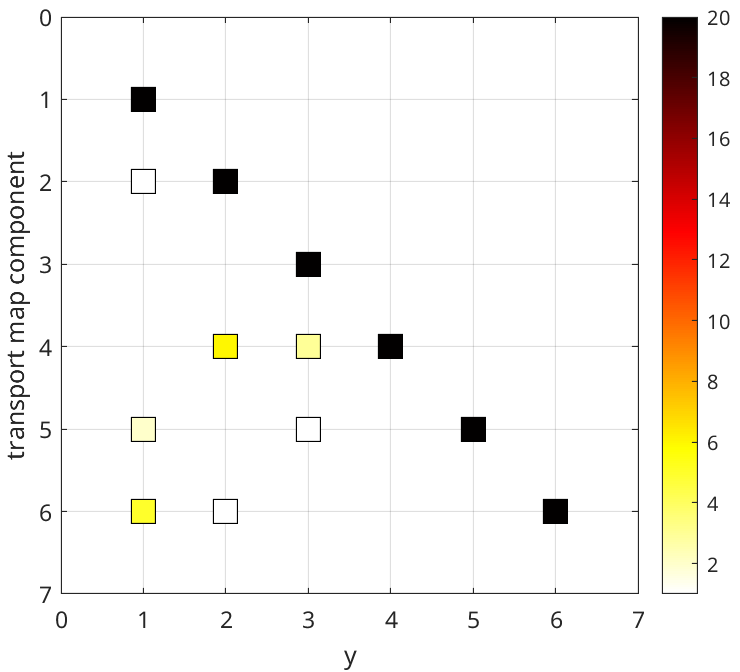}}}
    \label{fig:SP0_20}}
    \hspace{2ex}
    \subfloat[][]{{{\includegraphics[trim=0.0cm 0.0cm 0.0cm 0.0cm, clip=true, width=0.30\textwidth]{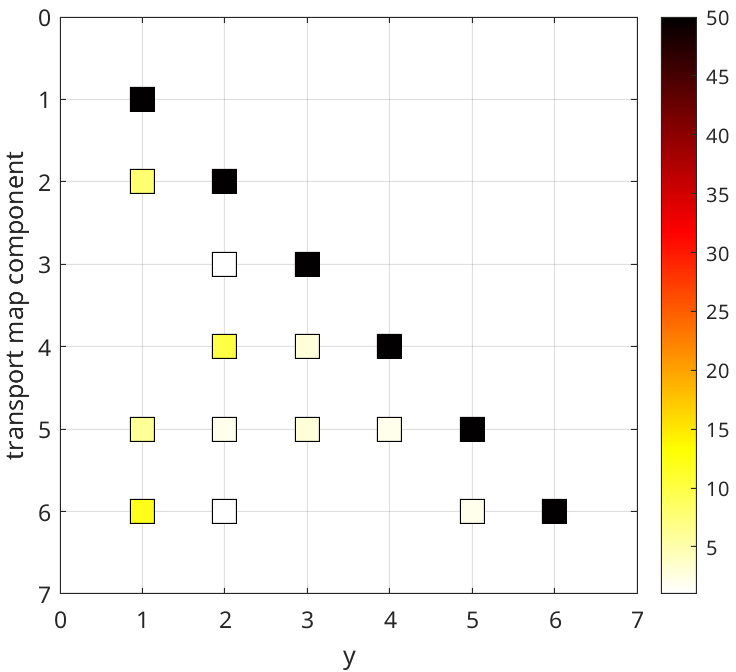}}}
    \label{fig:SP0_50}}
    \\
    \subfloat[][]{{{\includegraphics[trim=0.0cm 0.0cm 0.0cm 0.0cm, clip=true, width=0.30\textwidth]{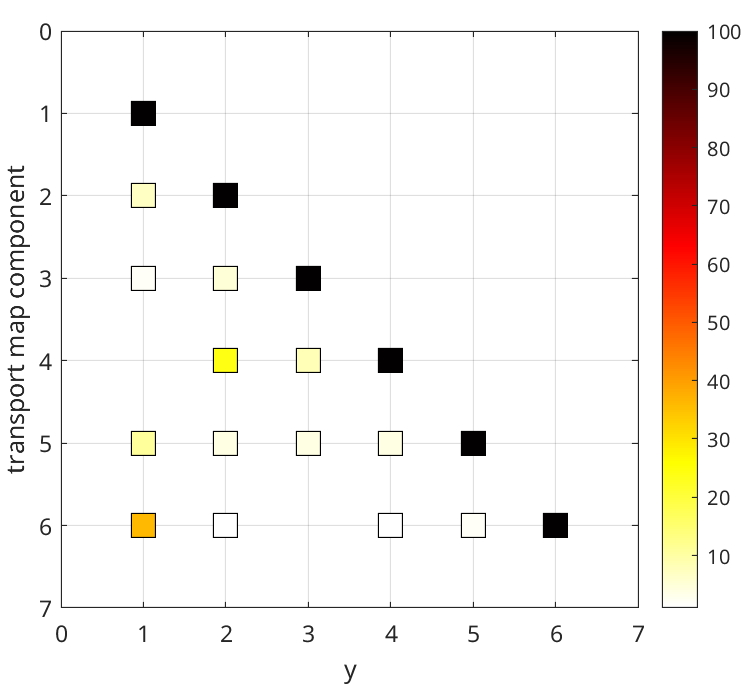}}}
    \label{fig:SP0_100}}
    \hspace{2ex}
    \subfloat[][]{{{\includegraphics[trim=0.0cm 0.0cm 0.0cm 0.0cm, clip=true, width=0.30\textwidth]{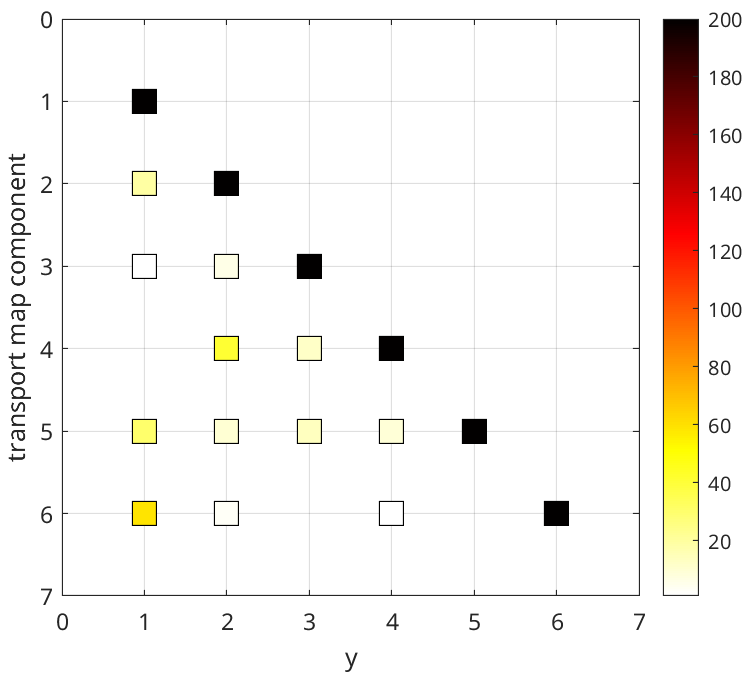}}}
    \label{fig:SP0_200}}
    \hspace{0ex} 
    \subfloat[][]{\footnotesize \addtolength{\tabcolsep}{-3.0pt} \renewcommand{\arraystretch}{1.2}
        \begin{tabular*}{0.32\textwidth}[b]{ l | r r r r r r } 
        $S_1$ & 200  &   0  &   0 &   0  &   0  &   0 \\
        $S_2$ &  19  & 200  &   0 &   0  &   0  &   0 \\
        $S_3$ &   1  &   6  & 200 &   0  &   0  &   0 \\
        $S_4$ &   0  &  41  &  12 & 200  &   0  &   0 \\
        $S_5$ &  30  &  10  &  14 &   9  & 200  &   0 \\
        $S_6$ &  59  &   3  &   0 &   1  &   0  & 200  \\
        \hline 
              & $\rv{Y}_{1}$ & $\rv{Y}_{2}$ & $\rv{Y}_{3}$ & $\rv{Y}_{4}$ & $\rv{Y}_{5}$ & $\rv{Y}_{6}$
        \\ \multicolumn{7}{c}{ }
        \end{tabular*} 
    \label{fig:SP0_200_freq}}
    \caption{Aggregated sparsity patterns from adaptive transport map approximations to data distributions for different subsets of the zero-dose radiation data samples.  
    The permutation on the ordered set $\Yt \equiv \rvsubset$ of random variables for the computational experiments summarized in Subfigures \ref{fig:SP0_10}--\ref{fig:SP0_200_freq} is the identity permutation $\ps{1} = (\rv{Y}_{1}\ \rv{Y}_{2}\ \rv{Y}_{3}\ \rv{Y}_{4}\ \rv{Y}_{5}\ \rv{Y}_{6})$ (see Table \ref{table:rvs_and_genes} from Appendix \ref{app:ATM_experiments}).
    Subsets of samples consisting of half the number of total samples (which for the zero-dose radiation class is \totalsampleszero) were generated from random permutations of the \gsedataset\ data set sample ordering.  In Subfigures \ref{fig:SP0_10}, \ref{fig:SP0_20}, \ref{fig:SP0_50}, \ref{fig:SP0_100}, and \ref{fig:SP0_200} are represented the sparsity patterns corresponding to, respectively, 10, 20, 50, 100, and 200 such sample subsets.  By aggregating results for increasing numbers of sample subsets, patterns of high-valued frequency counts are seen to emerge and persist.  
    The groups of runs represented in Subfigures \ref{fig:SP0_10}--\ref{fig:SP0_200} were independent, so low-valued frequency counts in the depicted sparsity patterns need not persist as we move from Subfigure \ref{fig:SP0_10} to Subfigure \ref{fig:SP0_200}. 
    This was the case for grid point (5,6) in Subfigures \ref{fig:SP0_100} and \ref{fig:SP0_200}, which shows a frequency count of two on the value $y_{5}$ of the random variable $\rv{Y}_{5}$ for map component $S_{6}$ in Subfigure \ref{fig:SP0_100}, but a frequency count of zero in Subfigure \ref{fig:SP0_200}.  
    For reference, Subfigure \ref{fig:SP0_200_freq} shows the frequency counts corresponding to the sparsity pattern from Subfigure \ref{fig:SP0_200}.  In other words, the matrix shown in Subfigure \ref{fig:SP0_200_freq} is the matrix $\FCperm{0}{(200,\sigmap{1})}$ from Algorithm \ref{alg:ATM_statistics_1}, lines \ref{alg:start_B_loop}--\ref{alg:end_B_loop}.
    To simplify notation in Subfigure \ref{fig:SP0_200_freq}, we denote $S \equiv T^{-1}$ for the inverse $T^{-1}$ of a transport map $T$.
    } 
    \label{fig:sparsity_aggregated_class0}
\end{figure}

\begin{figure}
    \centering
    \subfloat[][]{{{\includegraphics[trim=0.0cm 0.0cm 0.0cm 0.0cm, clip=true, width=0.30\textwidth]{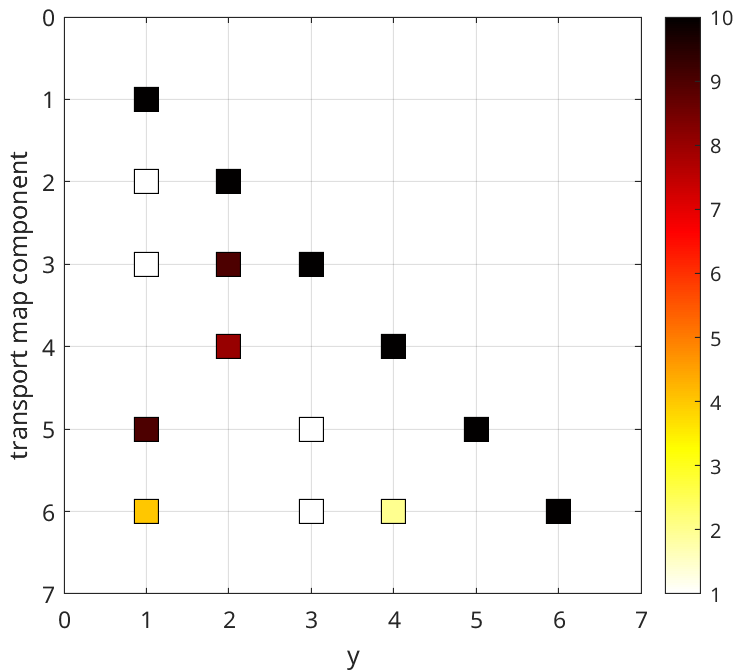}}}
    \label{fig:SP1_10}}
    \hspace{2ex}
    \subfloat[][]{{{\includegraphics[trim=0.0cm 0.0cm 0.0cm 0.0cm, clip=true, width=0.30\textwidth]{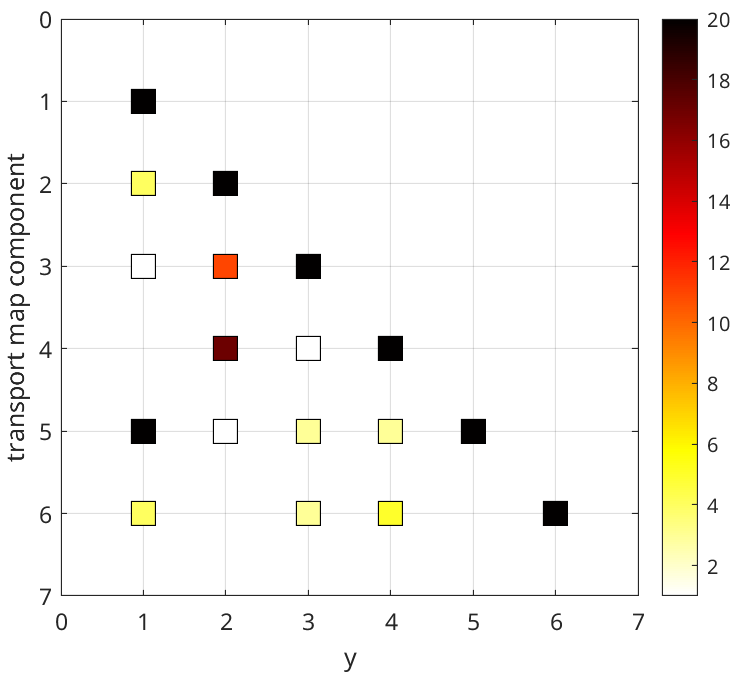}}}
    \label{fig:SP1_20}}
    \hspace{2ex}
    \subfloat[][]{{{\includegraphics[trim=0.0cm 0.0cm 0.0cm 0.0cm, clip=true, width=0.30\textwidth]{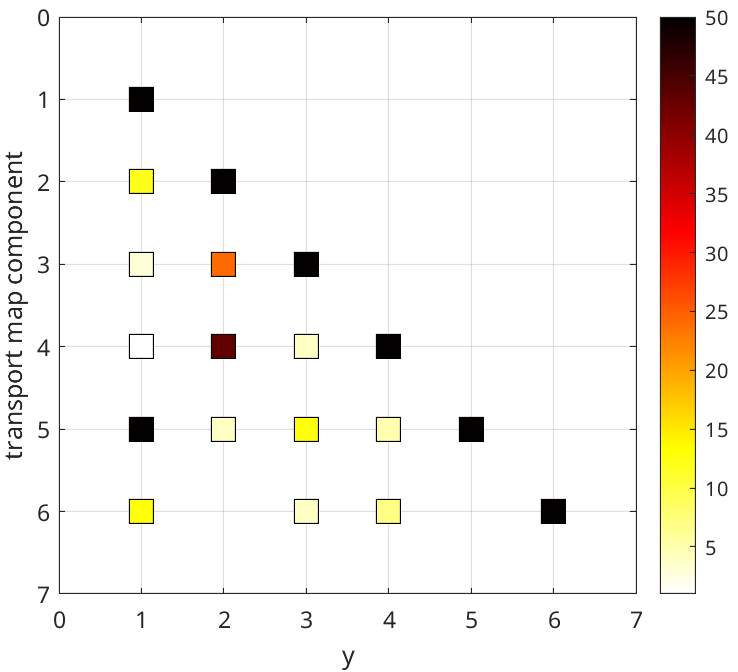}}}
    \label{fig:SP1_50}}
    \\
    \subfloat[][]{{{\includegraphics[trim=0.0cm 0.0cm 0.0cm 0.0cm, clip=true, width=0.30\textwidth]{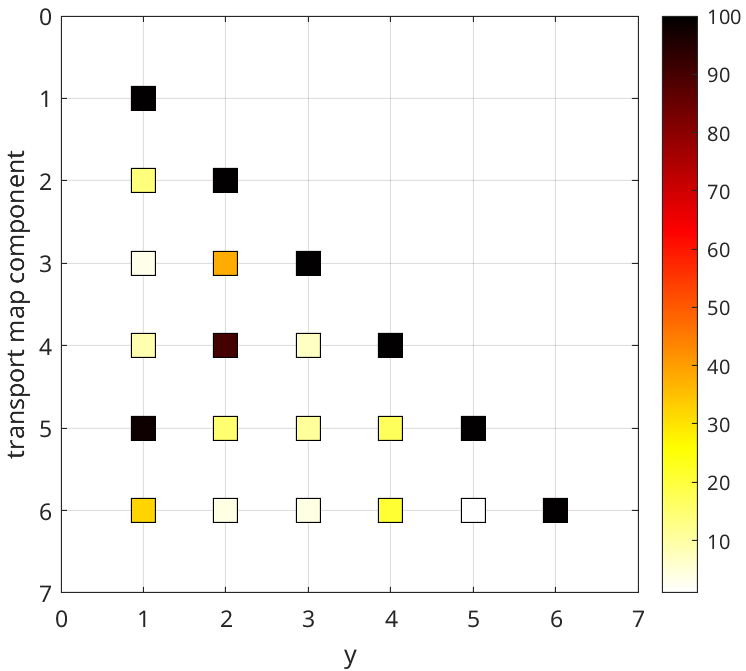}}}
    \label{fig:SP1_100}}
    \hspace{2ex}
    \subfloat[][]{{{\includegraphics[trim=0.0cm 0.0cm 0.0cm 0.0cm, clip=true, width=0.30\textwidth]{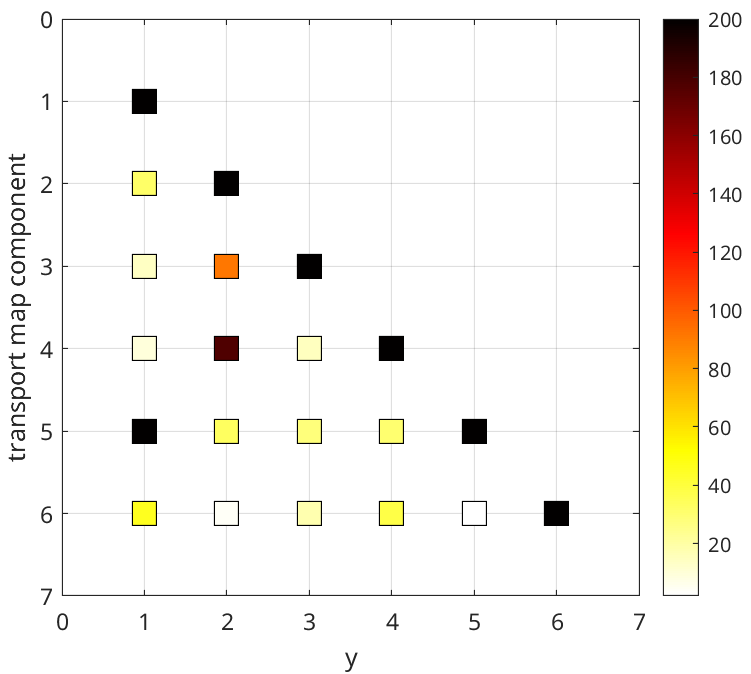}}}
    \label{fig:SP1_200}}
    \hspace{0ex} 
    \subfloat[][]{\footnotesize \addtolength{\tabcolsep}{-3.0pt} \renewcommand{\arraystretch}{1.2}
        \begin{tabular*}{0.32\textwidth}[b]{ l | r r r r r r } 
        $S_1$ & 200 &  0  &   0  &   0  &   0  &   0 \\
        $S_2$ &  32 & 200 &   0  &   0  &   0  &   0 \\
        $S_3$ &  14 &  91 &  200 &   0  &   0  &   0 \\
        $S_4$ &   9 & 177 &   15 & 200  &   0  &   0 \\
        $S_5$ & 200 &  33 &   28 &  30  & 200  &   0 \\
        $S_6$ &  46 &   4 &   18 &  38  &   2  & 200  \\
        \hline 
              & $\rv{Y}_{1}$ & $\rv{Y}_{2}$ & $\rv{Y}_{3}$ & $\rv{Y}_{4}$ & $\rv{Y}_{5}$ & $\rv{Y}_{6}$
        \\ \multicolumn{7}{c}{ }
        \end{tabular*}
    \label{fig:SP1_200_freq}}
    \caption{Aggregated sparsity patterns from adaptive transport map approximations to data distributions for different subsets of the low-dose radiation data samples.  
    The permutation on the ordered set $\Yt \equiv \rvsubset$ of random variables for the computational experiments summarized in Subfigures \ref{fig:SP0_10}--\ref{fig:SP0_200_freq} is the identity permutation $\ps{1} = (\rv{Y}_{1}\ \rv{Y}_{2}\ \rv{Y}_{3}\ \rv{Y}_{4}\ \rv{Y}_{5}\ \rv{Y}_{6})$ (see Table \ref{table:rvs_and_genes} from Appendix \ref{app:ATM_experiments}).
    Subsets  of samples consisting of half the number of total samples (which for the low-dose radiation class is \totalsampleslow) were generated from random permutations of the \gsedataset\ data set sample ordering.  In Subfigures \ref{fig:SP1_10}, \ref{fig:SP1_20}, \ref{fig:SP1_50}, \ref{fig:SP1_100}, and \ref{fig:SP1_200} are represented the sparsity patterns corresponding to, respectively, 10, 20, 50, 100, and 200 such sample  subsets.  As for the zero-dose radiation class in Figure \ref{fig:sparsity_aggregated_class0}, by aggregating results for increasing numbers of  sample subsets, patterns of high-valued frequency counts are seen to emerge and persist. 
    For reference, Subfigure \ref{fig:SP1_200_freq} shows the values of the frequency counts corresponding to the sparsity pattern from Subfigure \ref{fig:SP1_200}.  In other words, the matrix shown in Subfigure \ref{fig:SP1_200_freq} is the matrix $\FCperm{1}{(200,\sigmap{1})}$ from Algorithm \ref{alg:ATM_statistics_1}, lines \ref{alg:start_B_loop}--\ref{alg:end_B_loop}.
    To simplify notation in Subfigure \ref{fig:SP1_200_freq}, we denote $S \equiv T^{-1}$ for the inverse $T^{-1}$ of a transport map $T$.
    } 
    \label{fig:sparsity_aggregated_class1}
\end{figure}

\begin{figure}[!]
    \centering
    \subfloat[][zero-dose radiation class]{{{\includegraphics[trim=0.0cm 0.0cm 0.0cm 0.0cm, clip=true, width=0.39\textwidth]{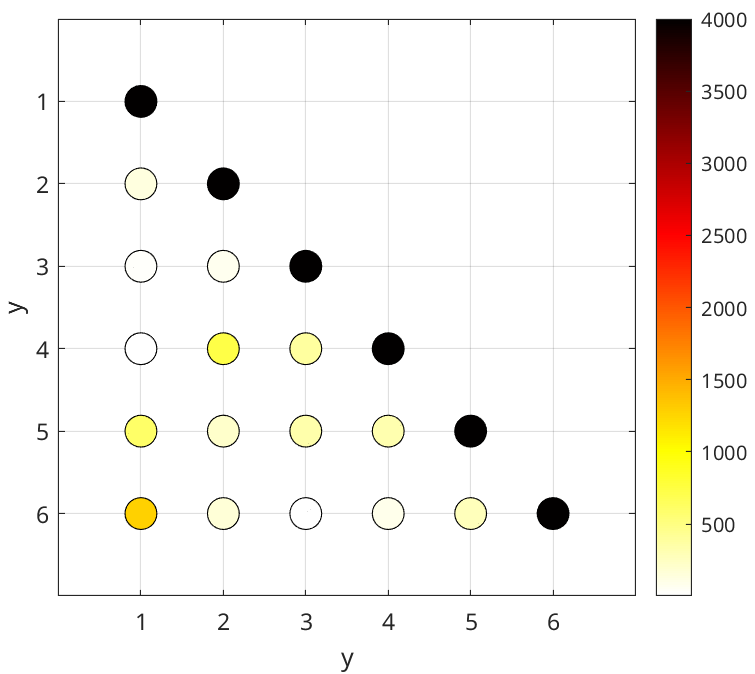}}}
    \label{fig:SP0_4000}}
    \hspace{2ex} 
    \subfloat[][$\FCall{0}{(4000)}$]{\footnotesize \addtolength{\tabcolsep}{-1.0pt} \renewcommand{\arraystretch}{1.2}
        \begin{tabular*}{0.46\textwidth}[b]{ l | r r r r r r } 
        $\rv{Y}_{1}$ & 4000 &    0 &     0 &    0  &    0  &    0 \\
        $\rv{Y}_{2}$ &  132 & 4000 &     0 &    0  &    0  &    0 \\
        $\rv{Y}_{3}$ &   20 &   69 &  4000 &    0  &    0  &    0 \\
        $\rv{Y}_{4}$ &   10 &  721 &   386 & 4000  &    0  &    0 \\
        $\rv{Y}_{5}$ &  600 &  212 &   337 &  321  & 4000  &    0 \\
        $\rv{Y}_{6}$ & 1273 &  160 &     1 &   85  &  279  & 4000 \\
        \hline 
              & $\rv{Y}_{1}$ & $\rv{Y}_{2}$ & $\rv{Y}_{3}$ & $\rv{Y}_{4}$ & $\rv{Y}_{5}$ & $\rv{Y}_{6}$
        \\ \multicolumn{7}{l}{ }
        \\ \multicolumn{7}{l}{ }
        \end{tabular*} 
    \label{fig:SP0_4000_freq}}
    \\
    \subfloat[][low-dose radiation class]{{{\includegraphics[trim=0.0cm 0.0cm 0.0cm 0.0cm, clip=true, width=0.39\textwidth]{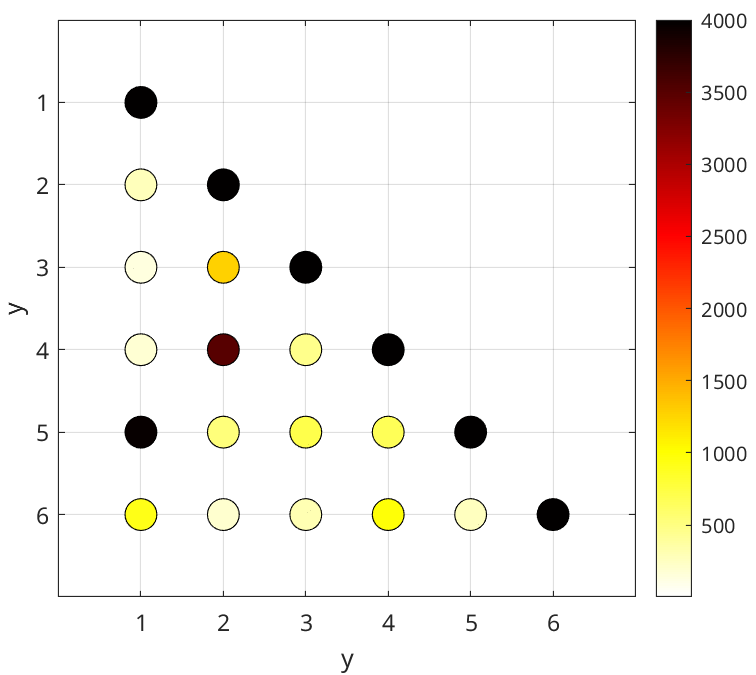}}}
    \label{fig:SP1_4000}}
    \hspace{2ex} 
    \subfloat[][$\FCall{1}{(4000)}$]{\footnotesize \addtolength{\tabcolsep}{-1.0pt} \renewcommand{\arraystretch}{1.2}
        \begin{tabular*}{0.46\textwidth}[b]{ l | r r r r r r } 
        $\rv{Y}_{1}$ & 4000 &    0 &     0 &    0  &    0  &    0 \\
        $\rv{Y}_{2}$ &  277 & 4000 &     0 &    0  &    0  &    0 \\
        $\rv{Y}_{3}$ &  137 & 1275 &  4000 &    0  &    0  &    0 \\
        $\rv{Y}_{4}$ &  185 & 3486 &   455 & 4000  &    0  &    0 \\
        $\rv{Y}_{5}$ & 3966 &  529 &   712 &  661  & 4000  &    0 \\
        $\rv{Y}_{6}$ &  922 &  203 &   313 &  985  &  258  & 4000 \\
        \hline 
              & $\rv{Y}_{1}$ & $\rv{Y}_{2}$ & $\rv{Y}_{3}$ & $\rv{Y}_{4}$ & $\rv{Y}_{5}$ & $\rv{Y}_{6}$
        \\ \multicolumn{7}{l}{ }
        \\ \multicolumn{7}{l}{ }
        \end{tabular*} 
    \label{fig:SP1_4000_freq}}
    \caption{\textit{Subfigures \ref{fig:SP0_4000}--\ref{fig:SP0_4000_freq}}: Frequency counts gathered (see Appendix \ref{app:ATM_experiments}) from sparsity patterns of adaptive transport map approximations to data distributions for different subsets of the zero-dose radiation data samples.
    \textit{Subfigures \ref{fig:SP1_4000}--\ref{fig:SP1_4000_freq}}: As in Subfigures \ref{fig:SP0_4000}--\ref{fig:SP0_4000_freq}, but for the low-dose radiation data samples.
    The above subfigures report the results obtained by executing Algorithm \ref{alg:ATM_statistics_1} for each of the $20$ different permutations $\{\ps{p}\}_{p=1}^{20}$ listed in Table \ref{table:rvs_and_genes} from Appendix \ref{app:ATM_experiments}, specifically for the case $N=200$ (\cf. Figures \ref{fig:SP0_200} and \ref{fig:SP1_200}; also Algorithm \ref{alg:ATM_statistics_1}).  To account collectively for the number of $4,\!000$ runs summarized in the above subfigures, the results obtained for each of the permutations $\{\ps{p}\}_{p=1}^{20}$ were mapped to the random variable ordering corresponding to the identity permutation $\ps{1}$, as described in Algorithm \ref{alg:ATM_statistics_2}.  Specifically, the matrices shown in Subfigures \ref{fig:SP0_4000_freq} and \ref{fig:SP1_4000_freq} resulted from line \ref{alg:line_accumulate_all} of Algorithm \ref{alg:ATM_statistics_2}.
    The random variable to gene ID mapping is $(\rv{Y}_{1}\ \rv{Y}_{2}\ \rv{Y}_{3}\ \rv{Y}_{4}\ \rv{Y}_{5}\ \rv{Y}_{6}) \mapsto (5533\ 5534\ 5530\ 5532\ 4772\ 4773)$. (Refer to Tables \ref{table:nih_genes} and \ref{table:rvs_and_genes} from Appendix \ref{app:ATM_experiments}.)
    Dependencies/relations among genes suggested by the above results are further depicted in Figure \ref{fig:sparsity_aggregated_genes}.
    } 
    \label{fig:sparsity_aggregated_4000_freq}
\end{figure}

\begin{figure}[htbp]
    \centering
    \subfloat[][zero-dose radiation class]{{{\includegraphics[trim=0.0cm 0.0cm 22.0cm 0.0cm, clip=true, width=0.44\textwidth]{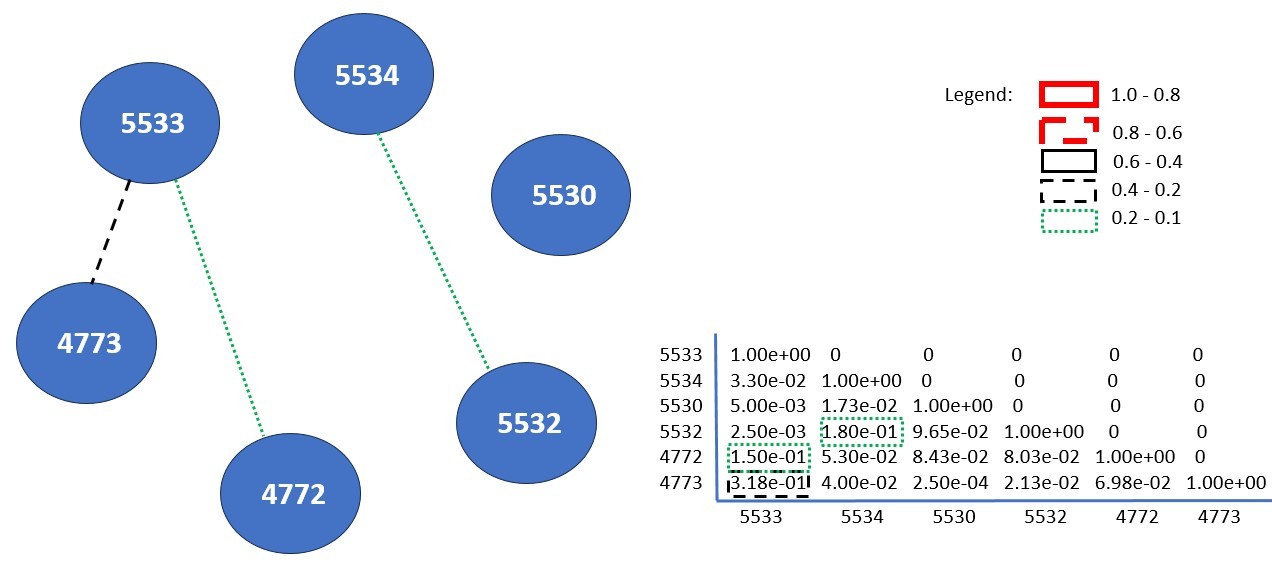}}}
    \label{fig:SP0_4000_genes}}
    \hspace*{2ex}
    \subfloat[][$\FCall{0}{(4000)}$, as a fraction of a total number $4\!,\!000$ of runs]{{{\includegraphics[trim=22.5cm 0.0cm 0.0cm 2.8cm, clip=true, height=0.40\textwidth]{subset_6_genes_all_sigmas_zero_dose.jpg}}}
    \label{fig:SP0_4000_genes_matrix}}
    \\
    \subfloat[][low-dose radiation class]{{{\includegraphics[trim=0.0cm 0.0cm 22.0cm 0.0cm, clip=true, width=0.44\textwidth]{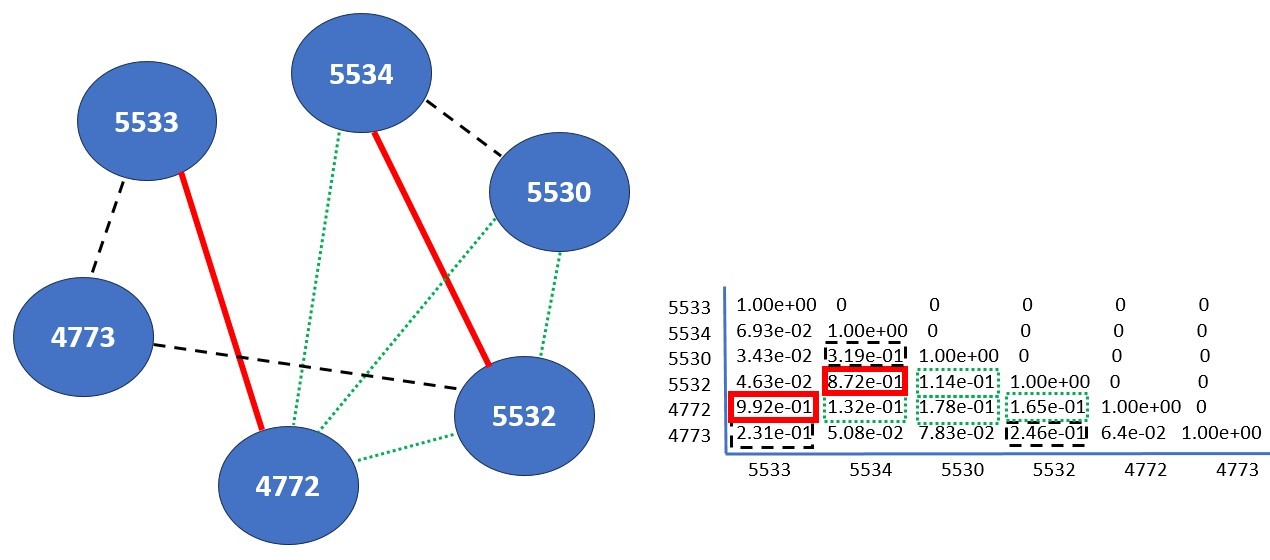}}}
    \label{fig:SP1_4000_genes}}
    \hspace*{2ex}
    \subfloat[][$\FCall{1}{(4000)}$, as a fraction of a total number $4\!,\!000$ of runs]{{{\includegraphics[trim=22.5cm 0.75cm 0.0cm 2.8cm, clip=true, height=0.38\textwidth]{subset_6_genes_all_sigmas_low_dose.jpg}}}
    \label{fig:SP1_4000_genes_matrix}}
    \caption{\textit{Subfigures \ref{fig:SP0_4000_genes}--\ref{fig:SP0_4000_genes_matrix}}: Dependencies among the genes for the zero-dose radiation class, as implied by the sparsity pattern from Figure \ref{fig:SP0_4000} and the associated frequency counts from Figure \ref{fig:SP0_4000_freq}.  
    \textit{Subfigures \ref{fig:SP1_4000_genes}--\ref{fig:SP1_4000_genes_matrix}}: Dependencies among the genes for the low-dose radiation class, as implied by the sparsity pattern from Figure \ref{fig:SP1_4000} and the associated frequency counts from Figure \ref{fig:SP1_4000_freq}.
    Information about the subset of six genes selected for the computational experiments, including the gene IDs, appears in Table \ref{table:nih_genes} from Appendix \ref{app:ATM_experiments}.
    The entries in the tables shown in Subfigures \ref{fig:SP0_4000_genes_matrix} and \ref{fig:SP1_4000_genes_matrix} above contain, respectively, the frequency counts from Figures  \ref{fig:SP0_4000_freq} and \ref{fig:SP1_4000_freq} as a fraction of a total number of $4,\!000$ runs, and mapped to the gene IDs.  These fractional values and the (user defined) thresholds from the legend at the top right corner of Subfigure \ref{fig:SP0_4000_genes_matrix} are used in the above figures to represent a ``strength'' of the relationship between genes, as implied by the sparsity patterns from Figures \ref{fig:SP0_4000} and \ref{fig:SP1_4000}.
    A natural question is whether the above reveal biological information, such as relationships among genes with no radiation exposure (Subfigure \ref{fig:SP0_4000_genes}) that change (appear, disappear) or that persist upon exposure to low-dose radiation (Subfigure \ref{fig:SP1_4000_genes}). 
    } 
    \label{fig:sparsity_aggregated_genes}
\end{figure}

More precisely, we propose a randomized technique designed to gather statistics by computing a series of adaptive transport map density approximations using different (randomly selected) subsets of the sample data as training sets, and then identifying common dominating patterns appearing in the resulting collection of transport map sparsity patterns. We discuss this line of research next and in Figures \ref{fig:sparsity_individual}--\ref{fig:sparsity_aggregated_genes}, where results are shown from our computational experiments. 
In order to develop some intuition into the dependence structure of the genes from the \gsedataset\ data set, without assuming any knowledge that fixes the set of transport map active variables a priori, we consider a subset of six genes from the \gsedataset\ data set as the random variables jointly distributed according to the unknown target measure. 
That is, in what follows we have $\atm\!:\mathbb{R}^{6} \rightarrow \mathbb{R}^{6}$ (analogously, $m=6$ in equation \eqref{eq:triangular_TM_inverse}). 
For reference, information about the selected subset of genes is provided in Table \ref{table:nih_genes} and Figure \ref{fig:kegg_hsa04650_segment}, both appearing in Appendix \ref{app:ATM_experiments}.

Recall from Figure \ref{fig:TM_sparsity_example} that in our pictorial representation of transport map sparsity patterns, the horizontal axis indexes the random variables $\{\rv{Y}_{j}\}_{j=1}^{6}$ and the vertical axis indexes the map components $\{\atmcomponent_{i}\}_{i=1}^{6}$.  A square appears at the intersection of grid point $(j,i)$ if the $i$-th map component $\atmcomponent_{i}$ depends on the value $y_{j}$ of the $j$-th random variable $\rv{Y}_{j}$. For any given map component, the set of active variables is the set of random variables that the map component depends on.  The active variables define a sparsity pattern for the transport map.  We emphasize again that the adaptive transport map computation methods from \cite{baptista2022_ATM} (summarized in Appendix \ref{app:ATM_framework} of the present paper) determine the set of active variables, without specification of it prior to learning the transport map.

Due to the small number of total samples available in the \gsedataset\ data set (18 in the zero-dose category, 87 in the low-dose category, and 16 in the high-dose category -- see the beginning of Section \ref{sec:application_GSE_dataset}), transport map sparsity patterns resulting from adaptive transport map approximations to the target measure may differ when different subsets of the total samples set are used to learn transport maps.  Figure \ref{fig:sparsity_individual} exemplifies this for the zero-dose and low-dose radiation classes.  However, as already noted, we hypothesize that statistics collected from sequences of learned adaptive transport maps may nevertheless allow for adaptive transport map sparsity patterns to be used in order to infer structure in the target measure from information hidden (or encoded) in the data.  The computations we perform to collect such statistics are described in Algorithm \ref{alg:ATM_statistics_1} and outlined next.

\begin{algorithm}
\newcommand{\samplesSet}{Z}
\begin{algorithmic}
\State \textbf{Notation:}
\State \hspace*{2ex} $N$: number of transport maps in the series of adaptive transport maps to be learned 
    \State \hspace*{5ex} (see line \ref{alg:start_r_loop} below)
\State \hspace*{2ex} $\TkNr$: $r$-th map in the series of $N$ adaptive transport maps learned for Class $k$ 
\State \hspace*{2ex} $m \ge 1$: number of genes in subset of genes extracted from \gsedataset\ data set
\State \hspace*{2ex} $\sigmap{p}$: permutation on $m$ elements, with inverse $\sigmapinv{p}$, where $p\in\{1,2,\ldots,m!\}$ 
    \State \hspace*{5ex} and $\sigmap{1}=I$ is the identity permutation
\State \hspace*{2ex} $\ps{p}$: result of the action of $\sigmap{p}$ on the vector $\rv{Y} = (\rv{Y}_1, \ldots, \rv{Y}_m)$ of random
    \State \hspace*{5ex} variables, for which 
    $\ps{p} =  \left(\rv{Y}_{\sigmapinv{p}(j)}\right)_{j=1}^{m}$ (\cf. Table \ref{table:rvs_and_genes} from Appendix \ref{app:ATM_experiments})
\State \hspace*{2ex} $z\in\mathbb{R}^{m}$: values (gene expression data) of a random vector $\rv{Z} \equiv \ps{p}$
\State \hspace*{2ex} $K$: number of classes
\State \hspace*{2ex} $\targetDensity_{k}$: (unknown) target probability density for Class $k \in \{0, \ldots, K-1\}$
\State \hspace*{2ex} $\samplesSet_{k} = \{z \in\mathbb{R}^{m}\,:\, z \sim {\nu_{\targetDensity_{k}}}\}$: set of all samples for class $k \in \{0,1,\ldots,K-1\}$
\State \hspace*{2ex} $n_{k}=\mathrm{Card}\left(\samplesSet_{k}\right)$: number of samples in ${\samplesSet_{k}}$, $k \in \{0,1,\ldots,K-1\}$
\State \hspace*{2ex} $\mathcal{S}_{n_{k}}$: group of permutations on $n_{k}$ elements, acting on $\samplesSet_{k}$, with $\mathrm{Card}\left(\mathcal{S}_{n_{k}}\right)=n_{k}!$
\State \hspace*{2ex} $\FCperm{k}{(N,\sigmap{p})} \,$: $m \times m$ lower triangular matrix storing frequency counts from sparsity patterns for
    \State \hspace*{5ex} the $N$ adaptive transport maps learned for class $k$ and permutation $\ps{p}$
\State \textbf{Algorithm:}
\end{algorithmic}
\begin{algorithmic}[1]
\State Choose a permutation $\sigmap{p}\in\mathcal{S}_{m}$ on $m$ elements.  \Comment{\acomment{see Table \ref{table:rvs_and_genes}}} \label{alg:fix_sigmap}
\State Let $(\mathcal{Z}_{1},\mathcal{Z}_{2},\dots,\mathcal{Z}_{m}) \equiv \ps{p}$  \label{alg:act_on_Y}
\For{$r=1$ \mto\ $N$}  \label{alg:start_r_loop}
    \For{\textbf{each} $k\in\{0,1,\ldots,K-1\}$}
        \State Initialize $\FCperm{k}{(N,\sigmap{p})}$ to the $m \times m$ (lower triangular) zero matrix.
    \EndFor  \label{alg2_line_break}
    \For{\textbf{each} $k\in\{0,1,\ldots,K-1\}$} 
        \State Choose randomly an element $\sigma\in\mathcal{S}_{n_{k}}$.  \label{alg:line_perm_act1}
        \State Compute $\sigma\left(\samplesSet_{k}\right)$, i.e., act on $\samplesSet_{k}$ by $\sigma$ and reorder it.   \label{alg:line_perm_act2}
        \State Split the set $\sigma\left(\samplesSet_{k}\right)=\sigma\left(\samplesSet_{k}\right)^{\mathrm{tr}}\sqcup\sigma\left(\samplesSet_{k}\right)^{\mathrm{inf}}$ along this order into a disjoint union of  \label{alg:line_perm_act3}
        \NoNumber the set $\sigma\left(\samplesSet_{k}\right)^{\mathrm{tr}}$ of its first $\lfloor\frac{n_{k}}{2}\rfloor$ elements (designated as training set) and the 
        \NoNumber set $\sigma\left(\samplesSet_{k}\right)^{\mathrm{inf}}$ of the remaining $n_{k}-\lfloor\frac{n_{k}}{2}\rfloor$ elements (designated as inference set).
        \State Learn an adaptive transport map approximation $\TkNr$ from the 
        \NoNumber reference $\referenceDensity$ to the target density ${\targetDensity_{k}}$ using only  $\sigma\left(\samplesSet_{k}\right)^{\mathrm{tr}}$ as a training set. 
        \NoNumber That is, $\TkNr$ is computed to satisfy the condition 
        \NoNumber \ $\nu_{\targetDensity_{k}}\left(\sigma\left(\samplesSet_{k}\right)^{\mathrm{tr}}\right)=\referenceMeasure\left(\invTkNr\!\left(\sigma\left(\samplesSet_{k}\right)^{\mathrm{tr}}\right)\right)$. (\cf. equation \eqref{eq:measure_transport})
        \For{$j=1$ \mto\ $m$}  \label{alg:start_B_loop}
            \For{$i=1$ \mto\ $j$}
                \If{$i$-th component of $\invTkNr$ depends on the value of the \label{alg:start_freq_count_if} \\ \hskip\algorithmicindent\hskip\algorithmicindent\hskip\algorithmicindent\hspace*{5ex} $j$-th random variable $\rv{Z}_{j}$ (\cf. Figures \ref{fig:TM_sparsity_example}, \ref{fig:sparsity_individual})}
                    \State $\FCperm{k}{(N,\sigmap{p})}(j,i) = \FCperm{k}{(N,\sigmap{p})}(j,i) \ + \ 1$ \Comment{\acomment{increase frequency count}}  \label{alg:line_freq_count}
                \EndIf  \label{alg:end_freq_count_if}
            \EndFor  
        \EndFor \label{alg:end_B_loop} 
    \EndFor  
\EndFor  
\end{algorithmic}
\caption{Gathering statistics from adaptive transport maps sparsity patterns}
\label{alg:ATM_statistics_1}
\end{algorithm}

Namely, like in Algorithm \ref{alg:ATM_classify} before, we work in Algorithm \ref{alg:ATM_statistics_1} with subsets (as training sets) extracted from the set of data samples after random permutations of the ordered set of \gsedataset\ data samples (lines \ref{alg:line_perm_act1}--\ref{alg:line_perm_act3} in Algorithm \ref{alg:ATM_statistics_1}).  First, we select at random an ordering
\begin{eqnarray}  \label{eq:fix_ordering_sigmap}
    \ps{p}  & = &  \left(\rv{Y}_{\sigmapinv{p}(j)}\right)_{j=1}^{6}
\end{eqnarray}
on the set $\{\rv{Y}_{j}\}_{j=1}^{6}$ of random variables and then fix this ordering (i.e., we fix an ordering on the arguments of the transport maps to be learned -- this corresponds to lines \ref{alg:fix_sigmap}--\ref{alg:act_on_Y} in Algorithm \ref{alg:ATM_statistics_1}), where in \eqref{eq:fix_ordering_sigmap} above we have $\sigmap{p}\in\mathcal{S}_{6}$ -- the permutation group on six ($m=6$) elements. Recall that we consider only triangular maps throughout (see equations \eqref{eq:triangular_TM}--\eqref{eq:triangular_TM_inverse}).  Also, for the purpose of illustration, we restrict our attention to only two ($K=2$) classes: the zero-dose ($k=0$) and low-dose ($k=1$) radiation classes. 

For each class $k\in\{0,1\}$ and a given number $N$, (see loop starting in line \ref{alg:start_r_loop} of Algorithm \ref{alg:ATM_statistics_1}) we learn a series $\{\TkNr\}_{r=1}^{N}$ of adaptive transport maps $\TkNr\!\!\!:\!\mathbb{R}^{6} \!\rightarrow\! \mathbb{R}^{6}$,  $r=1,\ldots,N$, rendering approximations $\targetkNr$ to the target density $\targetDensity_{k}$ on $\mathbb{R}^{6}$ describing the  joint density distribution of the six random variables $\{\rv{Y}_{j}\}_{j=1}^{6}$ in question.  
More precisely, for each class $k\in\{0,1\}$ we performed five \emph{independent} computations of series $\{\TkNr\}_{r=1}^{N}$ of transport maps by executing Algorithm \ref{alg:ATM_statistics_1}, independently, with $N \in \{10,20,50,100,200\}$ (see line \ref{alg:start_r_loop} in Algorithm \ref{alg:ATM_statistics_1}), and with the same fixed order $\ps{p}$ \eqref{eq:fix_ordering_sigmap} of the random variables. 

In doing so, we observed that for each class $k\in\{0,1\}$ (and more so with an increasing $N$) the elements in the series $\{\TkNr\}_{r=1}^{N}$ of adaptive transport maps cluster around a map $\overline{T}_{k}$ the components $\{(\overline{T}_{k}^{-1})_{i}\}_{i=1}^{6}$ of whose  inverse show a particular persistent and dominant pattern of dependence on its arguments, or active variables, which define a sparsity pattern for the inverse of the map $\overline{T}_{k}$.

In order to exhibit this dominant sparsity pattern quantitatively, for each of the zero-dose and low-dose radiation classes, we recorded the frequency (line \ref{alg:line_freq_count} in Algorithm \ref{alg:ATM_statistics_1}) with which each random variable in $\{\rv{Y}_{j}\}_{j=1}^{6}$ appeared as an active variable in the components $\{(\invTkNr)_{i}\}_{i=1}^{6}$, separately within each of the five independently computed series $\{\TkNr\}_{r=1}^{N}$ of adaptive transport maps, for $N=10,20,50,100,200$ as explained above. This corresponds to lines \ref{alg:start_B_loop}--\ref{alg:end_B_loop} in Algorithm \ref{alg:ATM_statistics_1}.  Using these frequency counts we then constructed an ``aggregated'' sparsity pattern for each of the five series  $\{\TkNr\}_{r=1}^{N}$, $N=10,20,50,100,200$, for each one of the two classes $k\in\{0,1\}$.  The results are shown in Figures \ref{fig:sparsity_aggregated_class0} and \ref{fig:sparsity_aggregated_class1}, respectively, for the zero-dose and low-dose radiation classes. To simplify notation in what follows, whenever there is no danger of confusion, we drop the subscript $k$ and superscript $(N,r)$ in the transport map notation $\TkNr$ introduced above and in Algorithm \ref{alg:ATM_statistics_1}.


To explain the pictorial representation of such ``aggregated'' sparsity patterns, lets take one subfigure from Figure \ref{fig:sparsity_aggregated_class0}, for example, Subfigure \ref{fig:SP0_10}.  Note the grid point ordering on the plot axes: left to right for the horizontal axis, indexing the random variables $\{\rv{Y}_{j}\}_{j=1}^{6}$, and top to bottom for the vertical axis, indexing the map components $\{\atmcomponent_{i}\}_{i=1}^{6}$.  The plot color bar ranges from 1 to the number $N$, which in this case is 10.  Since, for each $i = 1,\ldots,6$, the $i$-th map component $\atmcomponent_{i}$ must depend on the value $y_{i}$ of the $i$-th random variable $\rv{Y}_{i}$, and we are aggregating results for 10 computed transport maps, the frequency count depicted on each of the diagonal grid points is 10.  Thus, each square on a diagonal grid point is identified with the highest frequency count, per the color bar.  Out of the 10 transport maps computed, three resulted with the 4-th map component having dependence on the value $y_{2}$ of the random variable $\rv{Y}_{2}$.  Hence the square at grid point (2,4) is identified with a frequency count of 3, per the color bar.  Similarly, two out of the 10 transport maps resulted with dependence on the value $y_{1}$ of the random variable $\rv{Y}_{1}$ for the 6-th map component and therefore the square at grid point (1,6) is identified with a frequency count of 2.  Finally, there were three instances of aggregated dependence with frequency count of 1.  These are identified by the white squares (lowest frequency count per the color bar) appearing at grid points (1,5), (1,2), and (4,5).  


Computing frequency counts for (independently computed) larger collections of adaptive transport maps, as in Figures \ref{fig:SP0_10}--\ref{fig:SP0_200} for the zero-dose radiation class and Figures \ref{fig:SP1_10}--\ref{fig:SP1_200} for the low-dose one, we observe that patterns of high-valued frequency counts (pointing to particular subsets of random variables) start to emerge and persist.
We note that a larger collection of adaptive transport maps need not contain the smaller previously independently learned collection of adaptive transport maps, since, per lines \ref {alg:line_perm_act1}--\ref{alg:line_perm_act2} from Algorithm \ref{alg:ATM_statistics_1}, we know that the learning of each adaptive transport map used as a training set a randomly selected subset of the set of all available samples for the given class -- that is indeed the case in Figures \ref{fig:sparsity_aggregated_class0} and \ref{fig:sparsity_aggregated_class1}.

Finally, in order to verify robustness of the results with respect to changes in the ordering of the random variables, the computational experiments described via Figures \ref{fig:sparsity_aggregated_class0}--\ref{fig:sparsity_aggregated_class1} were repeated using $20$ different permutations $\{\ps{p}\}_{p=1}^{20}$ of the ordered set $\Yt \equiv \rvsubset$ of random variables.  The different permutations used are listed in Table \ref{table:rvs_and_genes}, appearing in Appendix \ref{app:ATM_experiments}.  Details about the experiments for each of the permutations are provided in Appendix \ref{app:ATM_experiments}.   Here we note that the persistence of patterns of high-valued frequency counts resulting from the computational experiments, as described in Figures \ref{fig:sparsity_aggregated_class0}--\ref{fig:sparsity_aggregated_class1} for the identity permutation $\ps{1}$ case, was also observed in the case of each of the 19 permutations $\{\ps{p}\}_{p=2}^{20}$.

We thus take the results from the case of $N=200$ runs (\cf. Figures \ref{fig:SP0_200} and \ref{fig:SP1_200}) for each of the $20$ permutations $\{\ps{p}\}_{p=1}^{20}$, and map them (see Algorithm \ref{alg:ATM_statistics_2} from Appendix \ref{app:ATM_experiments}) to the random variable ordering corresponding to the identity permutation $\ps{1}$, so that we can report the aggregated results for these $4,\!000$ runs. 
This is summarized in Figures \ref{fig:sparsity_aggregated_4000_freq}--\ref{fig:sparsity_aggregated_genes}.
The resulting high-valued frequency counts in the aggregated sparsity patterns are suggestive of gene relationships encoded (or hidden) in the data which, as we have already mentioned, may aid in hypothesis generation (as was done in Figure \ref{fig:sparsity_aggregated_genes}) for subsequent validation.  

\newcommand{\geneids}{$\{5533, 5534, 5530, 5532, 4772, 4773\}$}

\section{Discussion}
\label{sec:discussion}

As mentioned in Section \ref{sec:introduction}, one among several advantages of measure transport methods is that they allow for a unified framework for processing and analysis of data distributed according to a wide range of probability measures.  This powerful property is one of the reasons why measure transport techniques are active areas of research up to the present day, despite the fact that the subject itself has a long history.   A framework for probability density estimation supported by measure transport techniques clearly offers benefits for many applications, including those in biology.  For instance, recent work \cite{deTorrente_2020} has shown that improved interpretation of cancer transcriptomic data can result by accounting for differences in the distribution densities of gene expressions.  This is just one example where a framework for probability distribution density estimation, grounded on measure transport methods, could have significant impact. 

Specific to the present manuscript, one of the ultimate goals when studying the \gsedataset\ data set is to understand the global dependence structure between the genes  represented there, through their expression  data interpreted as a set of values of a set of random variables with a joint $d$-variate probability density  $\globalTargetDensity_{k}$ on $\mathbb{R}^{d}$, where $d$ denotes the number of genes in the data set, and $k\in\{0,1,\ldots,K-1\}$ labels the different radiation level classes. The \gsedataset\ data set we used (available from \cite{xihaier_github}) contains samples for $d=10,\!874$ genes. The approach and methods we presented here, which are intended to be inclusive of, although not exclusively for, scenarios where the number of samples available for training is limited in amount (order $O(10)$ samples or less), are very general and to our knowledge have not been pursued in the biology domain. 
(We note, however, that application of other measure transport methods in the biology domain appear in the literature -- for example, see \cite{uhler_2022, cao_2022, 10.1093/bib/bbad130} for joint neural network and optimal transport applications.) 
Combined with present day computational capabilities, the methods being pursued in this paper make feasible the tasks needed to achieve goals such as that of understanding the global dependence structure among a large number of genes (or, more generally, a large number of random variables jointly distributed according to an unknown probability measure). 

For instance, within this framework, the computations and results from Section \ref{sec:ATM_tool_scientific_discovery} can be viewed as follows.  By choosing a subset of $m$ genes (where $m=6$ genes were considered in Section \ref{sec:ATM_tool_scientific_discovery}) we essentially chose a subspace $\mathbb{R}^{m}\subset\mathbb{R}^{d}$ and, by learning adaptive triangular transport maps $T_{k}:\mathbb{R}^{m} \rightarrow \mathbb{R}^{m}$, estimated  target probability densities $\targetDensity_{k}$ on $\mathbb{R}^{m}$.  These estimated target densities can be thought of as approximations to $m$-variate marginals of the (unknown) global $d$-variate probability densities $\globalTargetDensity_{k}$. In other words, we chose a (known) projection $P:\mathbb{R}^{d} \rightarrow \mathbb{R}^{m}$ and estimated the marginals $\nu_{\targetDensity_{k}}=P_{\sharp}{\nu_{\globalTargetDensity_{k}}}$ (transports of $\nu_{\globalTargetDensity_{k}}$ along this projection)  on $\mathbb{R}^{m}$ by constructing triangular transport maps satisfying the measure transport equation $\nu_{\targetDensity_{k}} = (T_k)_{\sharp}\referenceMeasure$, where the reference probability density $\referenceDensity$ on  $\mathbb{R}^{m}$ was chosen to be that of the $m$-variate normal (i.e., Gaussian) measure $\mathcal{N}(0,I)$.

The dependence structure between genes in this small subset (as implied by the computed estimates to the probability distribution densities) was then uncovered by analyzing the sparsity patterns exhibited by the learned adaptive  triangular transport maps. In general, assuming that $\mathbb{R}^{d}$ is equipped with a reference Gaussian measure $\nu_{\Gamma}$, one may address the question of computing  global triangular transport maps  $\mathcal{T}_{k}:\mathbb{R}^{d} \rightarrow \mathbb{R}^{d}$, satisfying the measure transport equation $\nu_{\globalTargetDensity_{k}} = (\mathcal{T}_{k})_{\sharp}\nu_{\Gamma}$ and estimating the (unknown) global probability density $\globalTargetDensity_{k}$, where $\Gamma$ is a $d$-variate Gaussian reference probability density. It will be interesting to understand the sparsity patterns of the global maps $\mathcal{T}_{k}^{-1}$, as their structure will reveal dependencies and strength of correlations between different genes represented in the entire data set.

Note that such an insight should reveal known and unknown information about biological pathways as well, since pathway information should be encoded into the dependency structure of the global target probability densities $\globalTargetDensity_{k}$  and therefore in the sparsity patterns of the global triangular transport maps $\mathcal{T}_{k}^{-1}$.  As already mentioned, in Section \ref{sec:ATM_experiments} we do not assume \emph{a priori} any particular structure of the gene network. For example, graphs that represent (portions of) a gene network are not necessarily acyclic, in general.  Thus, our method and analysis from Section \ref{sec:ATM_tool_scientific_discovery} was not restricted a priori to directed acyclic graphs assumed to represent gene regulatory networks.

The results of adopting this general approach are already  seen in Figure \ref{fig:sparsity_aggregated_genes}, where emerging  complex relations can be observed in the low-dose radiation class within  the small subset of six genes (i.e., random variables) that we studied in detail. Whether to ignore or not the impact of possible feedback loops in the gene network, as well as the overall interpretation of the relationships among genes uncovered from the adaptive transport maps sparsity patterns, will depend on an external expert knowledge and judgment. Extracting local information by estimating the  triangular transport maps ${T}_{k}$ and target densities $\targetDensity_{k}$ on a subset of genes and/or using an already established knowledge about the relations between genes  within a subset  (for example, a pathway known with certainty, which will predetermine the sparsity pattern of  local transport maps and the dependency structure of  local target probability densities) can help in learning the global transport maps $\mathcal{T}_{k}$ and thus in estimating the target densities $\globalTargetDensity_{k}$.  Namely, this prior knowledge can be incorporated into the original mathematical problem (i.e., first equation in the system \eqref{eq:pushforward_global_target} below) as added constraints (second equation in the system \eqref{eq:pushforward_global_target} below):
\begin{eqnarray} 
    \nu_{\globalTargetDensity_{k}} & = & (\mathcal{T}_{k})_{\sharp}\nu_{\Gamma}  \nonumber  \\ 
    P_{\sharp} \left((\mathcal{T}_{k})_{\sharp}\nu_{\Gamma}\right) \label{eq:pushforward_global_target}  & = &  (T_k)_{\sharp}\left(P_{\sharp}\nu_{\Gamma}\right)  \label{eq:pushforward_global_target}  \\
    & \vdots & \nonumber
\end{eqnarray}
where the local map ${T}_{k}$ in \eqref{eq:pushforward_global_target} has been previously learned  and the second equation in the system  \eqref{eq:pushforward_global_target} above (i.e., constraint for the global triangular transport map $\mathcal{T}_{k}$) follows from the commutativity condition $P\circ\mathcal{T}_{k} = T_{k}\circ{P}$.

Furthermore, the results depicted in Figure \ref{fig:sparsity_aggregated_genes} suggest a certain ``dynamics'' on the space of transport maps (reflected in the changes to their sparsity patterns), as the level of radiation for class $k\in\{0,1,\ldots,K-1\}$ changes, revealing also changes in correlations between genes in the subset of genes under consideration. Moreover, as described in \cite{GSE43151_dataset}, one could gather gene expression data within a fixed class $k\in\{0,1,\ldots,K-1\}$ at different instances of time $t=t_{1},t_{2},\ldots,t_{n}$ after an irradiation with a certain dose (labeled by the class $k$) has occurred. In other words, one observes the time dynamics on the space of random variables $\rv{Y}(t)=\left(\rv{Y}_{1}(t),\rv{Y}_{2}(t),\dots,\rv{Y}_{d}(t)\right)\in\mathbb{R}^{d}$, which implies time dynamics on the space of target probability measures $ \nu_{\globalTargetDensity_{k}}(t)$ (or equivalently, on the space of target probability densities ${\globalTargetDensity_{k}}(t)$), which itself induces time dynamics on the space of triangular transport maps $\mathcal{T}_{k}(t)$ estimating the target densities ${\globalTargetDensity_{k}}(t)$, all within the class $k$.

Using such time-dependent gene expression data, one can learn then a sequence of triangular transport maps $\{\mathcal{T}_{k}(t_{l})\}_{l=1}^{n}$, and use them as validation points for a stochastic model describing the dynamics of $\mathcal{T}_{k}(t)$ and thus the correlation time dynamics of the entire gene network after a radiation dose has been applied. Of course, all of the above could be done locally if one restricts the study to a subset of genes, that is, to the time dynamics of a subset of random variables $\rv{Y}(t)=\left(\rv{Y}_{1}(t),\rv{Y}_{2}(t),\dots,\rv{Y}_{m}(t)\right)\in\mathbb{R}^{m}\subset\mathbb{R}^{d}$ and the corresponding local target probability measure $\nu_{\targetDensity_{k}}(t)$ (or target probability density ${\targetDensity_{k}}(t)$), and the estimating local triangular transport map ${T}_{k}(t)$.

In summary, the work reported in this manuscript shows some of the benefits that measure transport techniques will provide as part of the computational workflow from Section \ref{sec:motivating_application}, intended to support research in the biological sciences.  In domains for which it is common to have limited number of samples available for model training, such as in the radiation biology case study from Section \ref{sec:application_GSE_dataset}, engagement with domain scientists will aid in identifying types of prior knowledge (in addition to biological pathway knowledge) which can be incorporated into the formulation of the mathematical problems being solved.   How to incorporate prior knowledge will obviously depend on the type of prior knowledge and problem being addressed but, generally, we mention as common approaches feature selection for model training as well as incorporation of additional constraints in optimization problems being solved.  

Immediate tasks to pursue, in collaboration with domain scientists, include generation of synthetic data sets that will allow design of more controlled computational experiments with the goal of gaining a better (quantitative) understanding of the dependence of the methodologies proposed in Section \ref{sec:ATM_experiments} on the number of samples available for training.  This will aid in carrying out a formal mathematical analysis of the method proposed in Section \ref{sec:ATM_tool_scientific_discovery} as well.  Additional (non-synthetic) data sets for validation of the proposed methodologies also will be sought.

\section*{Acknowledgements}
This work was supported by the U.S. Department of Energy, Office of Science, RadBio program under Award KP1601011/FWP CC121.
VLM thanks Francis J. Alexander who, as a RadBio PI, supported the measure transport approach, as well as for helpful discussions while the research was being carried out.
VLM also thanks 
Ognyan Stoyanov for helpful discussions on measure transport theory and for very useful feedback on preliminary versions of this manuscript,
Ricardo Baptista for helpful discussions on adaptive transport maps and the TransportMaps library,  
Daniel Sharp for helpful discussions on the MParT library, 
and Shinjae Yoo and Shantenu Jha for some managerial assistance.

\section*{Statements and Declarations}
\noindent\textbf{Funding: } This work was supported by the U.S. Department of Energy, Office of Science, RadBio program under Award KP1601011/FWP CC121.
\\
\noindent\textbf{Financial interests: } The authors declare they have no financial interests.
\\
\noindent\textbf{Conflict of interests: } The authors declare they have no potential conflict of interests.
\\
\noindent\textbf{Data Availability: }  The data sets \cite{UCI_banknote_authentication_267, xihaier_github} used in this research are publicly available.

\appendix
\section{Appendix: Adaptive Transport Maps Framework}
\label{app:ATM_framework}

The computational transport maps framework proposed by \cite{baptista2022_ATM}, termed \textit{adaptive transport maps} by the authors, and which we use in the present manuscript in Section \ref{sec:ATM_experiments} and the example from Figure \ref{fig:banknote_data} of the main text as well, is outlined next.  For technical details, the reader is referred to the original publication \cite{baptista2022_ATM} and references therein.  

Recall from Section \ref{sec:triangular_TMs} of the main text that a transport map $T\!:\!\mathbb{R}^{m} \!\rightarrow\! \mathbb{R}^{m}$, such that the measure transport equation $\targetMeasure =  T_{\sharp}\referenceMeasure$ holds, is triangular if the $i$-th component $T_{i}$ of $T$ depends only on the values $\{x_{j}\}_{j=1}^{i}$ of the first $i$ random variables $\{\rv{X}_{j}\}_{j=1}^{i}$:
\begin{eqnarray} \label{eq:app_triangularTM2}
                   \hspace*{-2.0cm}
    T(x)  & = & \left [
            \begin{array}{l}
            T_{1}(x_{1}) \\
            T_{2}(x_{1},x_{2}) \\
              \qquad \vdots   \\
            T_{m}(x_{1},x_{2},\ldots,x_{m})
            \end{array}
            \right ] \, .
\end{eqnarray}
Conversely, and denoting the map components of the inverse $T^{-1}\!:\!\mathbb{R}^{m} \!\rightarrow\! \mathbb{R}^{m}$ of $T$ by $S_{i} \equiv (T^{-1})_{i}, i = 1, \ldots, m$, one has that the $i$-th map component $S_{i}$ depends only on the values $\{y_{j}\}_{j=1}^{i}$ of the first $i$ random variables $\{\rv{Y}_{j}\}_{j=1}^{i}$:
\begin{eqnarray} \label{eq:app_triangularTM2_inverse}
                   \hspace*{-2.0cm}
    S(y)  & = & \left [
            \begin{array}{l}
            S_{1}(y_{1}) \\
            S_{2}(y_{1},y_{2}) \\
              \qquad \vdots   \\
            S_{m}(y_{1},y_{2},\ldots,y_{m})
            \end{array}
            \right ] 
            \, \equiv \ T^{-1}(y) 
            \, .
\end{eqnarray}
Recall also that, to ensure positivity of the densities (refer to equations \eqref{eq:pullback_target} and \eqref{eq:pullback_reference} of the main text), each map component $T_{i}$ in equation \eqref{eq:app_triangularTM2} must be differentiable with respect to $x_{i}$, and we must have that $\partial_{x_{i}} T_{i} > 0$, for every $i=1,\ldots,m$.  Similarly, each component $S_{i}$ in equation  \eqref{eq:app_triangularTM2_inverse} of the inverse map $T^{-1}$ of $T$ must satisfy said properties with respect to the variables $y_{i}$ \cite{bogachev2005, marzouk2016}.

We now move on to describe briefly the method of constructing and computing (the components of) a triangular transport map as it has been implemented in  \cite{baptista2022_ATM}.  
First, we set the stage to define the space(s) of functions whose elements can be used as building blocks for the construction of the components of the transport map and whose bases will be used to construct approximations to the transport map components (via equation \eqref{eq:rectification_operator}, later in this section).

To begin, let $\gamma(\di y)=\frac{1}{\sqrt{2\pi}}e^{-{y^{2}}/{2}}\di y$ be the standard Gaussian measure on $\mathbb{R}$. Consider the space $\mathbb{R}^{\mathbb{N}}$ (sometimes denoted $\mathbb{R}^{\infty}$), where ${\mathbb{N}}=\{1,2,\ldots\}$ is the set of natural numbers, and the set $\mathbb{Z}_{+}^{\mathbb{N}}$, where $\mathbb{Z}_{+}=\{0,1,2,\ldots\}$ is the set of non-negative integers. For $y=(y_{1},y_{2},\ldots)\in\mathbb{R}^{\mathbb{N}}$, the canonical projections $P^{(i)}:\mathbb{R}^{\mathbb{N}}\rightarrow\mathbb{R}^{i}$ are defined as  $P^{(i)}(y)=(y_{1},y_{2},\ldots,y_{i})$.  Similarly, if $\alpha=(\alpha_{1},\alpha_{2},\ldots)\in\mathbb{Z}_{+}^{\mathbb{N}}$, and using the same notation for the canonical projection $P^{(i)}:\mathbb{Z}_{+}^{\mathbb{N}}\rightarrow\mathbb{Z}_{+}^{i}$, we have $P^{(i)}(\alpha)=(\alpha_{1},\alpha_{2},\ldots,\alpha_{i})\equiv\alpha^{(i)}$. 
 Let $\gamma^{\mathbb{N}}=\gamma\times\gamma\times\cdots$ be the infinite product Gaussian measure defined on the Borel $\sigma$-algebra $\mathcal{B}(\mathbb{R}^{\mathbb{N}})$. Then we obtain $P^{(i)}_{\sharp}\left(\gamma^{\mathbb{N}}\right)=\gamma^{i}=\underbrace{\gamma\times\gamma\times\cdots\times\gamma}_{i\, \mathrm{times}}$ to be the $i$-variate Gaussian product measure on the Borel $\sigma$-algebra $\mathcal{B}(\mathbb{R}^{i})$. Let $L^{2}(\mathbb{R}^{i},\gamma^{i})$ be the space of square integrable real valued functions on $\mathbb{R}^{i}$ with respect to $\gamma^{i}$. For all $i\in\mathbb{N}$ consider the Sobolev spaces
\begin{equation}
H^{1}(\mathbb{R}^{i},\gamma^{i})=\left\{f\in L^{2}(\mathbb{R}^{i},\gamma^{i})\mid\forall {s}: 1\le s\le i, \frac{\partial{f}}{\partial{y_{s}}}\in L^{2}(\mathbb{R}^{i},\gamma^{i})\right\}
\end{equation}
with norm,  for every $f\in H^{1}(\mathbb{R}^{i},\gamma^{i})$, defined by
\begin{equation}
\left\|f\right\|_{H^{1}(\mathbb{R}^{i},\gamma^{i})}=\left[\left\|f\right\|^{2}_{L^{2}(\mathbb{R}^{i},\gamma^{i})}+\sum^{i}_{s=1}\left\|\frac{\partial{f}}{\partial{y_{s}}}\right\|^{2}_{L^{2}(\mathbb{R}^{i},\gamma^{i})}\right]^{\frac{1}{2}},
\end{equation}
and the spaces
\begin{equation}  \label{eq:space_Vi}
V_{i}(\mathbb{R}^{i},\gamma^{i})=\left\{f\in L^{2}(\mathbb{R}^{i},\gamma^{i}) \,\mid\, \frac{\partial{f}}{\partial{y_{i}}}\in L^{2}(\mathbb{R}^{i},\gamma^{i}) \right\},
\end{equation}
with norm
\begin{equation}
\left\|f\right\|_{V_{i}(\mathbb{R}^{i},\gamma^{i})}=\left[\left\|f\right\|^{2}_{L^{2}(\mathbb{R}^{i},\gamma^{i})}+\left\|\frac{\partial{f}}{\partial{y_{i}}}\right\|^{2}_{L^{2}(\mathbb{R}^{i},\gamma^{i})}\right]^{\frac{1}{2}}.
\end{equation}
Clearly we have $H^{1}(\mathbb{R}^{i},\gamma^{i})\subset V_{i}(\mathbb{R}^{i},\gamma^{i})\subset L^{2}(\mathbb{R}^{i},\gamma^{i})$ for $i>1$ and $H^{1}(\mathbb{R},\gamma)=V_{1}(\mathbb{R},\gamma)$. For $i>1$ the space $V_{i}(\mathbb{R}^{i},\gamma^{i})$ can be constructed as a tensor product $L^{2}(\mathbb{R}^{i-1},\gamma^{i-1})\otimes H^{1}(\mathbb{R},\gamma)$ equipped with the Hilbert-Schmidt norm, that is, if $f=g\otimes h$, where $g\in L^{2}(\mathbb{R}^{i-1},\gamma^{i-1})$ and $h\in H^{1}(\mathbb{R},\gamma)$, then $\left\|f\right\|_{V_{i}(\mathbb{R}^{i},\gamma^{i})}=\left\|g\otimes h\right\|_{HS}=\left\|g\right\|_{L^{2}(\mathbb{R}^{i-1},\gamma^{i-1})}\left\|h\right\|_{H^{1}(\mathbb{R},\gamma)}$. 

Now, if $n\in\mathbb{Z}_{+}$, let $h_{n}$ be the $n$-th Hermite polynomial on $\mathbb{R}$:
\begin{equation}\label{eq:hermite_polynomials}
h_{n}(y)=\frac{(-1)^{n}}{\sqrt{n!}}e^{{y^{2}}/{2}}\frac{d^{n}}{dy^{n}}\left(e^{-{y^{2}}/{2}}\right).
\end{equation}
The set $\{h_{n}\}_{n\ge 0}$ forms an orthonormal basis on the Hilbert space $L^{2}(\mathbb{R},\gamma)$. Moreover, $\{h_{n}\}_{n\ge 0}$ forms  an orthogonal basis on the Hilbert space $H^{1}(\mathbb{R},\gamma)$ as well, with its inner product
\begin{equation}
\langle f,g\rangle_{H^{1}(\mathbb{R},\gamma)}=\langle f,g\rangle_{L^{2}(\mathbb{R},\gamma)}+\langle f^{\prime},g^{\prime}\rangle_{L^{2}(\mathbb{R},\gamma)},
\end{equation}
for every $f,g\in H^{1}(\mathbb{R},\gamma)$. More generally, the space of polynomials is dense in $L^{2}(\mathbb{R},\gamma)$ and this implies that the space of polynomials on $\mathbb{R}^{i}$ is dense in $L^{2}(\mathbb{R}^{i},\gamma^{i})$, for every $i\in \mathbb{N}$. Therefore, with the latter remark in mind, we take as an orthonormal basis for the Hilbert space $V_{i}(\mathbb{R}^{i},\gamma^{i})\subset L^{2}(\mathbb{R}^{i},\gamma^{i})$ from \eqref{eq:space_Vi} the set of polynomials
\begin{equation}\label{eq:basis_functions}
\left\{\mathcal{\widetilde H}_{\alpha^{(i)}}(y)=h_{\alpha_{1}}(y_{1})h_{\alpha_{2}}(y_{2})\cdots h_{\alpha_{i-1}}(y_{i-1}){\tilde h}_{\alpha_{i}}(y_{i})\mid\alpha^{(i)}\in\mathbb{Z}_{+}^{i},y\in\mathbb{R}^{i} \right\},
\end{equation}
where the polynomials $\prod_{s=1}^{i-1}h_{\alpha_{s}}(y_{s})$, $(\alpha_{1},\alpha_{2},\ldots,\alpha_{i-1})\in\mathbb{Z}_{+}^{i-1}$, form an orthonormal basis of $L^{2}(\mathbb{R}^{i-1},\gamma^{i-1})$ and ${\tilde h}_{\alpha_{i}}(y_{i})= h_{\alpha_{i}}(y_{i})/\sqrt{\alpha_{i}+1}$, $\alpha_{i}\in\mathbb{Z}_{+}$, form an orthonormal basis of $H^{1}(\mathbb{R},\gamma)$ (the latter normalization follows from the property $h_{n+1}^{\prime}=\sqrt{n+1}\,h_{n}$ of Hermite polynomials). 
Then, for each $i = 1,\ldots,m$, the elements of the space $V_{i}(\mathbb{R}^{i},\gamma^{i})$ with its basis \eqref{eq:basis_functions} can be used as  building blocks for constructing the $i$-th component $S_{i}$ of the triangular transport map as well as its approximations.

With the mathematical formalism outlined above, we can now describe the functional approximations used to represent the triangular transport maps.
The map components $\{S_{i}\}_{i=1}^{m}$ in equation \eqref{eq:app_triangularTM2_inverse} are sought via \textit{rectification operators} $\{\mathcal{R}_{i}\}_{i=1}^{m}$ of the form
\begin{equation}  \label{eq:rectification_operator}
    S_{i}(y_{1},\ldots,y_{i})  \, = \,   \mathcal{R}_{i}(f_{i})(y_{1},\ldots,y_{i}) \, = \,  f_{i}(y_{1},\ldots,y_{i-1},0) + \int_{0}^{y_{i}} g\left(\partial_{t}f_{i}(y_{1},\ldots,y_{i-1},t)\right) \, \di t ,
\end{equation}
that take sufficiently smooth (e.g., $f_{i} \in V_{i}(\mathbb{R}^{i},\gamma^{i})$ -- see \eqref{eq:space_Vi}) non-monotone functions $f_{i}: \mathbb{R}^{i} \rightarrow \mathbb{R}$, $i = 1, \ldots, m$, and transform them into monotone  increasing functions $S_{i}$ of their last argument $y_{i}$ (see \eqref{eq:app_triangularTM2_inverse}).  In equation \eqref{eq:rectification_operator} the function $g:\mathbb{R} \rightarrow \mathbb{R}^{+}$ is bijective and positive-valued.  This guarantees that the positivity of the densities (equations \eqref{eq:pullback_target} and \eqref{eq:pullback_reference} of the main text) is not violated, a condition that must be satisfied.  As discussed in \cite{baptista2022_ATM}, the choice of the function $g$ in equation \eqref{eq:rectification_operator} has significant impact on the properties of the optimization problem \eqref{eq:KL_divergence_optimization} (refer to Section \ref{sec:triangular_TMs}  of the main text) solved in order to compute, or learn, the transport map.   The soft-plus function $g(z) = \log(1 + \exp(z))$, whose inverse is $g^{-1}(z) = \log(\exp(z) - 1)$, is one appropriate choice.

For each $i = 1, \ldots, m$, the function $f_{i}$ in equation \eqref{eq:rectification_operator} is represented as an $M_{i}$-term linear combination of basis functions from the set \eqref{eq:basis_functions} (a finite $M_{i}$-dimensional approximation),
\begin{eqnarray}  \label{eq:f_expansion}
   f_{i}(y_{1},\ldots,y_{i})  & \ = \ & \sum_{\alpha^{(i)} \in \Lambda^{(i)}} c_{\alpha^{(i)}} \mathcal{\widetilde H}_{\alpha^{(i)}}(y_{1},\ldots,y_{i}),
\end{eqnarray}
where, for each $i = 1, \ldots, m$:
\begin{itemize}
    \item $\Lambda^{(i)}\subset\mathbb{Z}_{+}^{i}$ is a \emph{finite} set  of multi-indices $\left(\alpha_{1}^{(s)},\alpha_{2}^{(s)},\ldots,\alpha_{i}^{(s)}\right)_{s=1}^{M_{i}}$, where $M_{i}=\mathrm{Card}(\Lambda^{(i)})$, 
    \item $c_{\alpha^{(i)}} \in \mathbb{R}$ are real-valued coefficients (coordinates with respect to the basis), which are to be learned from data, and
    \item $\mathcal{\widetilde H}_{\alpha^{(i)}}:\mathbb{R}^{i} \rightarrow \mathbb{R}$ are the basis functions of the Hilbert space $V_{i}(\mathbb{R}^{i},\gamma^{i})$ previously described (refer to \eqref{eq:basis_functions}).
\end{itemize}
Finally, let us give an outline of the adaptive transport map algorithm applied in our computations to learn the transport maps. 
The sets (of active indices which identify the sets of active variables -- refer to Figure \ref{fig:TM_sparsity_example} of the main text) $\Lambda^{(i)}$ in equation \eqref{eq:f_expansion} are constructed by a greedy algorithm inspired by the work from \cite{10.1137/18M1198387}. Namely, starting with a set $\Lambda^{(i)}_{0}=\emptyset$ one constructs a sequence $\{\Lambda^{(i)}_{t}\}_{t\ge0}$ of \emph{downward closed} sets. A set $\Lambda^{(i)}_{t}\subset\mathbb{Z}_{+}^{i}$
is downward closed if $\alpha^{(i)} \in \Lambda^{(i)}_{t}$ and $\alpha^{\prime(i)}\le\alpha^{(i)}\Longrightarrow\alpha^{\prime(i)}\in\Lambda^{(i)}_{t}$, where $\alpha^{\prime(i)}\le\alpha^{(i)}$ means $\alpha^{\prime}_{s}\le\alpha^{}_{s}$, for $1\le s\le i$. The \emph{reduced margin} $\mathcal{RM}(\Lambda^{(i)}_{t})$ of $\Lambda^{(i)}_{t}$ is defined as
\begin{equation}
\mathcal{RM}(\Lambda^{(i)}_{t})=\left\{\alpha^{(i)}\notin\Lambda^{(i)}_{t}\mid\alpha^{(i)}-\epsilon^{(i)}_{(s)}\in\Lambda^{(i)}_{t},\forall s:1\le s\le i, \alpha_{s}\ne 0\right\},
\end{equation}
where the multi-index $\epsilon^{(i)}_{(s)}=(0,\ldots,1_{s},\ldots,0)\in\mathbb{Z}_{+}^{i}$ has the only non-zero entry equal to $1$ at position $s$. If $\Lambda^{(i)}_{t}$ is downward closed, then $\Lambda^{(i)}_{t}\cup\{\alpha^{(i)}_{t}\}$ is downward closed for any $\alpha^{(i)}_{t}\in\mathcal{RM}(\Lambda^{(i)}_{t})$. Then every consecutive iteration $\Lambda^{(i)}_{t+1}=\Lambda^{(i)}_{t}\cup\{\alpha^{*(i)}_{t}\}$ is constructed with $\alpha^{*(i)}_{t}\notin\Lambda^{(i)}_{t}$ and $\alpha^{*(i)}_{t}\in\mathcal{RM}(\Lambda^{(i)}_{t})$ such that it achieves the \emph{best} next step approximation $f_{i}^{t+1}$ to $f_{i}$ in \eqref{eq:f_expansion}. 

Cross-validation \cite{bk:hastie2009elements} is employed to determine the maximal cardinalities $M_{i}^{\mathrm{max}}$ of each set $\Lambda^{(i)}_{t}$ (that is, the step $t=M_{i}^{\mathrm{max}}$ at which to stop adaptation, for each $i=1,\ldots,m$), without overfitting the training data. 
This results in values $M_{i} = \mathrm{Card}(\Lambda^{(i)}) \le M_{i}^{\mathrm{max}}$ and approximations $f_{i}$ which are optimally \emph{adapted} to the size of the training data set. The full details are provided in \cite{baptista2022_ATM}. The greedy, incremental search for values $M_{i} = \mathrm{Card}(\Lambda^{(i)}) \le M_{i}^{\mathrm{max}}$ of the cardinalities of the sets $\Lambda^{(i)}$ leads to numerical efficiency in the algorithm (as the number of unknowns to determine in equation \eqref{eq:f_expansion} grows quickly with increasing $i$ and $\mathrm{Card}(\Lambda^{(i)})$), as well as numerical stability in the computations, including when training with a limited number of data samples.

\section{Appendix: Supplementary Material for Computational Experiments with Adaptive Transport Maps}
\label{app:ATM_experiments}

\begin{figure}[h]
    \centering
    \subfloat[][]{{{\includegraphics[trim=0.0cm 0.0cm 0.0cm 0.0cm, clip=true, width=0.47\textwidth]{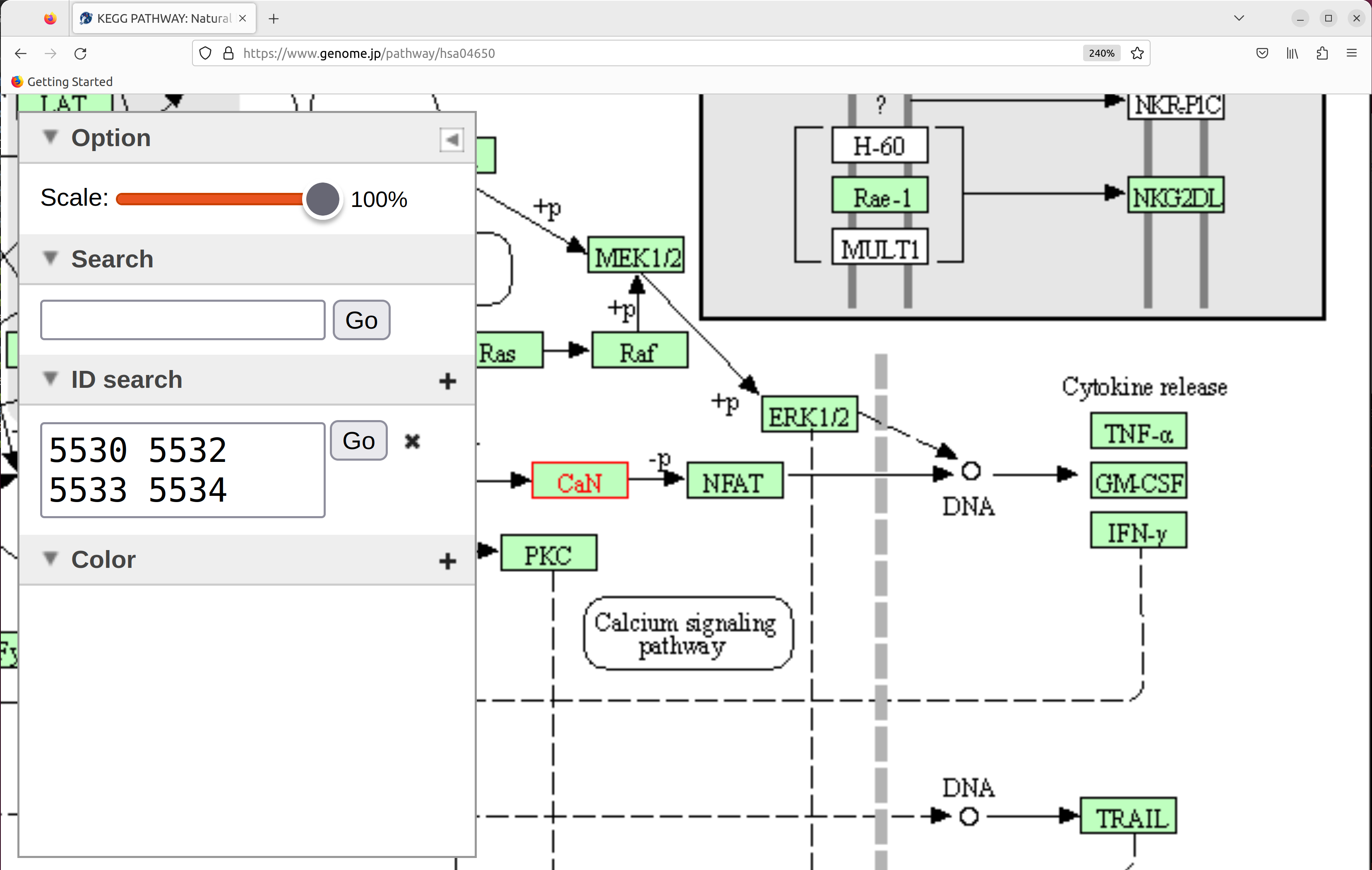}}}
    \label{fig:kegg_hsa04650_CaN}}
    \hspace{2ex}
    \subfloat[][]{{{\includegraphics[trim=0.0cm 0.0cm 0.0cm 0.0cm, clip=true, width=0.47\textwidth]{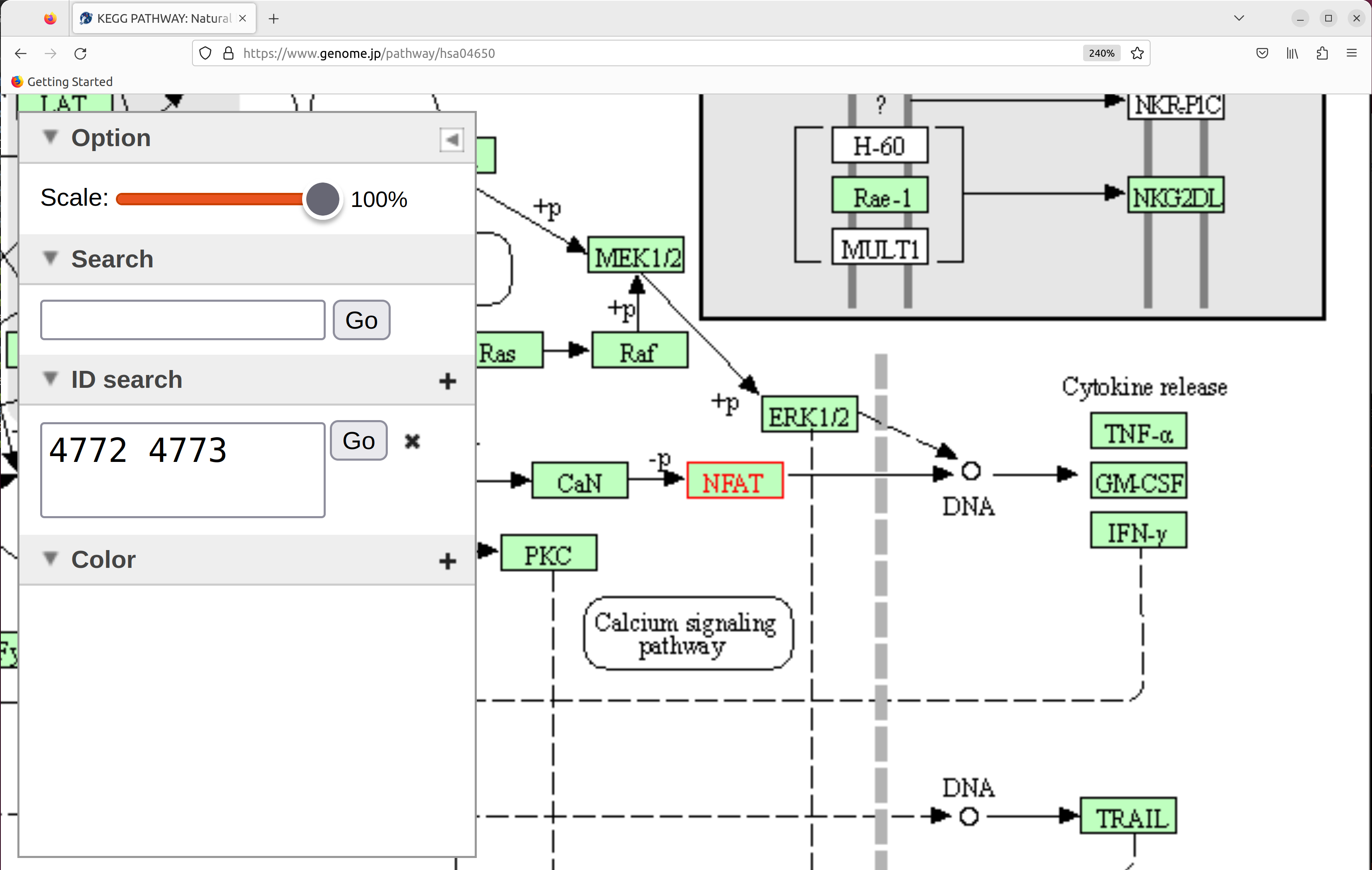}}}
    \label{fig:kegg_hsa04650_NFAT}}
    \caption{Segment of  pathway \pathwayzlSD\ containing genes 5530, 5532, 5533, 5534 (Subfigure \ref{fig:kegg_hsa04650_CaN}, CaN gene product, highlighted in red) and genes 4772, 4773 (Subfigure \ref{fig:kegg_hsa04650_NFAT}, NFAT gene product, highlighted in red).  Source of search tool and screenshots: https://www.genome.jp/pathway/hsa04650.  KEGG pathway map notation: https://www.genome.jp/kegg/document/help\_pathway.html
    } 
    \label{fig:kegg_hsa04650_segment}
\end{figure}

For the computational experiments reported in this Appendix and in Section \ref{sec:ATM_tool_scientific_discovery} of the main text, which employ adaptive transport maps \cite{baptista2022_ATM,ATM_library}, we work with a subset of six genes from those present both in the \gsedataset\ data set available from \cite{xihaier_github} and pathway \pathwayzlSD\ from the KEGG database \cite{KEGG_database}.  This pathway also ranked as one of the top pathways showing significant differential activation in response to low-dose radiation, as compared to zero-dose, based on the pathway ranking procedure from \cite{luo2023_frontiers}.  We hence use it for selecting genes (associated with the random variables) for the present computational experiments.  The selected subset of genes and additional information about them appear in Table \ref{table:nih_genes} and Figure \ref{fig:kegg_hsa04650_segment}.
\begin{table}[h]
\centering
\begin{tabular} { | c | c | l |  }
\hline
Gene ID  &  Official Symbol &  \multicolumn{1}{c|} {Official Full Name} \\
\hline \hline
5530 & PPP3CA & protein phosphatase 3 catalytic subunit alpha \\
5532 & PPP3CB & protein phosphatase 3 catalytic subunit beta \\
5533 & PPP3CC & protein phosphatase 3 catalytic subunit gamma \\
5534 & PPP3R1 & protein phosphatase 3 regulatory subunit B, alpha \\
4472 & NFATC1 & nuclear factor of activated T cells 1 \\
4473 & NFATC2 & nuclear factor of activated T cells 2 \\
\hline
\end{tabular}
\caption{Gene IDs with the corresponding official symbol and full name.  Source of information: https://www.ncbi.nlm.nih.gov/gene}
\label{table:nih_genes}
\end{table}


%
\begin{table}[h!]
\newcommand{\tblentry}[7]{$\ps{{#1}}$ & $\rv{Y}_{{#2}}$ & $\rv{Y}_{{#3}}$ & $\rv{Y}_{{#4}}$ & $\rv{Y}_{{#5}}$ & $\rv{Y}_{{#6}}$ & $\rv{Y}_{{#7}}$ \\}
\centering
\begin{tabular*}{0.9\textwidth}{@{\extracolsep\fill} | c || c | c | c | c | c | c | }
\hline
& \multicolumn{6} {  c | }{Gene ID} \\
\multicolumn{1} { | l || }{$\Yt \equiv \rvsubset$} & 5533 & 5534 & 5530 & 5532 & 4472 & 4473 \\
\hline \hline
\tblentry{ 1}{1}{2}{3}{4}{5}{6}
\tblentry{ 2}{4}{3}{2}{1}{5}{6}
\tblentry{ 3}{1}{2}{3}{4}{6}{5}
\tblentry{ 4}{4}{3}{2}{1}{6}{5}
\tblentry{ 5}{6}{5}{4}{3}{2}{1}
\tblentry{ 6}{5}{6}{1}{2}{3}{4}
\tblentry{ 7}{1}{2}{5}{3}{6}{4}
\tblentry{ 8}{6}{5}{3}{4}{1}{2}
\tblentry{ 9}{3}{2}{6}{1}{5}{4}
\tblentry{10}{3}{1}{2}{5}{6}{4}
\tblentry{11}{6}{5}{4}{2}{3}{1}
\tblentry{12}{6}{4}{2}{1}{3}{5}
\tblentry{13}{4}{5}{3}{1}{6}{2}
\tblentry{14}{5}{2}{1}{6}{4}{3}
\tblentry{15}{3}{5}{2}{4}{1}{6}
\tblentry{16}{4}{3}{1}{6}{5}{2}
\tblentry{17}{2}{4}{1}{6}{5}{3}
\tblentry{18}{6}{4}{2}{5}{3}{1}
\tblentry{19}{6}{3}{4}{2}{1}{5}
\tblentry{20}{4}{2}{5}{6}{1}{3}
\hline
\end{tabular*}
\caption{Different permutations $\{\ps{p}\}_{p=1}^{20}$ of the gene IDs from Table \ref{table:nih_genes}, used  for the computational experiments reported in this Appendix and in Section \ref{sec:ATM_tool_scientific_discovery} of the main text.  For each $p=1,2,\ldots,20$, we have $\ps{p} = \rvSubset{\psInv{p}{j}}{j}$.  Note that $\sigma_{1}$ is the identity.
}  
\label{table:rvs_and_genes}
\end{table}

As indicated in Section \ref{sec:ATM_tool_scientific_discovery} of the main text, in order to assess the robustness of the results obtained from Algorithm \ref{alg:ATM_statistics_1} with respect to ordering of the random variables, we carried out the computational experiments described in Section \ref{sec:ATM_tool_scientific_discovery} and, in particular, summarized via Figures \ref{fig:sparsity_aggregated_class0}--\ref{fig:sparsity_aggregated_class1} of the main text, using $20$ different permutations $\{\ps{p}\}_{p=1}^{20}$ of the ordered set $\Yt \equiv \rvsubset$ of random variables.  Specifically, Algorithm \ref{alg:ATM_statistics_1} was executed for each of the $20$ permutations $\{\ps{p}\}_{p=1}^{20}$ listed in Table \ref{table:rvs_and_genes}.  The results were then analyzed individually for each permutation, as well as collectively after mapping the results obtained for each of the $20$ permutations $\{\ps{p}\}_{p=1}^{20}$ to the random variable ordering corresponding to the identity permutation $\ps{1}$.  The outcome of this exercise is summarized next and in Figure \ref{fig:sparsity_aggregated_4000_freq} of the main text.

\begin{figure}[h!]
    \centering
    \subfloat[][]{{{\includegraphics[trim=0.0cm 0.0cm 0.0cm 0.0cm, clip=true, width=0.30\textwidth]{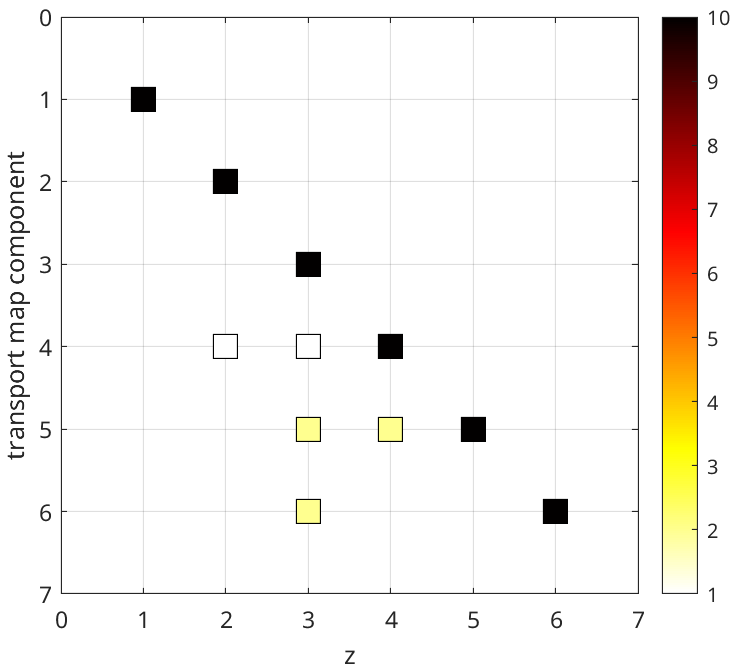}}}
    \label{fig:sig20_SP0_10}}
    \hspace{2ex}
    \subfloat[][]{{{\includegraphics[trim=0.0cm 0.0cm 0.0cm 0.0cm, clip=true, width=0.30\textwidth]{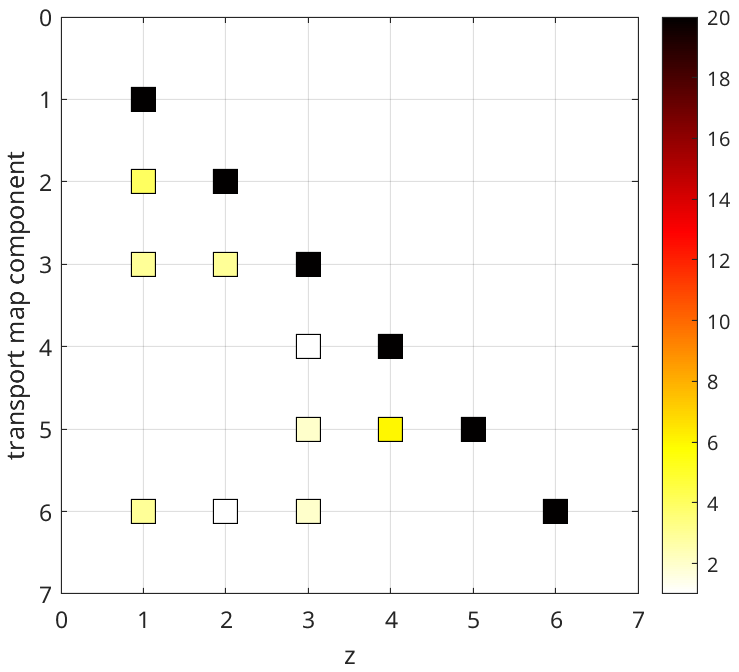}}}
    \label{fig:sig20_SP0_20}}
    \hspace{2ex}
    \subfloat[][]{{{\includegraphics[trim=0.0cm 0.0cm 0.0cm 0.0cm, clip=true, width=0.30\textwidth]{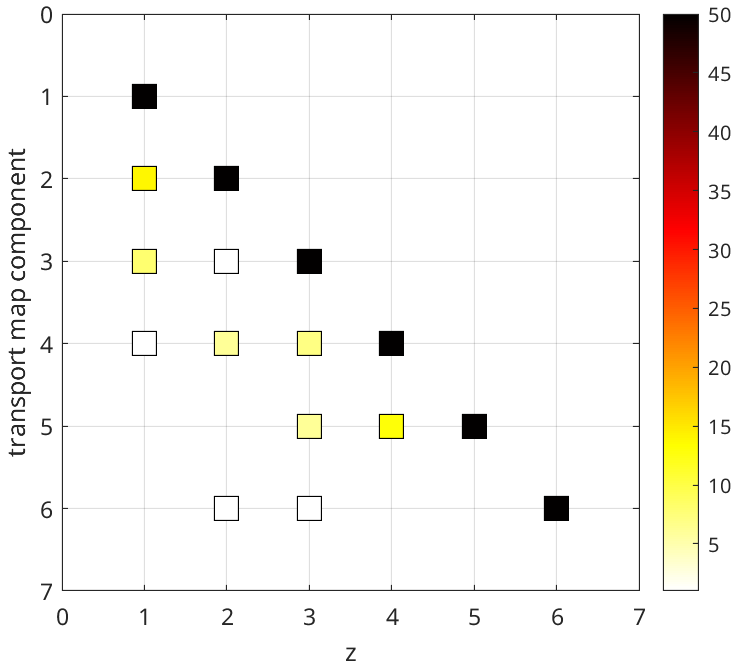}}}
    \label{fig:sig20_SP0_50}}
    \\
    \subfloat[][]{{{\includegraphics[trim=0.0cm 0.0cm 0.0cm 0.0cm, clip=true, width=0.30\textwidth]{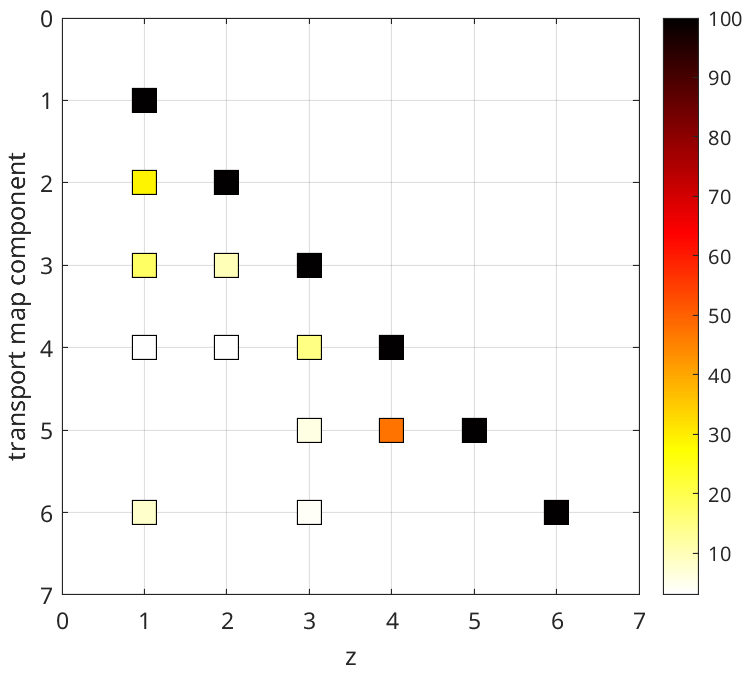}}}
    \label{fig:sig20_SP0_100}}
    \hspace{2ex}
    \subfloat[][]{{{\includegraphics[trim=0.0cm 0.0cm 0.0cm 0.0cm, clip=true, width=0.30\textwidth]{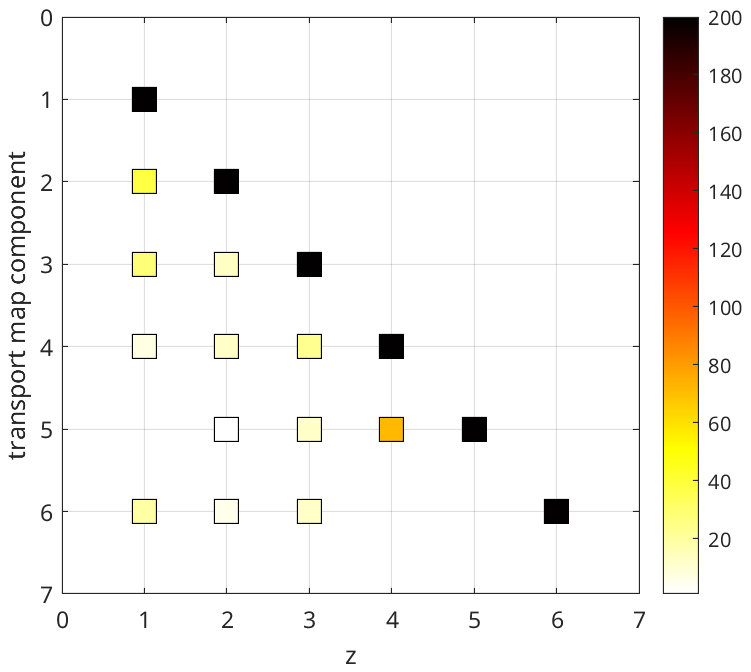}}}
    \label{fig:sig20_SP0_200}}
    \hspace{0ex} 
    \subfloat[][]{\footnotesize \addtolength{\tabcolsep}{-3.0pt} \renewcommand{\arraystretch}{1.2}
        \begin{tabular*}{0.32\textwidth}[b]{ l | r r r r r r } 
        $S_1$ & 200 &   0 &    0 &   0  &   0  &   0 \\
        $S_2$ &  38 & 200 &    0 &   0  &   0  &   0 \\
        $S_3$ &  28 &  13 &  200 &   0  &   0  &   0 \\
        $S_4$ &   7 &  12 &   23 & 200  &   0  &   0 \\
        $S_5$ &   0 &   1 &   12 &  71  & 200  &   0 \\
        $S_6$ &  19 &   5 &   12 &   0  &   0  & 200 \\
        \hline 
              & $\rv{Z}_{1}$ & $\rv{Z}_{2}$ & $\rv{Z}_{3}$ & $\rv{Z}_{4}$ & $\rv{Z}_{5}$ & $\rv{Z}_{6}$
        \\ \multicolumn{7}{c}{$\mathcal{Z} \equiv \ps{20}$}
        \end{tabular*} 
    \label{fig:sig20_SP0_200_freq}}
    \caption{Aggregated sparsity patterns from adaptive transport map approximations to data distributions for different subsets of the zero-dose radiation data samples. 
    The permutation on the ordered set $\Yt \equiv \rvsubset$ of random variables for the computational experiments summarized in Subfigures \ref{fig:sig20_SP0_10}--\ref{fig:sig20_SP0_200_freq} is that listed as permutation $\ps{20}$ in Table \ref{table:rvs_and_genes}.  As in Figure \ref{fig:sparsity_aggregated_class0} of the main text, subsets of samples consisting of half the number of total samples (which for the zero-dose radiation class is \totalsampleszero) were generated from random permutations of the \gsedataset\ data set sample ordering.  In Subfigures \ref{fig:sig20_SP0_10}, \ref{fig:sig20_SP0_20}, \ref{fig:sig20_SP0_50}, \ref{fig:sig20_SP0_100}, and \ref{fig:sig20_SP0_200} are represented the sparsity patterns corresponding to, respectively, 10, 20, 50, 100, and 200 such sample subsets.  By aggregating results for increasing numbers of sample subsets, patterns of high-valued frequency counts are seen to emerge and persist.  
    For reference, Subfigure \ref{fig:sig20_SP0_200_freq} shows the values for the frequency counts corresponding to the sparsity pattern from Subfigure \ref{fig:sig20_SP0_200}.
    In other words, the matrix shown in Subfigure \ref{fig:sig20_SP0_200_freq} is the matrix $\FCperm{0}{(200,\sigmap{20})}$ from Algorithm \ref{alg:ATM_statistics_1}, lines \ref{alg:start_B_loop}--\ref{alg:end_B_loop}, resulting from the execution of Algorithm \ref{alg:ATM_statistics_1} using permutation $\ps{20}$.
    To simplify notation in Subfigure \ref{fig:sig20_SP0_200_freq}, we denote $S \equiv T^{-1}$ for the inverse $T^{-1}$ of a transport map $T$. 
    } 
    \label{fig:sparsity_aggregated_sigma20_class0}
\end{figure}

\begin{figure}[h!]
    \centering
    \subfloat[][]{{{\includegraphics[trim=0.0cm 0.0cm 0.0cm 0.0cm, clip=true, width=0.30\textwidth]{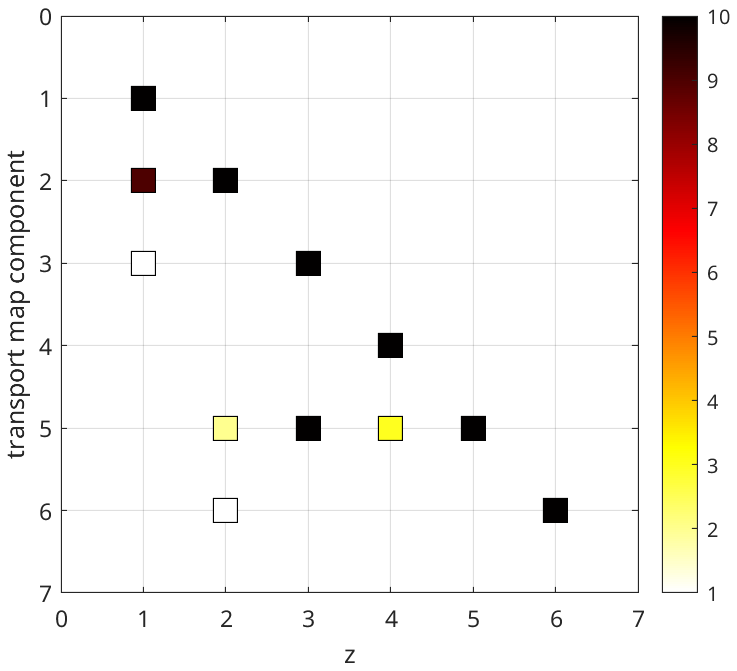}}}
    \label{fig:sig20_SP1_10}}
    \hspace{2ex}
    \subfloat[][]{{{\includegraphics[trim=0.0cm 0.0cm 0.0cm 0.0cm, clip=true, width=0.30\textwidth]{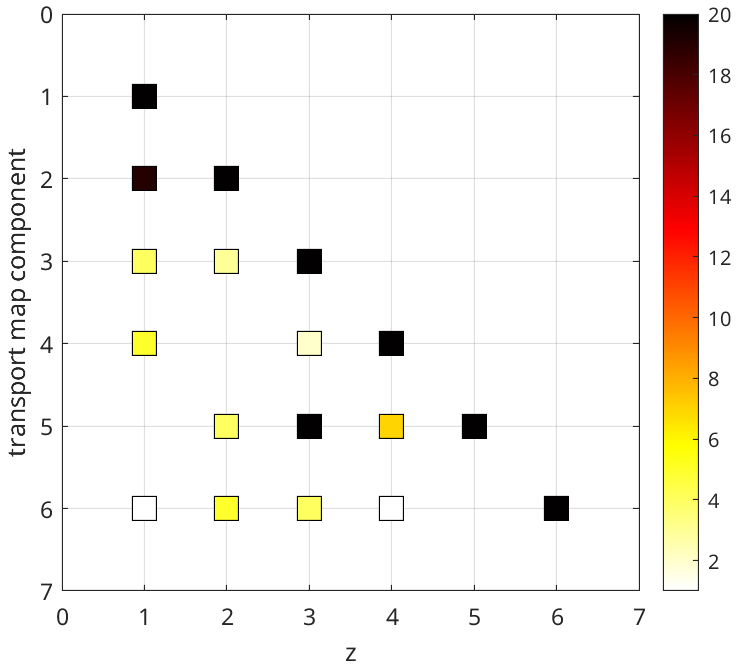}}}
    \label{fig:sig20_SP1_20}}
    \hspace{2ex}
    \subfloat[][]{{{\includegraphics[trim=0.0cm 0.0cm 0.0cm 0.0cm, clip=true, width=0.30\textwidth]{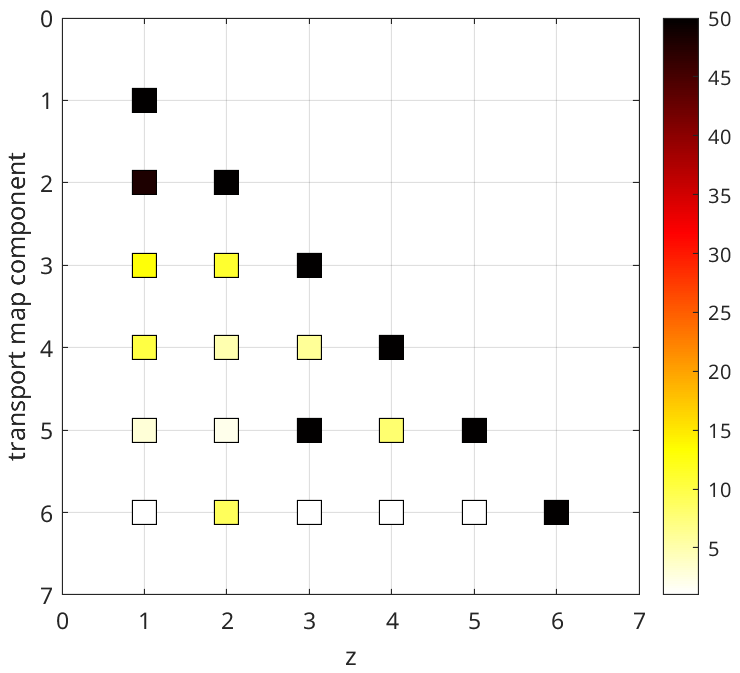}}}
    \label{fig:sig20_SP1_50}}
    \\
    \subfloat[][]{{{\includegraphics[trim=0.0cm 0.0cm 0.0cm 0.0cm, clip=true, width=0.30\textwidth]{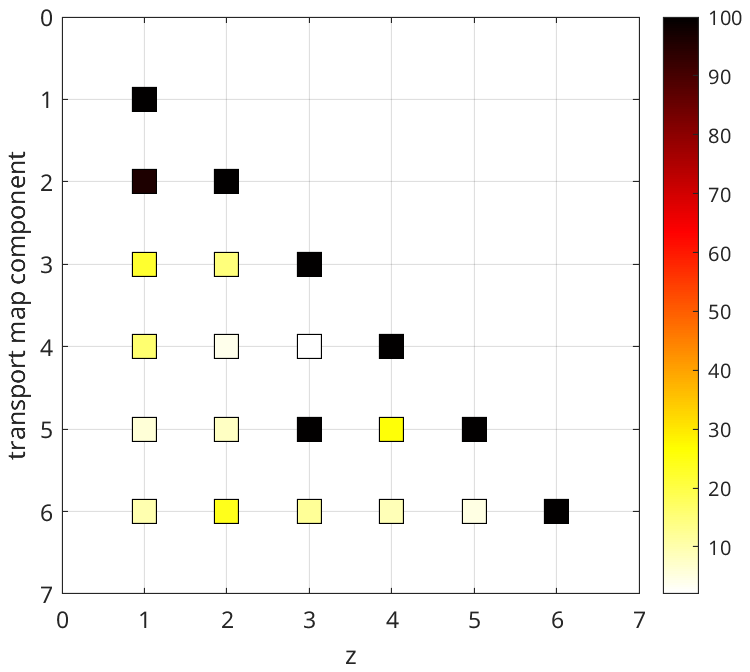}}}
    \label{fig:sig20_SP1_100}}
    \hspace{2ex}
    \subfloat[][]{{{\includegraphics[trim=0.0cm 0.0cm 0.0cm 0.0cm, clip=true, width=0.30\textwidth]{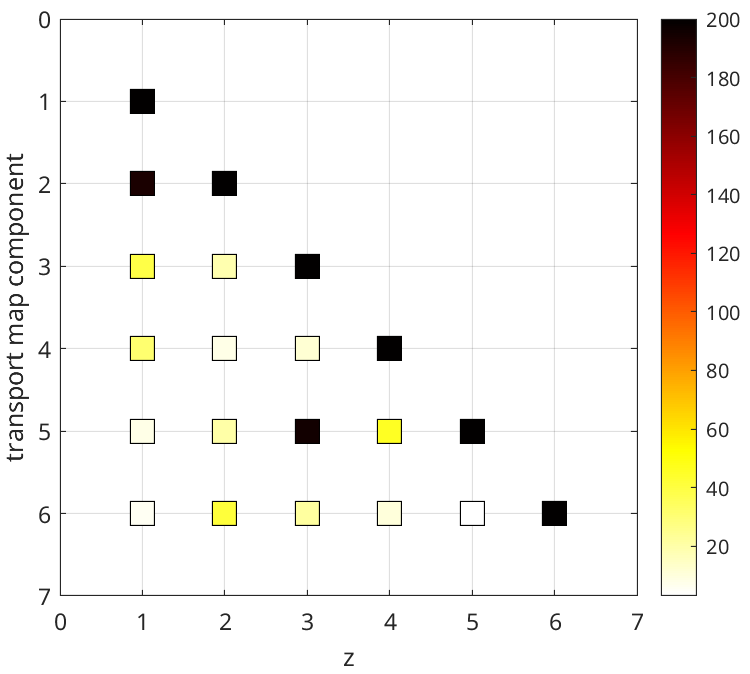}}}
    \label{fig:sig20_SP1_200}}
    \hspace{0ex} 
    \subfloat[][]{\footnotesize \addtolength{\tabcolsep}{-3.0pt} \renewcommand{\arraystretch}{1.2}
        \begin{tabular*}{0.32\textwidth}[b]{ l | r r r r r r } 
        $S_1$ & 200 &   0 &    0 &   0  &   0  &   0 \\
        $S_2$ & 193 & 200 &    0 &   0  &   0  &   0 \\
        $S_3$ &  39 &  19 &  200 &   0  &   0  &   0 \\
        $S_4$ &  31 &   8 &   12 & 200  &   0  &   0 \\
        $S_5$ &   8 &  20 &  195 &  46  & 200  &   0 \\
        $S_6$ &   6 &  41 &   22 &  10  &   3  & 200 \\
        \hline 
              & $\rv{Z}_{1}$ & $\rv{Z}_{2}$ & $\rv{Z}_{3}$ & $\rv{Z}_{4}$ & $\rv{Z}_{5}$ & $\rv{Z}_{6}$
        \\ \multicolumn{7}{c}{$\mathcal{Z} \equiv \ps{20}$}
        \end{tabular*} 
    \label{fig:sig20_SP1_200_freq}}
    \caption{Aggregated sparsity patterns from adaptive transport map approximations to data distributions for different subsets of the low-dose radiation data samples.  
    The permutation on the ordered set $\Yt \equiv \rvsubset$ of random variables for the computational experiments summarized in Subfigures \ref{fig:sig20_SP1_10}--\ref{fig:sig20_SP1_200_freq} is that listed as permutation $\ps{20}$ in Table \ref{table:rvs_and_genes}.  
    As in Figure \ref{fig:sparsity_aggregated_class1} of the main text, subsets of samples consisting of half the number of total samples (which for the low-dose radiation class is \totalsampleslow) were generated from random permutations of the \gsedataset\ data set sample ordering.  In Subfigures \ref{fig:sig20_SP1_10}, \ref{fig:sig20_SP1_20}, \ref{fig:sig20_SP1_50}, \ref{fig:sig20_SP1_100}, and \ref{fig:sig20_SP1_200} are represented sparsity patterns corresponding to, respectively, 10, 20, 50, 100, and 200 such sample subsets.  As for the zero-dose radiation class in Figure \ref{fig:sparsity_aggregated_sigma20_class0}, by aggregating results for increasing numbers of sample subsets, patterns of high-valued frequency counts are seen to emerge and persist. 
    For reference, Subfigure \ref{fig:sig20_SP1_200_freq} shows the values for the frequency counts corresponding to the sparsity pattern from Subfigure \ref{fig:sig20_SP1_200}.
    In other words, the matrix shown in Subfigure \ref{fig:sig20_SP1_200_freq} is the matrix $\FCperm{1}{(200,\sigmap{20})}$ from Algorithm \ref{alg:ATM_statistics_1}, lines \ref{alg:start_B_loop}--\ref{alg:end_B_loop}, resulting from the execution of Algorithm \ref{alg:ATM_statistics_1} using permutation $\ps{20}$.
    To simplify notation in Subfigure \ref{fig:sig20_SP1_200_freq}, we denote $S \equiv T^{-1}$ for the inverse $T^{-1}$ of a transport map $T$.
    } 
    \label{fig:sparsity_aggregated_sigma20_class1}
\end{figure}

Firstly, persistence of patterns of high-valued frequency counts resulting from the computational experiments, as described in Figures \ref{fig:sparsity_aggregated_class0}--\ref{fig:sparsity_aggregated_class1} for the permutation $\ps{1}$ case, was also observed when executing Algorithm \ref{alg:ATM_statistics_1} for each of the $19$ permutations $\{\ps{p}\}_{m=2}^{20}$.  As one example illustrating this, we display in Figures \ref{fig:sparsity_aggregated_sigma20_class0}--\ref{fig:sparsity_aggregated_sigma20_class1}, for the permutation $\ps{20}$ case, the analogues of Figures \ref{fig:sparsity_aggregated_class0}--\ref{fig:sparsity_aggregated_class1} from the main text.  Note, however, that the patterns from Figure \ref{fig:sparsity_aggregated_class0} versus the corresponding ones from Figure  \ref{fig:sparsity_aggregated_sigma20_class0} (similarly, Figure \ref{fig:sparsity_aggregated_class1} versus Figure \ref{fig:sparsity_aggregated_sigma20_class1}) are different because each were computed for different permutations (namely, $\ps{1}$ and $\ps{20}$) of the random variables.

So, to more easily compare and collect the results obtained for the $20$ different permutations $\{\ps{p}\}_{p=1}^{20}$, we map the statistics gathered for each of the permutations to the random variable ordering corresponding to the identity permutation $\ps{1} = \rvsubset$.  Precisely, given the matrices $\FCperm{k}{(N,\sigmap{p})}$ (see line \ref{alg:line_freq_count} of Algorithm \ref{alg:ATM_statistics_1}) containing the frequency counts resulting from the transport maps sparsity patterns in a series of $N$ adaptive transport maps learned for each class $k \in \{0,1\}$ and, within each class, for the permutations $\{\ps{p}\}_{p=1}^{20}$, we define new matrices, as described next.

First, for each of the matrices $\FCperm{k}{(N,\sigmap{p})}$ (from line \ref{alg:line_freq_count} of Algorithm \ref{alg:ATM_statistics_1}), we create a matrix
$\FCpermId{k}{(N,\sigmap{p})}=\sigmapinv{p}\left(\FCperm{k}{(N,\sigmap{p})}\right)$,which is a lower triangular matrix storing the frequency counts from the given matrix $\FCperm{k}{(N,\sigmap{p})} \,$, rearranged after mapping from the ordering of the random variables corresponding to the permutation $\ps{p}$ to that for the identity permutation $\ps{1} = \rvsubset$. That is, 
\begin{eqnarray}  \label{eq:freq_counts_mapped_to_id}
    \FCpermId{k}{(N,\sigmap{p})}(j,i)  & = &  \FCperm{k}{(N,\sigmap{p})}\left(\min(\sigmap{p}(j),\sigmap{p}(i)),\max(\sigmap{p}(j),\sigmap{p}(i)\right)
\end{eqnarray}
(see lines \ref{alg:line_jtilde}--\ref{alg:line_strength_rvs} of Algorithm \ref{alg:ATM_statistics_2}).  An entry $\FCpermId{k}{(N,\sigmap{p})}(j,i)$ in the lower triangular portion of the matrix $\FCpermId{k}{(N,\sigmap{p})}$ thus contains a frequency count reflecting a dependency ``strength'' between the random variables $\rv{Y}_{j}$ and $\rv{Y}_{i}$, obtained after learning a series of $N$ transport maps with the (particular) ordering $\ps{p}$ of the random variables, after being mapped back (i.e., rearranged) so that it can be compared with the matrix of frequency counts obtained from the learning of $N$ transport maps with the ordering $\ps{1}$ of the random variables, or with another matrix of frequency counts corresponding to another ordering  $\ps{q}$, $q \ne p,$ of the random variables,  after an analogous rearrangement (via the matrices $\FCpermId{k}{(N,\sigmap{p})}$) from the ordering $\ps{q}$ to the ordering $\ps{1}$ of the random variables.  This rearrangement simplifies comparison of results between separate sets of runs for each one of the $20$ different permutations $\{\ps{p}\}_{p=1}^{20}$.  

\begin{figure}[h!]
    \centering
    \subfloat[][$\FCpermId{0}{(200,\sigmap{1})}$]{{{\includegraphics[trim=0.0cm 0.0cm 0.0cm 0.0cm, clip=true, width=0.40\textwidth]{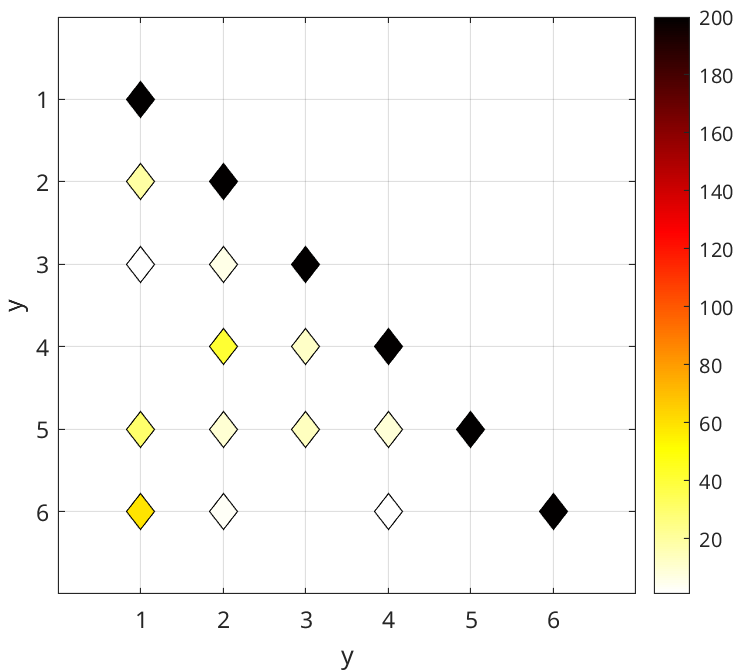}}}
    \label{fig:sig1_SP0_200}}
    \hspace{2ex} 
    \subfloat[][$\FCpermId{0}{(200,\sigmap{1})}$]{\footnotesize \addtolength{\tabcolsep}{-1.0pt} \renewcommand{\arraystretch}{1.2}
        \begin{tabular*}{0.42\textwidth}[b]{ l | r r r r r r } 
        $\rv{Y}_{1}$ & 200 &   0 &   0 &   0 &   0 &   0 \\
        $\rv{Y}_{2}$ &  19 & 200 &   0 &   0 &   0 &   0 \\
        $\rv{Y}_{3}$ &   1 &   6 & 200 &   0 &   0 &   0 \\
        $\rv{Y}_{4}$ &   0 &  41 &  12 & 200 &   0 &   0 \\
        $\rv{Y}_{5}$ &  30 &  10 &  14 &   9 & 200 &   0 \\
        $\rv{Y}_{6}$ &  59 &   3 &   0 &   1 &   0 & 200 \\
        \hline 
              & $\rv{Y}_{1}$ & $\rv{Y}_{2}$ & $\rv{Y}_{3}$ & $\rv{Y}_{4}$ & $\rv{Y}_{5}$ & $\rv{Y}_{6}$
        \\ \multicolumn{7}{l}{ }
        \\ \multicolumn{7}{l}{ }
        \end{tabular*} 
    \label{fig:sig1_SP0_200_freqcnt}}
    \\
    \subfloat[][$\FCpermId{0}{(200,\sigmap{20})}$]{{{\includegraphics[trim=0.0cm 0.0cm 0.0cm 0.0cm, clip=true, width=0.40\textwidth]{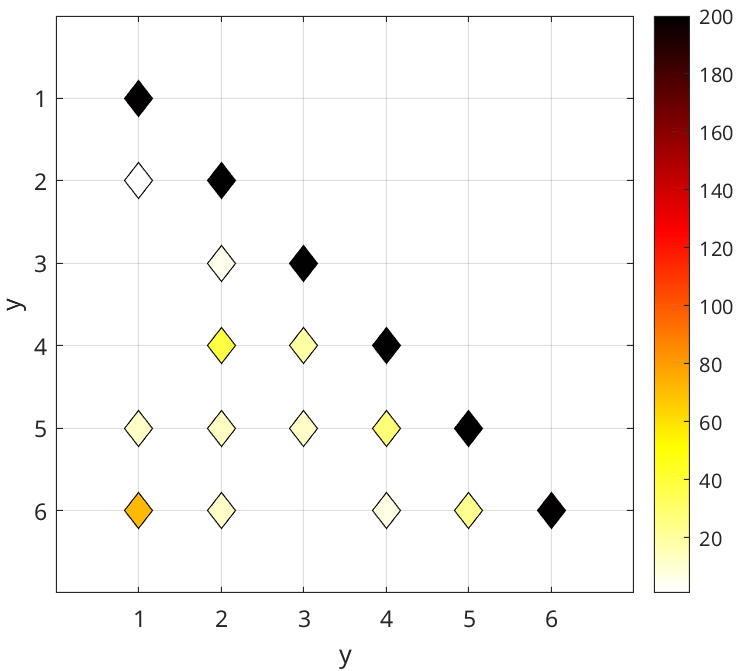}}}
    \label{fig:sig20_invSP0_200}}
    \hspace{2ex} 
    \subfloat[][$\FCpermId{0}{(200,\sigmap{20})}$]{\footnotesize \addtolength{\tabcolsep}{-1.0pt} \renewcommand{\arraystretch}{1.2}
        \begin{tabular*}{0.42\textwidth}[b]{ l | r r r r r r } 
        $\rv{Y}_{1}$ & 200 &   0 &   0 &   0 &   0 &   0 \\
        $\rv{Y}_{2}$ &   1 & 200 &   0 &   0 &   0 &   0 \\
        $\rv{Y}_{3}$ &   0 &   5 & 200 &   0 &   0 &   0 \\
        $\rv{Y}_{4}$ &   0 &  38 &  19 & 200 &   0 &   0 \\
        $\rv{Y}_{5}$ &  12 &  13 &  12 &  28 & 200 &   0 \\
        $\rv{Y}_{6}$ &  71 &  12 &   0 &   7 &  23 & 200 \\
        \hline 
              & $\rv{Y}_{1}$ & $\rv{Y}_{2}$ & $\rv{Y}_{3}$ & $\rv{Y}_{4}$ & $\rv{Y}_{5}$ & $\rv{Y}_{6}$
        \\ \multicolumn{7}{l}{ }
        \\ \multicolumn{7}{l}{ }
        \end{tabular*} 
    \label{fig:sig20_invSP0_200_freqcnt}}
    \caption{Aggregated sparsity patterns and associated frequency counts from adaptive transport map approximations to data distributions for different subsets of the zero-dose radiation data samples.  
    The assignment of genes to the ordered set $\Yt \equiv \rvsubset$ of random variables for the computational experiments summarized in Subfigures \ref{fig:sig1_SP0_200}--\ref{fig:sig1_SP0_200_freqcnt} is that listed as permutation $\ps{1}$ in Table \ref{table:rvs_and_genes}, whereas for Subfigures \ref{fig:sig20_invSP0_200}--\ref{fig:sig20_invSP0_200_freqcnt} it is that listed as permutation $\ps{20}$.
    To facilitate comparison, the frequency counts shown above were mapped to the random variable ordering corresponding to the identity permutation $\ps{1}$, as described in Algorithm \ref{alg:ATM_statistics_2}.  Specifically, the matrices shown in Subfigures \ref{fig:sig1_SP0_200_freqcnt} and \ref{fig:sig20_invSP0_200_freqcnt} resulted from line \ref{alg:line_strength_rvs} of Algorithm \ref{alg:ATM_statistics_2}.
    Good qualitative agreement is observed between the high-valued frequency counts for $\FCpermId{0}{(200,\sigmap{1})}$ (Subfigure \ref{fig:sig1_SP0_200}) and those for $\FCpermId{0}{(200,\sigmap{20})}$ (Subfigure \ref{fig:sig20_invSP0_200}), in spite of quantitative differences (particularly for low-valued frequency counts).
    } 
    \label{fig:sparsity_aggregated_sigma1_sigma20_zero_dose}
\end{figure}

\begin{figure}[h!]
    \centering
    \subfloat[][$\FCpermId{1}{(200,\sigmap{1})}$]{{{\includegraphics[trim=0.0cm 0.0cm 0.0cm 0.0cm, clip=true, width=0.40\textwidth]{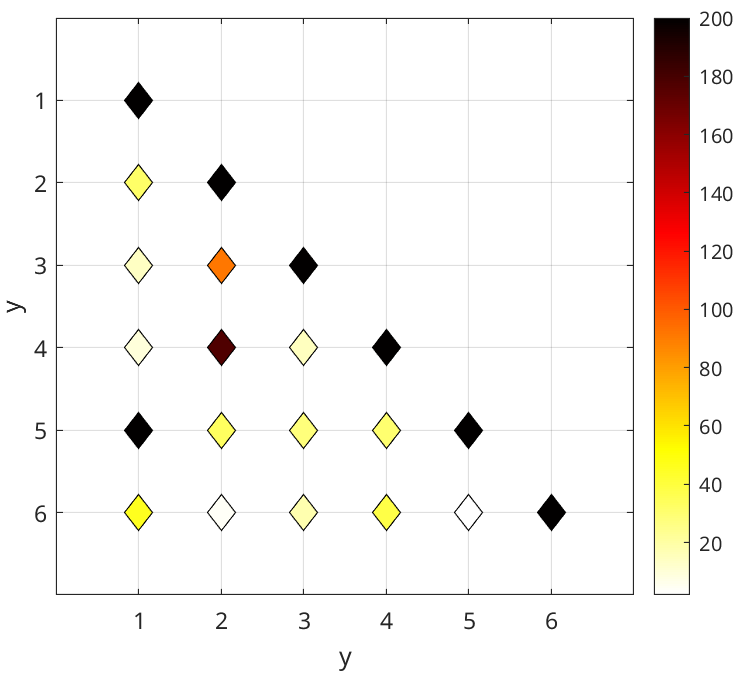}}}
    \label{fig:sig1_SP1_200}}
    \hspace{2ex} 
    \subfloat[][$\FCpermId{1}{(200,\sigmap{1})}$]{\footnotesize \addtolength{\tabcolsep}{-1.0pt} \renewcommand{\arraystretch}{1.2}
        \begin{tabular*}{0.42\textwidth}[b]{ l | r r r r r r } 
        $\rv{Y}_{1}$ & 200 &   0 &   0 &   0 &   0 &   0 \\
        $\rv{Y}_{2}$ &  32 & 200 &   0 &   0 &   0 &   0 \\
        $\rv{Y}_{3}$ &  14 &  91 & 200 &   0 &   0 &   0 \\
        $\rv{Y}_{4}$ &   9 & 177 &  15 & 200 &   0 &   0 \\
        $\rv{Y}_{5}$ & 200 &  33 &  28 &  30 & 200 &   0 \\
        $\rv{Y}_{6}$ &  46 &   4 &  18 &  38 &   2 & 200 \\
        \hline 
              & $\rv{Y}_{1}$ & $\rv{Y}_{2}$ & $\rv{Y}_{3}$ & $\rv{Y}_{4}$ & $\rv{Y}_{5}$ & $\rv{Y}_{6}$
        \\ \multicolumn{7}{l}{ }
        \\ \multicolumn{7}{l}{ }
        \end{tabular*} 
    \label{fig:sig1_SP1_200_freqcnt}}
    \\
    \subfloat[][$\FCpermId{1}{(200,\sigmap{20})}$]{{{\includegraphics[trim=0.0cm 0.0cm 0.0cm 0.0cm, clip=true, width=0.40\textwidth]{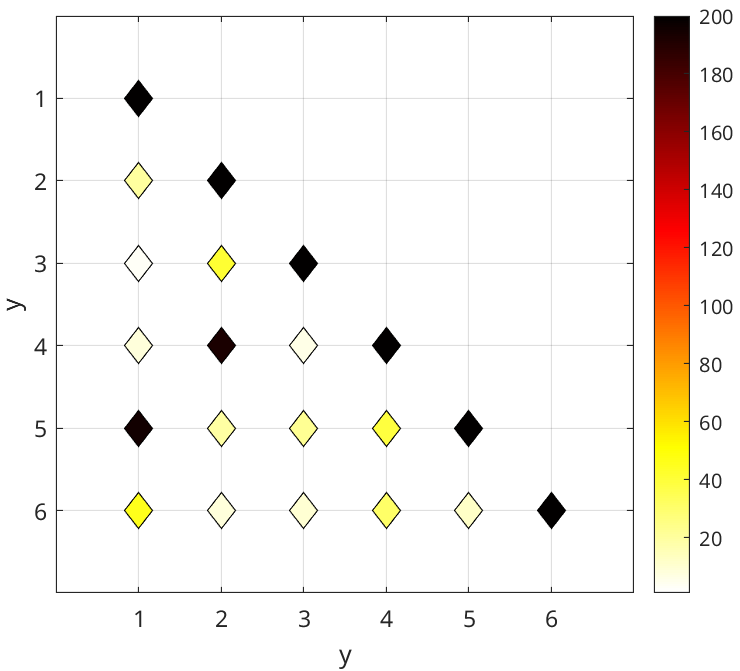}}}
    \label{fig:sig20_invSP1_200}}
    \hspace{2ex} 
    \subfloat[][$\FCpermId{1}{(200,\sigmap{20})}$]{\footnotesize \addtolength{\tabcolsep}{-1.0pt} \renewcommand{\arraystretch}{1.2}
        \begin{tabular*}{0.42\textwidth}[b]{ l | r r r r r r } 
        $\rv{Y}_{1}$ & 200 &   0 &   0 &   0 &   0 &   0 \\
        $\rv{Y}_{2}$ &  20 & 200 &   0 &   0 &   0 &   0 \\
        $\rv{Y}_{3}$ &   3 &  41 & 200 &   0 &   0 &   0 \\
        $\rv{Y}_{4}$ &   8 & 193 &   6 & 200 &   0 &   0 \\
        $\rv{Y}_{5}$ & 195 &  19 &  22 &  39 & 200 &   0 \\
        $\rv{Y}_{6}$ &  46 &   8 &  10 &  31 &  12 & 200 \\
        \hline 
              & $\rv{Y}_{1}$ & $\rv{Y}_{2}$ & $\rv{Y}_{3}$ & $\rv{Y}_{4}$ & $\rv{Y}_{5}$ & $\rv{Y}_{6}$
        \\ \multicolumn{7}{l}{ }
        \\ \multicolumn{7}{l}{ }
        \end{tabular*} 
    \label{fig:sig20_invSP1_200_freqcnt}}
    \caption{Aggregated sparsity patterns and associated frequency counts from adaptive transport map approximations to data distributions for different subsets of the low-dose radiation data samples.  
    The assignment of genes to the ordered set $\Yt \equiv \rvsubset$ of random variables for the computational experiments summarized in Subfigures \ref{fig:sig1_SP1_200}--\ref{fig:sig1_SP1_200_freqcnt} is that listed as permutation $\ps{1}$ in Table \ref{table:rvs_and_genes}, whereas for Subfigures \ref{fig:sig20_invSP1_200}--\ref{fig:sig20_invSP1_200_freqcnt} it is that listed as permutation $\ps{20}$.
    To facilitate comparison, the frequency counts shown above were mapped to the random variable ordering corresponding to the identity permutation $\ps{1}$, as described in Algorithm \ref{alg:ATM_statistics_2}.  Specifically, the matrices shown in Subfigures \ref{fig:sig1_SP1_200_freqcnt} and \ref{fig:sig20_invSP1_200_freqcnt} resulted from line \ref{alg:line_strength_rvs} of Algorithm \ref{alg:ATM_statistics_2}.
    Good qualitative agreement is observed between the high-valued frequency counts for $\FCpermId{1}{(200,\sigmap{1})}$ (Subfigure \ref{fig:sig1_SP1_200}) and those for $\FCpermId{1}{(200,\sigmap{20})}$ (Subfigure \ref{fig:sig20_invSP1_200}), in spite of quantitative differences (particularly for low-valued frequency counts).
    } 
    \label{fig:sparsity_aggregated_sigma1_sigma20_low_dose}
\end{figure}

An example of such rearrangement is provided in Figures \ref{fig:sparsity_aggregated_sigma1_sigma20_zero_dose}--\ref{fig:sparsity_aggregated_sigma1_sigma20_low_dose} for the instance $N\!=\!200$ of runs for permutations $\ps{1}$ and $\ps{20}$, where one can observe (qualitatively) good  agreement between the resulting patterns of high-valued frequency counts. We observed (qualitatively) the same patterns with each one of the remaining  permutations of random variables listed in Table \ref{table:rvs_and_genes}, after learning series of $N\!\in\!\{10,20,50,100,200\}$ adaptive transport maps and the subsequent rearrangement (in the matrices $\FCpermId{k}{(N,\sigmap{p})}$ -- see line \ref{alg:line_strength_rvs} of Algorithm \ref{alg:ATM_statistics_2}) to the random variable ordering corresponding to the identity permutation $\ps{1}$. This shows robustness of the numerical method and its \emph{invariance} with respect to the action of elements of the group of permutations $\mathcal{S}_{6}$ when identifying the dependence structure of the set of random variables $\{\rv{Y}_{j}\}_{j=1}^{6}$ as inferred from the sparsity patterns of the learned adaptive transport maps or, equivalently, from the structure of the corresponding estimated target densities.

Finally, for each class $k \in \{0,1\}$, we create a matrix
$\FCall{k}{(\widetilde{N})}$, which is a lower triangular matrix accumulating the frequency counts from the matrices $\FCpermId{k}{(N,\sigmap{p})}$ defined above.  That is, for a given value $N$ of the number of transport maps in a series of learned adaptive transport maps and for each class $k \in \{0, 1\}$, 
     \begin{eqnarray}  \label{eq:freq_counts_all}
         \FCall{k}{(\widetilde{N})}  & = &  \sum_{p=1}^{\nperm} \FCpermId{k}{(N,\sigmap{p})} \, ,
     \end{eqnarray}
     where $\nperm$ denotes the number of permutations on the ordering of the random variables (see Table \ref{table:rvs_and_genes}) and $\widetilde{N} = N \times \nperm$.
     This summarizes frequency counts resulting from execution of Algorithm \ref{alg:ATM_statistics_1} for all permutations $\{\ps{p}\}_{p=1}^{20}$ listed in Table \ref{table:rvs_and_genes}, collectively (see line \ref{alg:line_accumulate_all} from Algorithm \ref{alg:ATM_statistics_2}), as illustrated in Figure \ref{fig:sparsity_aggregated_4000_freq} of the main text.
The above described matrices $\FCpermId{k}{(N,\sigmap{p})}$ \eqref{eq:freq_counts_mapped_to_id} and $\FCall{k}{(\widetilde{N})}$ \eqref{eq:freq_counts_all} are created as outlined in Algorithm \ref{alg:ATM_statistics_2}.


\newpage
\begin{algorithm}
\begin{algorithmic}[1]
\State \textbf{Notation:}
\State \hspace*{2ex} $m \ge 1$: number of random variables
\State \hspace*{2ex} $\nperm$: number of permutations $\ps{p}$ of the random variables
\State \hspace*{2ex} $N$: number of transport maps in a series of learned adaptive transport maps
\State \hspace*{2ex} $\widetilde{N}$: the value $N \times \nperm$
\State \hspace*{2ex} $K$: number of classes
\State \hspace*{2ex} $\FCperm{k}{(N,\sigmap{p})}$: $m \times m$ lower triangular matrix storing frequency counts from sparsity patterns for 
    \State \hspace*{5ex} $N$ adaptive transport maps learned for class $k$ and permutation $\ps{p}$
    \State \hspace*{5ex} (see line \ref{alg:line_freq_count} of Algorithm \ref{alg:ATM_statistics_1})
\State \textbf{Algorithm:}
\For{$k=0$ \mto\ $K-1$}   
    \State Initialize $\FCall{k}{(\widetilde{N})}$ to the $m \times m$ zero matrix
    \For{$p=1$ \mto\ $\nperm$}
        \State Initialize $\FCpermId{k}{(N,\sigmap{p})}$ to the $m \times m$ zero matrix
        \For{$j=1$ \mto\ $m$}
            \For{$i=1$ \mto\ $j$}
               \State $\jj = \min(\sigmap{p}(j),\sigmap{p}(i))$  \label{alg:line_jtilde}
               \State $\ii = \max(\sigmap{p}(j),\sigmap{p}(i))$  \label{alg:line_itilde}
               \State $\FCpermId{k}{(N,\sigmap{p})}(j,i) \ = \ \FCperm{k}{(N,\sigmap{p})}(\jj,\ii)$  \label{alg:line_strength_rvs}
                       \Comment{\acomment{frequency count reflecting dependency \\ \hspace*{3.85in}
                       ``strength'' between random variables \\ \hspace*{3.85in}
                       $\rv{Y}_{j}$ and $\rv{Y}_{i}$, out of $N$ runs for \\ \hspace*{3.85in}
                       permutation $\ps{p}$}}
               \State $\FCall{k}{(\widetilde{N})}(j,i) \ = \ \FCall{k}{(\widetilde{N})}(j,i) + \FCpermId{k}{(N,\sigmap{p})}(j,i)$  \label{alg:line_accumulate_all}
                       \Comment{\acomment{accumulate frequency counts \\ \hspace*{4.45in}
                       from line \ref{alg:line_strength_rvs}}}
            \EndFor  
        \EndFor  
    \EndFor  
\EndFor  
\end{algorithmic}
\caption{Mapping statistics gathered from results for a number of permutations $\{\ps{p}\}_{p=1}^{\nperm}$ of the random variables to the random variable ordering corresponding to the identity permutation $\ps{1}$}
\label{alg:ATM_statistics_2}
\end{algorithm}

\clearpage
\bibliography{refs}
\bibliographystyle{plainnat}

\end{document}